\documentclass[
 superscriptaddress,
reprint,
 aps,
 prb,
 nofootinbib, 
floatfix,
]{revtex4-2}
\usepackage{graphicx}
\usepackage{dcolumn}
\usepackage{bm}
\usepackage{natbib}
\usepackage{graphicx}
\usepackage{epsfig}
\usepackage{amsfonts} 
\usepackage{mathtools}
\usepackage{amsmath} 
\usepackage{amssymb}
\usepackage{diagbox}
\usepackage{dsfont}
\usepackage[dvipsnames]{xcolor}
\usepackage{bm}
\usepackage{braket}
\usepackage{color}
\usepackage{physics} 
\usepackage{bbm}
\usepackage{hyperref}
\usepackage{verbatim}
\usepackage{cancel}
\hypersetup{
    colorlinks=true,
    linkcolor=blue,
    citecolor=magenta,
    filecolor=magenta,      
    urlcolor=blue,
    pdftitle={},
    pdfpagemode=FullScreen,
    }

\usepackage{comment}
\usepackage[export]{adjustbox}  
\usepackage{enumitem}
\usepackage[titletoc, title]{appendix}
\usepackage{titletoc}
\usepackage[normalem]{ulem}

\usepackage{graphicx} 
\usepackage[margin=1in]{geometry} 
\usepackage{amsthm}
\usepackage{float}
\usepackage{soul}
\usepackage{bbm}
\usepackage[english]{babel}

\usepackage{tikz}
\usetikzlibrary{arrows.meta,decorations.pathmorphing,decorations.markings,backgrounds,positioning,fit,petri}

\newcommand{\Z}{{\mathbb Z}}

\newcommand{\beq}{\begin{equation}\begin{aligned}}
\newcommand{\eeq}{\end{aligned}\end{equation}}

\begin{document}


\title{Classification of Thouless pumps with non-invertible symmetries and implications for Floquet phases}
\author{Yabo Li}
\affiliation{
Center for Quantum Phenomena, Department of Physics,
New York University, 726 Broadway, New York, New York, 10003, USA
}
\author{Matteo Dell'Acqua}
\affiliation{
Center for Cosmology and Particle Physics, Department of Physics,
New York University, 726 Broadway, New York, New York, 10003, USA
}
\author{Aditi Mitra}
\affiliation{
Center for Quantum Phenomena, Department of Physics,
New York University, 726 Broadway, New York, New York, 10003, USA
}

\begin{abstract}
We study symmetry preserving adiabatic and Floquet dynamics of one-dimensional systems. Using quasiadiabatic evolution, we establish a correspondence between adiabatic cycles and invertible defects generated by spatially truncated Thouless pump operators. Employing the classification of gapped phases by module categories, we show that the Thouless pumps are classified by the group of autoequivalences of the module category. We then explicitly construct Thouless pump operators for minimal lattice models with $\text{Vec}_G$, Rep($G$), and Rep($H$) symmetries, and show how the Thouless pump operators have the group structure of autoequivalences. The Thouless pump operators, together with Hamiltonians with gapped ground states, are then used to construct Floquet drives. An analytic solution for the Floquet phase diagram characterized by winding numbers is constructed when the Floquet drives obey an Onsager algebra. Our approach points the way to a general connection between distinct Thouless pumps and distinct families of Floquet phases.  
\end{abstract}

\maketitle
{
  \hypersetup{linkcolor=black}
  \tableofcontents
}

\section{Introduction}

Since the classic work by Thouless~\cite{thouless1983quantization}, the 
``Thouless pump'' has been used to describe the general phenomenon in which quantized charges are transported without a net external force as the potential undergoes an adiabatic cyclic evolution. For particle-number conserving $U(1)$ symmetry, the charge transported per cycle is related to the Chern number~\cite{simon1983holonomy}, hence the adiabatically driven system can be thought of as a dynamical realization of the integer quantum Hall effect~\cite{thouless1982quantized,niu1985quantized,hastings2015quantization,bachmann2020many,kapustin2020hall}. Due to this topological nature, the Thouless pump is robust against perturbations and disorders~\cite{niu1984quantised,nakajima2021competition}, enabling demonstrations in the laboratory~\cite{kraus2012topological,verbin2015topological,lohse2016thouless,nakajima2016topological,lu2016geometrical}; see Ref.~\cite{citro2023thouless} for a review. The studies of Thouless pumps have since been generalized to various dimensions and symmetries~\cite{fu2006time,teo2010topological,berg2011quantized,kapustin2020higher,kuno2021topological,hsin2020berry,bachmann2024classification,ohyama2022generalized,shiozaki2022adiabatic,aasen2022adiabatic,wen2023flow,debray2023long,jones2025charge}. 

The Thouless pump has a natural consequence on the homotopy of the adiabatic cycle. That is, if a non-zero charge is transported as the Hamiltonian undergoes an adiabatic evolution along some cycle, this cycle cannot be smoothly shrunk into a constant  point in the parameter space of the gapped Hamiltonian. Kitaev used such a pumping picture to propose that the inequivalent adiabatic cycles of short-range entangled (SRE) states with symmetry $G$ in $d+1$ dimension are given by SRE states with symmetry $G$ in $d$ dimension~\cite{kitaev2011SRE,kitaev2013SRE}. In particular, the spaces of SRE states $\{M_{d}^{G}\}$ form an $\Omega$-spectrum (the loop space of $M_{d+1}^{G}$ corresponds to $M_{d}^{G}$), leading to a classification theory of invertible states~\cite{xiong2018minimalist,gaiotto2019symmetry,freed2021reflection}. Following this path, Ref.~\cite{wen2023flow} obtained a generalized cohomology theory of invertible
phases over a general parameter space. In an explicit and rigorous setting, Ref.~\cite{bachmann2024classification} proved the classification of Thouless pumps in invertible states with on-site compact group $G$ symmetry in 1d bosonic systems, which agrees with Kitaev's proposal.

In this work, we will focus on the adiabatic cycles of gapped Hamiltonians in 1d bosonic systems whose ground states are not necessarily SRE symmetry protected topological (SPT) states. To be more precise, our setup is a one-dimensional chain $\mathcal{H} = \bigotimes_i \mathcal{H}_{\rm loc}^{(i)}$ with a finite dimensional local Hilbert space at each site, and a smooth family of Hamiltonians that depend on a parameter $\theta\in [0,2\pi]$,
\beq
    H(\theta) = \sum_i H_i(\theta),
\eeq
where each term $H_i(\theta)$ is supported locally near site $i$, and is smoothly dependent on $\theta$. The Hamiltonian $H(\theta)$ has a gap in the spectrum above the ground states that is uniformly larger than a constant $\Delta>0$ for any $\theta\in [0,2\pi]$. The Hamiltonian $H(0)$ and $H(2\pi)$ have the \textit{same ground subspace}. We note that although we use the adiabatic change of gapped Hamiltonians to define a cycle, it is the ground subspaces $\{P(\theta)\}$ that are closed under the adiabatic evolution, rather than the whole spectrum because the Hamiltonians $H(0)\neq H(2\pi)$ in general.

Furthermore, we require that $H(\theta)$ respect some non-invertible symmetry $\mathcal{D}$. The non-invertible symmetries are described by fusion categories~\cite{2006Kitaev,etingof2015tensor}, which generalize the group-like symmetries when some symmetry operators do not have their inverse. In short, a fusion category $\mathcal{D}$ is specified by a set of simple objects $\{a\}$ with fusion rules
\beq
    a\cdot b = \sum_{c}N^{ab}_c c,
\eeq
where $N^{ab}_c$ are non-negative integers. The fusion of simple objects satisfies associativity up to an $F$-matrix, and these obey the pentagon identities for consistency. The fusion rules and $F$-matrices are the characterizing data for a fusion category. The realization of a fusion category as a non-invertible symmetry implies a (locality preserving) representation of these data on the Hilbert space $\mathcal{H}$. A fusion category $\mathcal{D}$ can be realized if and only if it is integral~\cite{evans2025operator}.
Usually, the realizations of the fusion category symmetries are given by matrix product operators (MPO)~\cite{hauru2016topological,bultinck2017anyons,lootens2021matrix,inamura2022lattice,garre2023classifying,lootens2023dualities,meng2024non}, which will also be used for the lattice models in this work. See Ref.~\cite{Fendley20,schafer2024ictp,shao2023s} for more thorough reviews of non-invertible symmetries.

Unlike group-like symmetries, the realizations of non-invertible symmetries are sensitive to the spatial dimension; therefore, the spaces of symmetric states in different dimensions do not form an $\Omega$-spectrum as for group-like symmetries. In Ref.~\cite{inamura20241+}, a Thouless pump is defined for the matrix product state (MPS) representations of SPT phases via certain violations of the $2\pi$-periodicity of the action tensors for the non-invertible symmetry. A classification of inequivalent Thouless pumps in gapped phases is conjectured. 

In the following, we will start from the adiabatic cycles of gapped Hamiltonians on the spin chain, and define the spatially truncated quasi-adiabatic evolution, whose action on the ground states gives rise to some invertible defect at the cut. The defects are generated by spatially truncated unitaries, and therefore they are invertible because inverse of the unitaries always removes them. This will be made precise later. This approach aligns with  Kitaev's pumping picture, and is similar to the `edge' characterization of charge transport used in Ref.~\cite{bachmann2024classification}. We are able to show that the homotopy classes of adiabatic cycles are in one-to-one correspondence with the pumping of invertible defects. With the help of the correspondence between the gapped phases under fusion category $\mathcal{D}$ symmetry and the module categories over $\mathcal{D}$, which follows from MPS methods~\cite{garre2023classifying} and topological field theory~\cite{thorngren2024fusion,inamura2021topological}, we obtain the result that classification of Thouless pumps in a gapped phase under non-invertible symmetries is given by the group of $\mathcal{D}$-autoequivalences for the corresponding module category.

To be more explicit, for the SPT phases protected by a group-like symmetry $G$, the Thouless pumps are classified by group cohomology $\Gamma = H^1(G,U(1))$. For the SPT phases protected by a non-invertible Rep($G$) symmetry corresponding to the pair $(K,\alpha)$, where $K$ is a subgroup of the finite group $G$ and $\alpha$ is a 2-cocycle of $K$, the classification of Thouless pumps is $\Gamma = G'$, a group extension of $N_G(K,[\alpha])/K$ by $H^1(K,U(1))$. Here $N_G(K,[\alpha])$ is the subgroup of $N_G(K)$, the normalizer of $K$ in $G$, whose conjugation action leaves the cohomology class of $\alpha$ invariant. A general anomaly-free non-invertible symmetry can be described by the representation category of a semisimple Hopf algebra $H$, with its canonical SPT phase having Thouless pumps classified by $\Gamma=\mathcal{G}(H)$. Our results agree with the conjecture in Ref.~\cite{inamura20241+}. We then present the lattice models for the gapped phases, and study the SPT phases. We construct the adiabatic cycles of Hamiltonians on a lattice and show that the groups formed by pumped invertible defects agree with the known result in mathematics about the group of $\mathcal{D}$-autoequivalences. Finally, we study the dynamics beyond adiabatic evolutions. Namely, we use the lattice SPT models and the ``Thouless pump Hamiltonians'' to construct periodically driven (Floquet) models with various symmetries. We employ the Onsager algebra to show that in these Floquet models, the notion of Thouless pumps extends beyond the ground subspace, such that we can define distinct Floquet phases with (non-invertible) symmetries due to different behaviors of the quasi-spectra.

The paper is structured as follows. In Sec.~\ref{sec:qa evolution}, employing quasiadiabatic evolution to define Thouless pump operators, we construct localized defects from adiabatic cycles
of the gapped Hamiltonian, and show the one-to-one correspondence between homotopy classes of adiabatic cycles and equivalent invertible defects. In Sec.~\ref{sec:gapped phases}, we use the gapped phases classification to complete the classification of adiabatic cycles as the group of $\mathcal{D}$-autoequivalences. In Sec.~\ref{sec:lattice}, we discuss SPT lattice models for different non-invertible symmetries, and construct adiabatic cycles in each homotopy class. In Sec.~\ref{sec:floquet}, we use the Thouless pump operators and the SPT lattice models to construct Floquet models, and study the Floquet phases for drives that obey the Onsager algebra. In Section \ref{Conc} we present our conclusions, while seven appendices provide the intermediate steps of derivations.

\section{Quasiadiabatic evolution}
\label{sec:qa evolution}
Given an adiabatic evolution of a gapped Hamiltonian $H(\theta)$, a quasiadiabatic evolution or spectral flow ~\cite{hastings2004lieb,hastings2005quasiadiabatic} can be defined as a family of unitaries $U(\theta)$~\cite{bachmann2012automorphic},
\beq
    U(\theta) &= \mathcal{S}\exp{i \int_0^{\theta} ds K(s)},\\
    K(\theta) &= \int_{-\infty}^{\infty}dt W_{\gamma}(t)\cdot e^{itH(\theta)}\biggl(\partial_{\theta} H(\theta)\biggr)e^{-itH(\theta)},
\eeq
where $\mathcal{S}$ denotes that the exponential is $s$-ordered, in analogy to the usual time ordered or path ordered exponential. $W_{\gamma}(t)$ is a cutoff function with a control parameter $\gamma$ introduced in Ref.~\cite{bachmann2012automorphic}, such that $U(\theta)$ is quasi-local. According to Proposition 2.4 of Ref.~\cite{bachmann2012automorphic}, 
\beq
    P(\theta) = U(\theta) P(0) U(\theta)^{\dagger},
\eeq
where $P(\theta)$ is the projection into the ground subspace of $H(\theta)$. Furthermore, since we require the Hamiltonians $\{H(\theta)\}$ to be symmetric under $\mathcal{D}$, by definition the quasiadiabatic evolution satisfies
\beq
    U(\theta) A = A U(\theta),
    \label{eq:commutation}
\eeq
for any symmetry operator $A\in\mathcal{D}$. We define the \textit{Thouless pump operator} given by an adiabatic cycle of a gapped Hamiltonian as $U_{\rm TP} \equiv U(2\pi)$. From the above identity, the Thouless pump operator $U_{\rm TP}$ preserves the ground subspace and commutes with the symmetry operators. Hence, $U_{\rm TP}$ is an emergent symmetry in the infrared which extends the original symmetry $\mathcal{D}$, this will be elaborated in the next section. Given two adiabatic cycles $H(\theta)$ and $H'(\theta)$, we can concatenate them by defining an adiabatic evolution of the ground subspace as follows,
\beq
    H''(\theta) = \begin{cases}H(2\theta),\quad  &0\leq \theta <\pi,\\
 H'(2\theta-2\pi),\quad  &\pi\leq \theta<2\pi.\end{cases}
\eeq
The corresponding Thouless pump operator for this cycle is $U''_{\rm TP} = U'_{\rm TP} U_{\rm TP}$. Furthermore, if the second cycle is the inverse of the first one, i.e., $H'(\theta)=H(2\pi - \theta)$, the Thouless pump operator given by the composite cycle is just identity $\openone$.

Although $U(\theta)$ preserves the ground state sector, it does not necessarily preserve all the ground states. In general, the action of $U_{\rm TP}$ may permute the ground states. For example, we can consider the following family of Hamiltonians on a $\mathbb{Z}_2$ spin chain with each site associated with a local Hilbert space $\mathbb{C}(\mathbb{Z}_2)$,
\beq
    &H(\theta) = \sum_j -g X_j \\
    &\quad -Je^{\frac{i \theta (X_j+X_{j+1})}{4}}  Z_j Z_{j+1} e^{-\frac{i \theta (X_j+X_{j+1})}{4}},
    \label{eq:Z_2 SSB cycle}
\eeq
where $X_j$ and $Z_j$ denote the Pauli-$X$ and $Z$ matrices in the Hilbert space on site $j$. When $J>g$, the Hamiltonian $H(0)$ is gapped and in the spontaneously $\mathbb{Z}_2$ symmetry broken (SSB) phase with two degenerate ground states. The corresponding Thouless pump operator $U_{\rm TP}$ interchanges the two ground states, see Appendix \ref{app:berry} for details. 

In this work, we assume that there is no accidental ground state degeneracy in the Hamiltonian $H(\theta)$, i.e., the SRE ground states should always be related by some symmetry operator. Thus, $U_{\rm TP}$ acting on different SRE ground states should not give rise to different phase factors, because otherwise it would not commute with all symmetry operators. Furthermore, for fusion category symmetries, there exists a unique positive combination $\ket{\phi}$ of SRE ground states such that it is symmetric~\cite{etingof2015tensor,diatlyk2024gauging}
\beq
    A\ket{\phi} \propto \ket{\phi}.
\eeq
Due to the uniqueness of this positive symmetric combination, although $U_{\rm TP}$ may permute the SRE ground states, it always leaves $\ket{\phi}$ invariant, i.e., $U_{\rm TP}\ket{\phi} = \ket{\phi}$. 

When each local term in the Hamiltonian $H(s)$ is symmetric\footnote{If the realization of the non-invertible symmetry does not mix with lattice translation, (e.g. has trivial index~\cite{ma2024quantum}), it is true that a symmetric Hamiltonian can be written as a sum of symmetric local terms.}, the quasi-local terms $K_i(\theta)$ in the quasiadiabatic evolution 
\beq
    K(\theta) = \sum_i K_i(\theta)
\eeq
are also symmetric. Thus, we can define a truncated operator by including in the evolution only terms that overlap with the left half of the system $(-\infty, 0]$. Schematically, we can write $U_{\rm tr}(\theta)$ as
\beq
    U_{\rm tr}(\theta) = \mathcal{S}\exp{i \int_0^{\theta}ds \sum_{i<0} K_i(s)},
\eeq
which is quasi-local and supported on $(-\infty, R]$ for some finite $R$. The truncated pump operator $U_{\rm tr, TP}\equiv U_{\rm tr}(2\pi)$ acting on the symmetric ground state $\ket{\phi}$ gives rise to the state $\ket{\phi_{\rm defect}} = U_{\rm tr,TP} \ket{\phi}$, which has the same reduced density matrix as $\ket{\phi}$ away from the cut. The set of states that have the same reduced density matrix away from the cut define the set of \textit{localized defects} near the cut. For $\ket{\phi_{\rm defect}}$, the defect can be moved around by quasi-local symmetric unitaries defined from a suitable truncation of $U_{\rm TP}$. Given two adiabatic cycles, we can also define the fusion of corresponding defects by the product of two truncated evolutions. Since the quasiadiabatic evolution is invertible, the localized defects created from their truncation also have invertible fusion relations. The definition of truncation is ambiguous up to some quasi-local symmetric unitary supported near the cut, hence, we say that two defects are in the same equivalence class if there is a quasi-local symmetric unitary $V_{\rm loc}$ such that
$\ket{\phi_{\rm defect}} = V _{\rm loc}\ket{\phi_{\rm defect'}}$. 

The notion of based homotopy\textemdash homotopy with a based point\textemdash in the parameter space of this gapped Hamiltonian is introduced as follows. When there is a smooth family of uniformly gapped symmetric Hamiltonians $H(\theta, \lambda)$ with $\theta\in [0,2\pi]$ and $\lambda\in [0,1]$, such that $H(0,\lambda)$ and $H(2\pi,\lambda)$ share the same ground subspace as $H(0,0)$, then cycles $H(\theta,0)$ and $H(\theta,1)$ are homotopic. In the remainder of this section, we will try to establish the correspondence between the class of homotopic adiabatic cycles and inequivalent defects. 

First, given a smooth deformation of cycles $H(\theta,\lambda)$, we can construct a family of Thouless pump operators $U_{\rm TP}(\lambda)$, as well as the truncated operators $U_{\rm tr,TP}(\lambda)$ for $\lambda\in [0,1]$. Their actions on the symmetric ground state $\ket{\phi}$ give rise to a smooth set of states,
\beq
    \ket{\phi_{\rm defect}(\lambda)} = U_{\rm tr,TP}(\lambda)\ket{\phi}.
\eeq
As will be shown later, there are only discrete equivalent classes of invertible defects in the gapped phases with non-invertible symmetries. Therefore, the above continuous family of states all belong to the same equivalence class.

Now we argue the reverse direction of this correspondence. Consider two adiabatic cycles $H(\theta)$ and $H'(\theta)$, whose truncated pump operators give rise to the same defect state up to a local symmetric unitary supported near the cut. We can concatenate the cycle $H(\theta)$ and the inverse of cycle $H'(\theta)$, resulting in a composite cycle that corresponds to the trivial defect (up to a local symmetric unitary) near the cut. To see that two cycles are homotopic, it suffices to show that, given an adiabatic cycle $H_0(\theta)$, whose corresponding quasiadiabatic evolution $U_0(2\pi)$ gives rise to a trivial defect after truncation, this cycle can always be smoothly deformed to a constant point $H_{\rm const}(\theta)\equiv H_0(0)$. For that, we first notice that the following family of Hamiltonians,
\beq
    H_0(\theta, \lambda) = (1-\lambda)H_0(\theta) + \lambda U_0(\theta) H_0(0) U_0(\theta)^{\dagger},
\eeq
are uniformly gapped and symmetric since both terms have the same ground subspace for any $\theta$ and $\lambda$. Hence, from this deformation, the cycle $H_0(\theta)$ and the conjugation cycle $U_0(\theta) H_0(0) U_0(\theta)^{\dagger}$ are homotopic. 

We now show that $H_0(\theta,1)= U_0(\theta) H_0(0) U_0(\theta)^{\dagger}$ can be made independent of $\theta$. The key input is the assumption of a trivial defect: suitable truncations of $U_0(2\pi)$ act trivially on $\ket{\phi}$ up to a local symmetric unitary. To see how this is used, it is helpful to think of $U_0(2\pi)$ as a finite-depth local unitary (FDLU) circuit $U$ with the same property. The order of local gates in $U$ may be rearranged by conjugation, and gate-inverse pairs may be inserted simultaneously; for the continuously parametrized family $U_0(\theta)$, both operations are realized as smooth deformations of that family. Using these moves we bring $U$ to the two-layer form $U = \Bigl(\prod_{2k+1}U_{2k+1}\Bigr)\Bigl(\prod_{2k} U_{2k}\Bigr)$, in which the small circuits $\{U_{2k}\}$ are mutually non-overlapping, and the leftover circuits $\{U_{2k+1}\}$, each supported between the right boundary of $U_{2k}$ and the left boundary of $U_{2k+2}$, are likewise mutually non-overlapping. We choose this rearrangement so that the boundaries of the $U_{2k}$ coincide with the truncation of $U$ at certain location. Each $U_{2k}$ is then, by construction, a truncation of $U$, and therefore satisfies $U_{2k}\ket{\phi} = \ket{\phi}$ by the trivial-defect assumption. The odd blocks require a separate argument: since the $\{U_{2k+1}\}$ do not overlap one another, each acts as a quasi-local operator on $\ket{\phi}$, and a quasi-local operator cannot permute the ground states; hence each $U_{2k+1}$ preserves every ground state, and in particular $U_{2k+1}\ket{\phi} = \ket{\phi}$. Every factor in the above product therefore satisfies $U_j\ket{\phi} = \ket{\phi}$.

The deformation of the adiabatic cycle given by the above deformation of $U(\theta)$ takes the conjugation cycle $U_0(\theta) H_0(0) U_0(\theta)^{\dagger}$ to a composite of some cycles in which the Hamiltonian only has local changes, which allows us to construct a deformation to a constant point. For example, for local unitaries $U_j = e^{iK_j}$, where $\ket{\phi}$ is the eigenstate of $K_j$ with zero eigenvalue, one can deform the unitary as $U_j(\lambda) = e^{i(1-\lambda)K_j}$, such that it always preserves $\ket{\phi}$ and becomes identity when $\lambda\rightarrow 1$. Eventually we obtain from the deformation just the constant point $H_{\rm const}(\theta)\equiv H_0(0)$.

We have demonstrated a one-to-one correspondence between the group formed by homotopic cycles and that of inequivalent invertible defects in gapped phases. In the next section, we will obtain this group by mapping the problem to the classification of gapped phases with an extended symmetry. Before we move on, since the conjugation cycle of the form $U(\theta)H U(\theta)^{\dagger}$ will be used extensively in the following sections, we refer the readers to Appendix \ref{app:berry}, in which we derive the quasiadiabatic evolution for the conjugation cycles, and comment on its relation with the Berry/Wilczek-Zee holonomy in adiabatic problems~\cite{Berry:1984jv,simon1983holonomy,wilczek1984appearance,kato1950adiabatic}. When $U(\theta)$ is generated by some $\theta$-independent local pivot Hamiltonian, these conjugation cycles are referred to as pivot loops \cite{tantivasadakarn2023pivot,tantivasadakarn2023building,jones2025pivoting,jones2025charge}.

\section{Gapped phases with the extended symmetry}
\label{sec:gapped phases}
In this section, we will complete the classification of adiabatic cycles using the classification of gapped phases. We note that there are two levels of rigor when we use the term ``gapped phases under a fusion category symmetry''. First, if we can define a renormalization flow such that the lattice models flow to some fixed points in the infrared, then we expect the ground subspace to be described by certain topological field theories with fusion category symmetry. These theories are classified by the module categories over the fusion category, and in particular the SPT phases are associated with fiber functors~\cite{thorngren2024fusion,inamura2021topological}. On the other hand, gapped phases of quantum many-body systems are usually defined as an equivalence class of ground states on the lattice, with equivalence relations given by the existence of a continuous family of local gapped Hamiltonians~\cite{chen2010local,chen2011classification,bachmann2012automorphic}. Under the assumptions that gapped Hamiltonians are translationally invariant with periodic boundary conditions, and fusion category symmetries are realized by matrix product operators (MPO), a classification of gapped phases is obtained following this definition, which agrees with the classification of topological field theories~\cite{garre2023classifying}. We will use the latter gapped phases classification with the assumptions therein, while expecting the result to extend to more general settings due to the classification of topological field theories.
A module category $\mathcal{M}$ over a fusion category $\mathcal{D}$ consists of simple objects $\{m_i\}$ that allow for the (fusion) action of $A\in \mathcal{D}$,
\beq
    A\rhd m_i = \sum_{j}P^{A,i}_j m_j,
\eeq
where $P^{A,i}_j$ are non-negative integers. The fusions in the fusion category $\mathcal{D}$, and the fusion action between $\mathcal{D}$ and the module category $\mathcal{M}$, satisfy associativity up to some $^{\triangledown}F$-symbols. These obey the pentagon identities with the $F$-symbols in $\mathcal{D}$. The physical intuition is that the simple objects in $\mathcal{M}$ describe the SRE ground states in the corresponding gapped phase with $\mathcal{D}$ symmetry, while the fusion action describes the action of symmetry operators on the ground states. For example, in the $\mathbb{Z}_2$ SSB phase, there are two simple objects $m_{\uparrow}$ and $m_{\downarrow}$ in the associated module category, and the symmetry operator exchanges the two objects. When a module category contains only one simple object $m$, the module data give rise exactly to a tensor functor from $\mathcal{D}$ to the category of complex vector spaces $\text{Vec}_{\mathbb{C}}$, i.e., a fiber functor. Physically, it describes an SPT phase, since there is only one SRE ground state. See Appendix \ref{app:vecG} for details about module categories over group-like symmetry $\text{Vec}_G$.

We have argued that a Thouless pump operator $U_{\rm TP}$ preserves the ground subspace, and commutes with the symmetry operators. Hence, the emergent symmetry operators of the gapped system in the infrared are of the form $A\cdot U_{\rm TP}^k$ for any $A\in \mathcal{D}$ and $k\in \mathbb{Z}$. It is often the case that the action of $U_{\rm TP}$ on the ground subspace is trivial (i.e., preserves each ground state individually), especially in a $\mathcal{D}$ SPT phase where there is only one ground state. On the other hand, $U_{\rm TP}$ can in general generate a non-trivial emergent symmetry since a truncation of $U_{\rm TP}$ acting on the ground states could create some invertible defect. The extended symmetry is thus the (Deligne) tensor product of $\mathcal{D}$ and $\mathbb{Z}_n$, the group generated by the invertible defect created by $U_{\rm TP}$\footnote{The mathematically precise notation for this extended category is $\mathcal{C}=\mathcal{D}\boxtimes \text{Vec}^{\omega}_{\mathbb{Z}_n}$, where $\omega$ denotes the potential anomaly of $U_{\rm TP}$.} 
\beq
    \mathcal{C}=\mathcal{D}\times \mathbb{Z}_n=\bigoplus_{k\in\mathbb{Z}_n}\mathcal{C}_k,
\eeq
where $\mathcal{C}_0 = \mathcal{D}$, and the $k$-graded component $\mathcal{C}_k$ contains objects of the form $A\cdot U_{\rm TP}^k$ for any $A\in\mathcal{D}$. It is obvious that the objects in $\mathcal{C}_k$ and the objects in $\mathcal{C}_{k'}$ always fuse into the objects in $\mathcal{C}_{k+k'}$, i.e., $\mathcal{C}$ is $\mathbb{Z}_n$-graded. We note that $U_{\rm TP}$ is not mixed anomalous with $\mathcal{D}$, but $U_{\rm TP}$ itself can be anomalous. Given an adiabatic cycle of gapped Hamiltonians, we can define a gapped phase with an extended $\mathcal{C}$ symmetry by including the Thouless pump operator $U_{\rm TP}$. Although $H(0)$ does not commute with $U_{\rm TP}$ outside of the ground subspace, we can always construct a symmetrized parent Hamiltonian~\cite{perez2006matrix} whose continuous family defines this gapped phase on a lattice. 

A Thouless pump operator in a gapped phase with $\mathcal{D}$ symmetry defines a localized defect, whose equivalence class (up to local symmetric unitaries) corresponds to a homotopy class of adiabatic cycles. On the other hand, localized defects are described by action tensors in the ground matrix product state (MPS). Their fusion with other symmetry defects give rise to the $^{\triangledown}F$-symbols ($L$-symbols)~\cite{garre2023classifying}, which characterize the module categories over $\mathcal{C}$. Since the gapped phases with a fusion category $\mathcal{C}$ symmetry correspond  one-to-one to inequivalent module categories, \textit{we conclude that homotopy classes of adiabatic cycles correspond  one-to-one to lifts from a $\mathcal{D}$-module category to inequivalent module categories over the extended category $\mathcal{C}$.} We will give an argument below for the classification of the lifts and refer the readers to Appendix \ref{app:proof} for a more detailed treatment following Ref.~\cite{meir2012module}. 

Before discussing the lifts, we first introduce the notion of autoequivalence. Let $\mathcal{L}$ denote the $\mathcal{D}$-module category describing the gapped phase of our Hamiltonian. A $\mathcal{D}$-autoequivalence of $\mathcal{L}$ is an equivalence (``bijection'') $\gamma$ from the module category $\mathcal{L}$ to itself that preserves the fusion action of $\mathcal{D}$ with a natural isomorphism~\cite{etingof2015tensor}. Namely,
\beq
    \gamma:\ \mathcal{L}\rightarrow\mathcal{L},
\eeq
such that $\gamma(A\rhd m)$ and $A\rhd \gamma(m)$ are related by a natural isomorphism that obeys the associativity constraint. Concretely, an autoequivalence consists of two pieces of data. The first is the map $\gamma$ itself, which invertibly relabels the ground states. The second is the natural isomorphism identifying two routes: symmetry action followed by a relabeling $\gamma(A\rhd m)$, and a relabeling followed by a symmetry action $A\rhd \gamma(m)$. In general, a $\mathcal{D}$-autoequivalence may permute the simple objects in $\mathcal{L}$. However, non-trivial $\mathcal{D}$-autoequivalences exist even when there is only one simple object in $\mathcal{L}$. Due to the bijective nature of these autoequivalences, their compositions form a group we denote as $\Gamma=Aut_{\mathcal{D}}(\mathcal{L})$. See Appendix \ref{app:vecG} for details about the autoequivalences of  $\text{Vec}_G$-module categories as an example.   

By including a Thouless pump operator to extend the symmetry, $\mathcal{L}$ is lifted to a $\mathcal{C}$-module category. To complete the lift, we need to specify the fusion action of the object $U_{\rm TP}\in\mathcal{C}$ on the objects in $\mathcal{L}$. As a $\mathcal{C}$-module category, $\mathcal{L}$ is equipped with a fusion action of $U_{\rm TP}^k$ for $k\in\mathbb{Z}_n$, denoted as
\beq
    \psi_k:\ \mathcal{L}\rightarrow\mathcal{L},\quad m\mapsto U_{\rm TP}^k \rhd m.
\eeq
Since $U_{\rm TP}^k$ is invertible, this fusion action defines an equivalence from $\mathcal{L}$ to itself. Since $U_{\rm TP}$ is symmetric under $\mathcal{D}$, for any objects $m\in\mathcal{L}$ and $A\in\mathcal{D}$ we have
\beq
    &\psi_k(A\rhd m ) = U_{\rm TP}^k \rhd (A\rhd m) \\
    &= A\rhd (U_{\rm TP}^k \rhd m) = A\rhd \psi_k(m).
\eeq
The second equality is because $\left[A,U_{\rm TP}\right]=0$.
Hence, the fusion action $\psi_k$ is in fact a $\mathcal{D}$-autoequivalence, i.e., $\psi_k\in \Gamma$. The composition of fusion actions are given by $\psi_k\circ\psi_{k'} = \psi_{k+k'}$. As a result, $\{\psi_k\}$ defines a group homomorphism from $\mathbb{Z}_n$ to $\Gamma$. The set of these group homomorphisms is given by the elements of $\Gamma$, because the choice of element $\psi_1\in\Gamma$ uniquely defines the group homomorphism. Therefore, spanning over all orders, we arrive at the conclusion that \textit{the homotopy classes of adiabatic cycles are classified by $\Gamma$.} Physically, the autoequivalence $\psi_1$ corresponds to (equivalence class of) the localized invertible defect created by the truncated Thouless pump operator for the adiabatic cycle. 

We note that our result agrees with the conjecture in Ref.~\cite{inamura20241+}. For SPT phases with a group-like symmetry $G$, the Thouless pumps are classified by $\Gamma = H^1(G,U(1))$, which agrees with Ref.~\cite{bachmann2024classification}; and can be rephrased as: $0$d 
$G$-SPT (a $G$-charge) can be pumped in a 1d $G$-SPT state, consistent with the dimensional-reduction picture of Refs.~\cite{kitaev2011SRE, bachmann2024classification}.

\section{Lattice models for adiabatic cycles}
\label{sec:lattice}
\subsection{Anyonic chain model for fusion category symmetry}
Now that we have established the classification of Thouless pumps in the gapped phases with fusion category symmetries, in this section we present the lattice models for various symmetries to obtain a more concrete understanding of the group $\Gamma$. All the adiabatic cycles we construct in the following are conjugation cycles (i.e., of the form $U(\theta)HU(\theta)^{\dagger}$). In Appendix \ref{app:berry}, we derive the quasiadiabatic evolution operators for these cycles and show that their actions on the ground states are the same $U(\theta)$ up to an overall phase. Therefore in this section we will use $U(\theta)$ to denote both the conjugation unitaries and the quasiadiabatic evolutions. 

\begin{figure}[t!]
  \centering
  \begin{tikzpicture}[thick,scale=0.6, every node/.style={scale=1},
  arrow/.style={
    line width=1.2pt, 
    postaction={decorate},
    decoration={markings, mark=at position 0.6 with {\arrow{latex}}}
  },
  midarrow/.style={
    line width=1.2pt, 
    postaction={decorate},
    decoration={markings, mark=at position 0.5 with {\arrow{latex}}}
  },
  modulearrow/.style={
    dashed,
    blue,
    postaction={decorate},
    decoration={markings, mark=at position 0.6 with {\arrow{latex}}}
  }
  ]

  \def\gap{2}     
  \def\h{1.5}     

  \foreach \i in {1,2,3,4,5,6} {
    \pgfmathsetmacro\x{\i*\gap}
    \pgfmathtruncatemacro{\xf}{2*\i-1}
    \fill[black] (\x,0) circle (3pt);  
    \node[below] at (\x,0) {$\alpha_{\frac{\xf}{2}}$};              
    \node[blue, above] at (\x,\h) {$\mu_{\frac{\xf}{2}}$};              
    \draw[arrow] (\x,0) -- (\x,\h);  
    \node[right] at (\x+0.1,0.5*\h-0.1) {$b_{\frac{\xf}{2}}$};
    \fill[blue] (\x,\h) circle (3pt);
  }

  \foreach \i in {1,2,3,4,5} {
    \pgfmathsetmacro\xA{\i*\gap}
    \pgfmathsetmacro\xB{\xA + \gap}
    \draw[arrow] (\xA,0) -- (\xB,0);                
    \node[below] at ({(\xA+\xB)/2},-0.1) {$a_{\i}$};
  }
  \draw[arrow] (\gap-1,0) -- (\gap,0);  
  \draw[arrow] (6*\gap,0) -- (6*\gap+1,0);  
  \foreach \i in {1,2,3,4,5} {
    \pgfmathsetmacro\xA{\i*\gap}
    \pgfmathsetmacro\xB{\xA + \gap}
    \draw[modulearrow] (\xB,\h) -- (\xA,\h); 
    \node[blue,above] at ({(\xA+\xB)/2},\h+0.1) {$m_{\i}$};
  }
  \draw[modulearrow] (\gap,\h) -- (\gap-1,\h);  
  \draw[modulearrow] (6*\gap+1,\h) -- (6*\gap,\h); 
  
  \draw [red, decorate,
      decoration={snake,amplitude=.6mm,segment length=5mm}]
    (\gap-1,\h+1.2) -- (6*\gap+1,\h+1.2)
    node [above,align=center,midway]
    {\textcolor{red}{$\gamma\in \Gamma$}
    };
  \draw[midarrow]
    (\gap-1,-1.2) -- (6*\gap+1,-1.2);
    \node[below] at (3.5*\gap, -1.3)
    {$A\in \mathcal{D}$
    };
\end{tikzpicture}
  \caption{The labels $a_i,b_{i+1/2}\in\mathcal{D}$, $m_i\in\mathcal{M}$. $\alpha_{i+1/2}$ and $\mu_{i+1/2}$ denote the basis vectors in the fusion spaces $V^{b_{i+1/2} a_{i+1}}_{a_{i}}$ and $\text{Hom}_{\mathcal{M}}(b_{i+1/2}\otimes m_{i+1},m_{i})$, $\alpha_{i+1/2}=1,\cdots,N^{b_{i+1/2} a_{i+1}}_{a_{i}}$. When $\mathcal{M}$ is a fiber functor over $\mathcal{D}$, $\mu_{i+1/2}=1,\cdots,\text{dim}(b_{i+1/2})$. The symmetry $\mathcal{D}$ of the model is defined via fusing a line $A\in \mathcal{D}$ from the bottom. The Thouless pump operators are defined via fusing a line $\gamma\in\Gamma$ from the top, which commute with both the Hamiltonian and the symmetry.}
  \label{fig:anyonic-chain}
\end{figure}
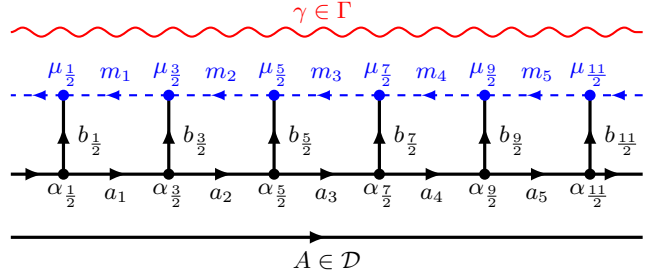

We present here the prototypical lattice model for a general gapped phase with non-invertible symmetry $\mathcal{D}$ inspired by the topological holography (symTFT) picture~\cite{kong2020algebraic,kong2020classification,kong2022one,chatterjee2023symmetry,ji2021unified,moradi2023topological,bhardwaj2024categorical,bhardwaj2025gapped,huang2025topological,evans2025operator}. It is a specific anyonic chain model given by the string-net model on a strip with two gapped boundaries at the top and bottom shown in Fig.~\ref{fig:anyonic-chain}. Similar to 1d gapped phases, gapped boundaries of the string-net models for a fusion category $\mathcal{D}$ correspond to its module categories. We choose the Dirichlet boundary at the bottom, whose corresponding module category is $\mathcal{D}$ itself. At the top boundary, we can choose an arbitrary $\mathcal{D}$ module category $\mathcal{M}$. For different choices of $\mathcal{M}$, this quasi-1d anyonic chain model realizes different 1d gapped phases with symmetry $\mathcal{D}$. The Hamiltonian is a sum of
mutually commuting projectors given by the string-net models~\cite{levin2005string,kitaev2012models,lin2021generalized},
\beq
    H = -\sum_i A^t_i-\sum_i A^d_i-\sum_i B_i,
\label{eq:anyonic chain}
\eeq
where $A^t_i$ and $A^d_i$ are projectors into fusion rule preserving vertices, and $B_i$ are the plaquette terms, proportional to projectors, obtained by fusing object $\sum_a d_a a$ to the edges of each plaquette.

This model is not defined on a tensor product Hilbert space, because the degrees of freedom at the vertices depend on the state on the neighboring edges. Nevertheless, as we will show below, for a large class of fusion categories, this model can be mapped to some spin chain lattice models. Since a fusion category can be realized in a strictly locality preserving manner (on-site) if and only if it admits fiber functors~\cite{meng2024non,evans2025operator}, we will focus on the corresponding fusion category symmetries and their SPT phases for simplicity. The construction and classification of adiabatic cycles should be straightforwardly extended to other gapped phases beyond SPT, such as the symmetry breaking phases.

\subsection{Vec$_G$: string-net}
When the symmetry is an ordinary group-like symmetry, i.e., $\mathcal{D}=\text{Vec}_G$ for some finite group $G$, the fusion rules are just the group multiplication rules, and a fiber functor $\mathcal{M}$ over $\mathcal{D}$ is given by a 2-cocycle $\omega:G\times G\rightarrow U(1)$, satisfying the cocycle condition
\beq
    \omega(h,k)\omega(g,hk)=\omega(gh,k)\omega(g,h).
\eeq
The degrees of freedom $\alpha_{i+1/2}$, $\mu_{i+1/2}$, and $m_i$ on the top edges in Figure \ref{fig:anyonic-chain} are all trivial, thus the model is built on a tensor product Hilbert space, where there is a local Hilbert space $\mathcal{H}_{\rm loc}=\mathbb{C}[G]$ on each black edge. 
We denote the degrees of freedom on the horizontal black edge $i$, as $a_i$, and degrees of freedom on the vertical edge $i+1/2$, by $b_{i+1/2}$. Then the total Hilbert space is given by $\mathcal{H} = \bigotimes_{i} \left(\mathcal{H}_{\rm loc}^{(i)} \otimes \mathcal{H}_{\rm loc}^{(i+1/2)}\right)$.
The $A^t_i$ terms are trivial, while the $A^d_i$ terms enforce the group multiplication rules on each black vertex,
\begin{align}
A^d_i
\ket{\begin{tikzpicture}[baseline={([yshift=-.5ex]current bounding box.center)},thick,scale=0.6, every node/.style={scale=1.0},
  arrow/.style={
    line width=1.2pt, 
    postaction={decorate},
    decoration={markings, mark=at position 0.6 with {\arrow{latex}}}
  }
  ]
  \def\n{5}       
  \def\gap{2}     
  \def\h{1.5}     
    \fill[black] (0,0) circle (3pt); 
    \draw[arrow] (0,0) -- (0,\h);  
    \node[right] at (0+0.1,0.5*\h) {$b_{i-\frac{1}{2}}$};  
  \draw[arrow] (0,0) -- (\h,0); 
    \node[below] at (\h/2,-0.1) {$a_i$};
    \draw[arrow] (-\h,0) -- (0,0);  
    \node[below] at (-\h/2,-0.1) {$a_{i-1}$};
\end{tikzpicture} }=
\delta_{a_{i-1},b_{i-\frac{1}{2}} a_i}\ket{\begin{tikzpicture}[baseline={([yshift=-.5ex]current bounding box.center)},thick,scale=0.6, every node/.style={scale=1.0},
  arrow/.style={
    line width=1.2pt, 
    postaction={decorate},
    decoration={markings, mark=at position 0.6 with {\arrow{latex}}}
  }
  ]
  \def\n{5}       
  \def\gap{2}     
  \def\h{1.5}     
    \fill[black] (0,0) circle (3pt); 
    \draw[arrow] (0,0) -- (0,\h);  
    \node[right] at (0+0.1,0.5*\h) {$b_{i-\frac{1}{2}}$};  
  \draw[arrow] (0,0) -- (\h,0); 
    \node[below] at (\h/2,-0.1) {$a_i$};
    \draw[arrow] (-\h,0) -- (0,0);  
    \node[below] at (-\h/2,-0.1) {$a_{i-1}$};
\end{tikzpicture} }.
\label{eq:A_i string-net}
\end{align}
The plaquette terms are given from the fusion of objects in the module category onto each plaquette via the $^{\triangledown}F$-moves,
\begin{widetext}
\beq
B_i \ket{\begin{tikzpicture}[baseline={([yshift=-.4ex]current bounding box.center)},thick,scale=0.6, every node/.style={scale=1.0},
  arrow/.style={
    line width=1.2pt, 
    postaction={decorate},
    decoration={markings, mark=at position 0.6 with {\arrow{latex}}}
  },
  modulearrow/.style={
    dashed,
    blue,
    postaction={decorate},
    decoration={markings, mark=at position 0.6 with {\arrow{latex}}}
  }
  ]
  \def\n{5}       
  \def\gap{2}     
  \def\h{1.5}     
    \fill[black] (0,0) circle (3pt); 
    \draw[arrow] (0,0) -- (0,\h);  
    \node[left] at (0-0.1,0.5*\h) {$b_{i-\frac{1}{2}}$};
    \fill[blue] (0,\h) circle (3pt);
    \fill[black] (\gap,0) circle (3pt); 
    \draw[arrow] (\gap,0) -- (\gap,\h);  
    \node[right] at (\gap+0.1,0.5*\h) {$b_{i+\frac{1}{2}}$};
    \fill[blue] (\gap,\h) circle (3pt);
  \draw[arrow] (0,0) -- (\gap,0); 
    \node[below] at (\gap/2,-0.1) {$a_i$};
  \draw[arrow] (-1,0) -- (0,0);  
  \draw[arrow] (\gap,0) -- (\gap+1,0);  
    \draw[modulearrow] (\gap,\h) -- (0,\h); 
  \draw[modulearrow] (0,\h) -- (-1,\h);  
  \draw[modulearrow] (\gap+1,\h) -- (\gap,\h); 
\end{tikzpicture}  }
=\sum_{g\in G}\omega(b_{i-\frac{1}{2}}, \overline{g})\omega(g,b_{i+\frac{1}{2}})
\ket{\begin{tikzpicture}[baseline={([yshift=-.4ex]current bounding box.center)},thick,scale=0.6, every node/.style={scale=1.0},
  arrow/.style={
    line width=1.2pt, 
    postaction={decorate},
    decoration={markings, mark=at position 0.6 with {\arrow{latex}}}
  },
  modulearrow/.style={
    dashed,
    blue,
    postaction={decorate},
    decoration={markings, mark=at position 0.6 with {\arrow{latex}}}
  }
  ]
  \def\n{5}       
  \def\gap{2}     
  \def\h{1.5}     
    \fill[black] (0,0) circle (3pt); 
    \draw[arrow] (0,0) -- (0,\h);  
    \node[left] at (0-0.1,0.5*\h) {$b_{i-\frac{1}{2}} \overline{g}$};
    \fill[blue] (0,\h) circle (3pt);
    \fill[black] (\gap,0) circle (3pt); 
    \draw[arrow] (\gap,0) -- (\gap,\h);  
    \node[right] at (\gap+0.1,0.5*\h) {$g b_{i+\frac{1}{2}}$};
    \fill[blue] (\gap,\h) circle (3pt);
  \draw[arrow] (0,0) -- (\gap,0); 
    \node[below] at (\gap/2,-0.1) {$g a_i$};
  \draw[arrow] (-1,0) -- (0,0);  
  \draw[arrow] (\gap,0) -- (\gap+1,0);  
    \draw[modulearrow] (\gap,\h) -- (0,\h); 
  \draw[modulearrow] (0,\h) -- (-1,\h);  
  \draw[modulearrow] (\gap+1,\h) -- (\gap,\h); 
\end{tikzpicture}  },
\label{eq:B_i string-net}
\eeq
\end{widetext}
where $\overline{g}$ denotes the inverse of $g$. For the derivation of plaquette terms and more details about the module categories, see Appendix \ref{app:vecG}. The symmetry operator $\mathcal{A}_g$ for $g\in G$ is given from the fusion of a $g$-line from the bottom,
\beq
    \mathcal{A}_g = \prod_i R_g^i,\text{ in which } R_g^i\equiv \sum_{a_i\in G}\ket{a_i \overline{g}}\bra{a_i}.
\eeq
Moreover, given a one-dimensional group representation $\rho\in H^1(G, U(1))$, we define a charge operator on $\mathcal{H}^{(i+1/2)}_{\rm loc}$ as
\beq
    Z_{\rho}^{(i+\frac{1}{2})}\equiv \sum_{h\in G} \rho(h)(\ket{h}\bra{h})_{i+\frac{1}{2}}.
\eeq
Then the Thouless pump operator is as follows,
\begin{widetext}
\begin{equation}
\begin{aligned}
U_{\rm TP}^{(\rho)} &\ket{\begin{tikzpicture}[baseline={([yshift=-.4ex]current bounding box.center)},thick,scale=0.5, every node/.style={scale=1.0},
  arrow/.style={
    line width=1.2pt, 
    postaction={decorate},
    decoration={markings, mark=at position 0.6 with {\arrow{latex}}}
  },
  modulearrow/.style={
    dashed,
    blue,
    postaction={decorate},
    decoration={markings, mark=at position 0.6 with {\arrow{latex}}}
  }
  ]
  \def\n{5}       
  \def\gap{2}     
  \def\h{1.5}     
  \foreach \i in {0,1,2,3} {
    \pgfmathsetmacro\x{\i*\gap}
    \fill[black] (\x,0) circle (3pt);
    \draw[arrow] (\x,0) -- (\x,\h);  
    \fill[blue] (\x,\h) circle (3pt);
  }
  \foreach \i in {0,1,2} {
    \pgfmathsetmacro\xA{\i*\gap}
    \pgfmathsetmacro\xB{\xA + \gap}
    \draw[arrow] (\xA,0) -- (\xB,0);                
  }
  \draw[arrow] (-1,0) -- (0,0);  
  \draw[arrow] (3*\gap,0) -- (3*\gap+1,0);  
  \foreach \i in {0,1,2} {
    \pgfmathsetmacro\xA{\i*\gap}
    \pgfmathsetmacro\xB{\xA + \gap}
    \draw[modulearrow] (\xB,\h) -- (\xA,\h); 
  }
  \draw[modulearrow] (0,\h) -- (-1,\h);  
  \draw[modulearrow] (3*\gap+1,\h) -- (3*\gap,\h); 
\end{tikzpicture}}
\equiv
\ket{\begin{tikzpicture}[baseline={([yshift=-.4ex]current bounding box.center)},thick,scale=0.5, every node/.style={scale=1.0},
  arrow/.style={
    line width=1.2pt, 
    postaction={decorate},
    decoration={markings, mark=at position 0.6 with {\arrow{latex}}}
  },
  modulearrow/.style={
    dashed,
    blue,
    postaction={decorate},
    decoration={markings, mark=at position 0.6 with {\arrow{latex}}}
  }
  ]
  \def\n{5}       
  \def\gap{2}     
  \def\h{1.5}     
  \foreach \i in {0,1,2,3} {
    \pgfmathsetmacro\x{\i*\gap}
    \fill[black] (\x,0) circle (3pt);
    \draw[arrow] (\x,0) -- (\x,\h);  
    \fill[blue] (\x,\h) circle (3pt);
  }
  \foreach \i in {0,1,2} {
    \pgfmathsetmacro\xA{\i*\gap}
    \pgfmathsetmacro\xB{\xA + \gap}
    \draw[arrow] (\xA,0) -- (\xB,0);                
  }
  \draw[arrow] (-1,0) -- (0,0);  
  \draw[arrow] (3*\gap,0) -- (3*\gap+1,0);  
  \foreach \i in {0,1,2} {
    \pgfmathsetmacro\xA{\i*\gap}
    \pgfmathsetmacro\xB{\xA + \gap}
    \draw[modulearrow] (\xB,\h) -- (\xA,\h); 
  }
  \draw[modulearrow] (0,\h) -- (-1,\h);  
  \draw[modulearrow] (3*\gap+1,\h) -- (3*\gap,\h); 
  \draw [red, decorate,
      decoration={snake,amplitude=.6mm,segment length=5mm}]
    (-1,\h+0.5) -- (3*\gap+1,\h+0.5)
    node [above,align=center,midway]
    {\textcolor{red}{$\rho$}
    };
\end{tikzpicture}}\\ 
=
&\ket{\begin{tikzpicture}[baseline={([yshift=-.4ex]current bounding box.center)},thick,scale=0.5, every node/.style={scale=1.0},
  arrow/.style={
    line width=1.2pt, 
    postaction={decorate},
    decoration={markings, mark=at position 0.6 with {\arrow{latex}}}
  },
  modulearrow/.style={
    dashed,
    blue,
    postaction={decorate},
    decoration={markings, mark=at position 0.6 with {\arrow{latex}}}
  }
  ]
  \def\n{5}       
  \def\gap{2}     
  \def\h{1.5}     
  \def\arc{\h+0.8}
  \foreach \i in {0,1,2,3} {
    \pgfmathsetmacro\x{\i*\gap}
    \fill[black] (\x,0) circle (3pt);
    \draw[arrow] (\x,0) -- (\x,\h);  
    \fill[blue] (\x,\h) circle (3pt);
  }
  \foreach \i in {0,1,2} {
    \pgfmathsetmacro\xA{\i*\gap}
    \pgfmathsetmacro\xB{\xA + \gap}
    \draw[arrow] (\xA,0) -- (\xB,0);                
  }
  \draw[arrow] (-1,0) -- (0,0);  
  \draw[arrow] (3*\gap,0) -- (3*\gap+1,0);  
  \foreach \i in {0,1,2} {
    \pgfmathsetmacro\xA{\i*\gap}
    \pgfmathsetmacro\xB{\xA + \gap}
    \draw[modulearrow] (\xB,\h) -- (\xA,\h); 
  }
  \draw[modulearrow] (0,\h) -- (-1,\h);  
  \draw[modulearrow] (3*\gap+1,\h) -- (3*\gap,\h); 
  \foreach \i in {0,1,2,3} {
    \pgfmathsetmacro\x{\i*\gap}
    \draw[red, decorate,
      decoration={snake,amplitude=.6mm,segment length=1.6mm}] (\x-0.3*\gap,\h) .. controls (\x,\arc) .. (\x+0.3*\gap,\h)
      node [above,align=center,midway]
    {\textcolor{red}{$\rho$}
    };
  }
\end{tikzpicture}} 
=\Big(\prod_i Z_{\rho}^{(i+1/2)}\Big)\ket{\begin{tikzpicture}[baseline={([yshift=-.4ex]current bounding box.center)},thick,scale=0.5, every node/.style={scale=1.0},
  arrow/.style={
    line width=1.2pt, 
    postaction={decorate},
    decoration={markings, mark=at position 0.6 with {\arrow{latex}}}
  },
  modulearrow/.style={
    dashed,
    blue,
    postaction={decorate},
    decoration={markings, mark=at position 0.6 with {\arrow{latex}}}
  }
  ]
  \def\n{5}       
  \def\gap{2}     
  \def\h{1.5}     
  \foreach \i in {0,1,2,3} {
    \pgfmathsetmacro\x{\i*\gap}
    \fill[black] (\x,0) circle (3pt);
    \draw[arrow] (\x,0) -- (\x,\h);  
    \fill[blue] (\x,\h) circle (3pt);
  }
  \foreach \i in {0,1,2} {
    \pgfmathsetmacro\xA{\i*\gap}
    \pgfmathsetmacro\xB{\xA + \gap}
    \draw[arrow] (\xA,0) -- (\xB,0);                
  }
  \draw[arrow] (-1,0) -- (0,0);  
  \draw[arrow] (3*\gap,0) -- (3*\gap+1,0);  
  \foreach \i in {0,1,2} {
    \pgfmathsetmacro\xA{\i*\gap}
    \pgfmathsetmacro\xB{\xA + \gap}
    \draw[modulearrow] (\xB,\h) -- (\xA,\h); 
  }
  \draw[modulearrow] (0,\h) -- (-1,\h);  
  \draw[modulearrow] (3*\gap+1,\h) -- (3*\gap,\h); 
\end{tikzpicture}}.
\label{eq:UTP}
\end{aligned}
\end{equation}
\end{widetext}

It is straightforward to see that $U_{\rm TP}^{(\rho)}$ is symmetric and commutes with the Hamiltonian. A corresponding adiabatic cycle can be given by a conjugation of the Hamiltonian by $U(\theta)=\prod_i u_{i+1/2}(\theta)$, such that $u_{i+1/2}(0)=I$ and $u_{i+1/2}(2\pi)=Z_{\rho}^{(i+1/2)}$. The Thouless pump operator $U_{\rm TP}^{(\rho)}$ has a clear physical meaning in the bulk where the topological order $D(G)$ implies that $U_{\rm TP}$ pumps a charge anyon, as shown in Fig.~\ref{fig:symTFT_vecG}. The multiplication of Thouless pump operators for different one-dimensional representations form an abelian group $H^1(G,U(1))$, which is exactly the group of autoequivalences $\Gamma$ for all the $\text{Vec}_G$ fiber functors. 

\begin{figure}[t!]
  \centering
  \begin{tikzpicture}[thick,scale=.5, every node/.style={scale=1.0}
  ]

  \def\n{5}       
  \def\gap{2}     
  \def\h{1.5}     
  \def\arc{\h-0.8}     
  \fill[cyan!30] (-1.5*\gap,0) rectangle (6.5*\gap,\h);
  \draw[line width=1.2pt] (-1.5*\gap,0) -- (6.5*\gap,0);
  \node[below] at (2.5*\gap,-.1) {symmetry boundary};
  \draw[blue,line width=1.2pt] (-1.5*\gap,\h) -- (6.5*\gap,\h);
  \node[blue,above] at (2.5*\gap,\h+.1) {physical boundary};
  \fill[red] (-\gap/2,\h) circle (3pt);
  \node[above] at (-\gap/2,\h+.1) {\textcolor{red}{charge $\overline{e_{\rho}}$}};
  \fill[red] (5.5*\gap,\h) circle (3pt);
  \node[above] at (5.5*\gap,\h+.1) {\textcolor{red}{charge $e_{\rho}$}};
  \foreach \i in {0,1,2,3,4,5} {
    \pgfmathsetmacro\x{\i*\gap}
    \draw[red] (\x-0.5*\gap,\h) .. controls (\x,\arc) .. (\x+0.5*\gap,\h);
    }
\end{tikzpicture}
  \caption{The Vec$_G$ anyonic chain model can be interpreted as a quasi-1d system defined on the sandwich. A topological order $D(G)$ lives in the bulk, and a representation $\rho$ corresponds to an anyon $e_{\rho}$. Truncating the $U_{\rm TP}$ operator creates the anyon $e_{\rho}$ at the two endpoints, which cannot be condensed on the physical boundary for an SPT phase. Thus, the adiabatic cycle pumps a charge $e_{\rho}$ along the quasi-1d system.}
  \label{fig:symTFT_vecG}
\end{figure}

\subsection{Minimal model for $\text{Vec}_G$ symmetry}
\label{sec:minimal VecG}
The quasi-1d lattice model above can be simplified to the standard group cohomology construction of the $G$-SPT model, by considering the following unitary operator
\beq
    S &= \prod_i \sum_{a,b,b'}\left(\ket{a}\bra{a}\right)_{i}\otimes \left(\ket{b a}\bra{b}\right)_{i-\frac{1}{2}}\otimes \left(\ket{\overline{a}b'}\bra{b'}\right)_{i+\frac{1}{2}}.
\eeq
It conjugates the vertex and plaquette terms in the string-net model in Eq.~\eqref{eq:A_i string-net} and Eq.~\eqref{eq:B_i string-net} to 
\beq
    S A_i^d S^{\dagger} &= (\ket{e}\bra{e})_{i-\frac{1}{2}},\quad
    S B_i S^{\dagger} = \sum_{g\in G} L_g^i\Omega_i(g) , \\
    &\qquad L_g^i\equiv \sum_{a\in G}(\ket{g a}\bra{a})_i,
\eeq
and the phase gate is
\beq
\Omega_i(g)\equiv \sum_{a, a', a''} &\omega(a\overline{a'},\overline{g})\omega(g,a'\overline{a''})\cdot\left(\ket{a}\bra{a}\right)_{i-1}\\
&\otimes \left(\ket{a'}\bra{a'}\right)_{i}\otimes \left(\ket{a''}\bra{a''}\right)_{i+1}.
\eeq
Since the $S A_i S^{\dagger}$ terms project the vertical edges to state $\ket{e}\in \mathbb{C}[G]$, and the rest of the system is decoupled from the vertical edges, we can consider the low-energy subspace of the model where $S A_i S^{\dagger}\equiv 1$, such that the system effectively is a $G$ spin chain of the horizontal edges with an effective Hamiltonian,
\beq
    H_{\rm SPT} &= -\sum_i \sum_{g\in G} L_g^i\Omega_i(g).
\eeq
The Thouless pump operator in this effective system is $U^{(\rho)}_{\rm TP}=\prod_i Z_{\rho}^{(i-1)}(Z_{\rho}^{(i)})^{\dagger}$. This can be seen by applying the vertex constraint $b_{i-1/2}= a_{i-1}\bar{a}_i$ to Eq.~\eqref{eq:UTP} and using that $\rho(a_{i-1}\bar{a}_i)= \rho(a_{i-1})\rho(\bar{a}_i)$. We can define the adiabatic cycle by the conjugation of $U'(\theta)=e^{i \theta H_{\rm TP}^{(\rho)}}$, where $H_{\rm TP}^{(\rho)}$ is composed of commuting Hermitian terms
\beq
H_{\rm TP}^{(\rho)} = \sum_i H_i^{(\rho)}, \,\, e^{i 2\pi H_i^{(\rho)}}=Z_{\rho}^{(i-1)}(Z_{\rho}^{(i)})^{\dagger},
\eeq
see more details in Sec.~\ref{sec:floquet}.

To be more explicit, we take $G=\mathbb{Z}_2^2$ as an example. The local Hilbert space on each edge is $\mathbb{C}^2 \otimes \mathbb{C}^2$ (two qubits). We denote the group element as $g = (g_1, g_2)$ where $g_1,g_2\in\{0,1\}$, and denote the Pauli operators on the two qubits as $X/Y/Z$ and $\Tilde{X}/\Tilde{Y}/\Tilde{Z}$ respectively. $\mathbb{Z}_2^2$ has two inequivalence classes of 2-cocycles, commonly used representatives of which are
\beq
    \omega_{\text{trivial}}(g,h)= 1,\quad \omega_{\text{nontrivial}}(g,h)=(-1)^{g_1 h_2}.
\eeq

The effective SPT Hamiltonians for these two cocycles of $\mathbb{Z}_2^2$ are
\beq
    H_{\text{trivial}}&=-\sum_i (1+X_i) (1+\tilde{X}_i),\\
    H_{\text{cluster}}&=-\sum_i (1+\Tilde{Z}_{i}X_i \Tilde{Z}_{i+1})(1+Z_{i-1}\Tilde{X}_{i}Z_{i}).
    \label{eq:hamiltonian cluster}
\eeq
The adiabatic cycles can be given by conjugations by the following operators,
\beq
    U(\theta)&=e^{i\theta H_{\text{TP}}}=e^{\frac{i\theta}{2}\sum_j \frac{1-Z_{j-1}Z_j}{2}},\\
    \tilde{U}(\theta)&=e^{i\theta \tilde{H}_{\text{TP}}}=e^{\frac{i\theta}{2}\sum_j\frac{1-\tilde{Z}_{j-1}\tilde{Z}_j}{2}}.
\eeq
When $\theta=2\pi$, truncating the Thouless pump operators gives rise to $Z_i Z_j$ and $\tilde{Z}_i \tilde{Z}_j$, which pump the charges of two $\mathbb{Z}_2$ symmetries from the site $i$ to the site $j$. This conjugation cycle is discussed in Ref.~\cite{shiozaki2022adiabatic}, including the definition of an invariant indicating Thouless pumps on the open chain. 

\subsection{Rep($G$): quantum double}
When the symmetry is the representation category of a finite group, i.e., $\mathcal{D}=\text{Rep}(G)$, the Hilbert space of the anyonic chain model does not, in general, factorize into a tensor product of local Hilbert spaces. Refs.~\cite{buerschaper2009mapping,kadar2010microscopic} construct a map from the Levin–Wen string-net model to the Kitaev quantum double, which is defined on a genuine tensor product Hilbert space, with the fusion constraints imposed energetically rather than kinematically. This map, however, covers only the bulk terms. Since anyonic chains are string-nets on a strip, the map must be extended to the boundary terms. We present the detailed map between quantum double and Levin–Wen string-net models with boundaries in Appendix~\ref{app:map}. In this section we work directly with the quantum double, which is defined on a tensor product Hilbert space.

The $\text{Rep}(G)$ SPT models depend on choices of the fiber functor $\mathcal{M}$. As we mentioned in Sec.~\ref{sec:gapped phases}, a fiber functor is a tensor functor from $\mathcal{D}$ to the category of complex vector spaces $\text{Vec}_{\mathbb{C}}$. A representation is given by a complex vector space that respects an action of the group element $g\in G$.  As a representation category, $\text{Rep}(G)$ contains finite-dimensional representations of $G$ as its objects. Once we forget about the group action in the representations, there is a canonical functor from objects in $\text{Rep}(G)$ to complex vector spaces. This canonical fiber functor for a representation category is called the forgetful functor. Therefore, there is also a canonical SPT phase for every $\text{Rep}(G)$ symmetry.
We now focus on this SPT phase, discussing other SPT phases later. 

The model for the canonical SPT is defined on the Hilbert space $\mathcal{H}=\mathbb{C}[G]^{\otimes \text{edges}}$. The basis states are given by assigning a group element $g_i, h_j\in G$ to label each edge,
\begin{align*}
    \begin{tikzpicture}[thick,scale=0.5, every node/.style={scale=1.0}
  ]
  \def\gap{2}     
  \def\h{1.5}     
  \foreach \i in {1,2,3,4,5} {
    \pgfmathsetmacro\x{\i*\gap}
    \pgfmathtruncatemacro{\xf}{2*\i-1}
    \draw[thick] (\x,0) -- (\x,\h);
    \node[right] at (\x+0.1,0.5*\h) {$g_{\frac{\xf}{2}}$};
    \fill[black] (\x,0.5*\h) circle (3pt);
  }
  \foreach \i in {0,1,2,3,4} {
    \pgfmathsetmacro\xA{\i*\gap}
    \pgfmathsetmacro\xB{\xA + \gap}
    \draw[thick] (\xA,0) -- (\xB,0);
    \fill[black] ({(\xA+\xB)/2},0) circle (3pt);
    \node[below] at ({(\xA+\xB)/2},-0.1) {$h_{\i}$};
  }
  \fill[black] (5.5*\gap,0) circle (3pt);
  \node[below] at (5.5*\gap,-0.1) {$h_5$};
  \draw[thick] (5*\gap,0) -- (6*\gap,0);
  \node at (6.5*\gap,0.25*\h) {$\cdots\cdots$};
  \node at (-0.5*\gap,0.25*\h) {$\cdots\cdots$};
\end{tikzpicture}
\end{align*}
The Hamiltonian is formed by the mutually commuting projector terms
\beq
    H^{\text{QD}}=-\sum_i A_i^{\text{QD}} - \sum_{i} B_i^{\text{QD}},
\eeq
where the vertex terms are given by $A^{\text{QD}}_i=\frac{1}{|G|}\sum_{k\in G}A^{\text{QD}}_{i,k}$ and
\begin{align}
A^{\text{QD}}_{i,k}
\ket{\begin{tikzpicture}[baseline={([yshift=-.5ex]current bounding box.center)},thick,scale=0.6, every node/.style={scale=1.0},
  arrow/.style={
    line width=1.2pt, 
    postaction={decorate},
    decoration={markings, mark=at position 0.6 with {\arrow{latex}}}
  }
  ]
  \def\gap{2}     
  \def\h{1.5}     
    \fill[black] (0.5*\h,0) circle (3pt); 
    \fill[black] (-0.5*\h,0) circle (3pt); 
    \fill[black] (0,0.5*\h) circle (3pt); 
    \draw[thick] (0,0) -- (0,\h);  
    \node[right] at (0+0.1,0.5*\h) {$g_{i-\frac{1}{2}}$};  
  \draw[thick] (0,0) -- (\h,0); 
    \node[below] at (\h/2,-0.1) {$h_i$};
    \draw[thick] (-\h,0) -- (0,0);  
    \node[below] at (-\h/2,-0.1) {$h_{i-1}$};
\end{tikzpicture} }
=\ket{\begin{tikzpicture}[baseline={([yshift=-.5ex]current bounding box.center)},thick,scale=0.6, every node/.style={scale=1.0}]
  \def\h{1.5}     
    \fill[black] (0.5*\h,0) circle (3pt); 
    \fill[black] (-0.5*\h,0) circle (3pt); 
    \fill[black] (0,0.5*\h) circle (3pt); 
    \draw[thick] (0,0) -- (0,\h);  
    \node[right] at (0+0.1,0.5*\h) {$k g_{i-\frac{1}{2}}$};  
  \draw[thick] (0,0) -- (\h,0); 
    \node[below] at (\h/2,-0.1) {$k h_i $};
    \draw[thick] (-\h,0) -- (0,0);  
    \node[below] at (-\h/2,-0.1) {$h_{i-1}\overline{k}$};
\end{tikzpicture} }.
\end{align}
The plaquette terms are given by 
\begin{align}
&B_i^{\text{QD}} \ket{\begin{tikzpicture}[baseline={([yshift=-.4ex]current bounding box.center)},thick,scale=0.6, every node/.style={scale=1.0}]
  \def\h{1.5}     
    \fill[black] (0.5*\h,0) circle (3pt); 
    \fill[black] (0,0.5*\h) circle (3pt); 
    \fill[black] (\h,0.5*\h) circle (3pt); 
    \draw[thick] (0,0) -- (0,\h);  
    \node[left] at (0-0.1,0.5*\h) {$g_{i-\frac{1}{2}}$};
    \draw[thick] (\h,0) -- (\h,\h);  
    \node[right] at (\h+0.1,0.5*\h) {$g_{i+\frac{1}{2}}$};
  \draw[thick] (0,0) -- (\h,0); 
    \node[below] at (\h/2,-0.1) {$h_i$};
  \draw[thick] (-\h/2,0) -- (0,0);  
  \draw[thick] (\h,0) -- (\h+\h/2,0);  
\end{tikzpicture} }\\
&=\delta_{g_{i-\frac{1}{2}}, h_i g_{i+\frac{1}{2}}}
\ket{\begin{tikzpicture}[baseline={([yshift=-.4ex]current bounding box.center)},thick,scale=0.6, every node/.style={scale=1.0}]
  \def\h{1.5}     
    \fill[black] (0.5*\h,0) circle (3pt); 
    \fill[black] (0,0.5*\h) circle (3pt); 
    \fill[black] (\h,0.5*\h) circle (3pt); 
    \draw[thick] (0,0) -- (0,\h);  
    \node[left] at (0-0.1,0.5*\h) {$g_{i-\frac{1}{2}}$};
    \draw[thick] (\h,0) -- (\h,\h);  
    \node[right] at (\h+0.1,0.5*\h) {$g_{i+\frac{1}{2}}$};
  \draw[thick] (0,0) -- (\h,0); 
    \node[below] at (\h/2,-0.1) {$h_i$};
  \draw[thick] (-\h/2,0) -- (0,0);  
  \draw[thick] (\h,0) -- (\h+\h/2,0);  
\end{tikzpicture} }.
\end{align}
The $\text{Rep}(G)$ symmetry operators are realized as matrix product operators (MPO) as follows,
\beq
    \mathcal{A}_{\mu} = &\begin{tikzpicture}[baseline=(current bounding box),scale=1.2]
        \node[text width=3cm] at (-5,2){$\cdots$};
        \node[text width=3cm] at (.1,2){$\cdots$};
        \filldraw[fill = white] (-5.3,1.7) rectangle node {$\mu$} (-4.7,2.3);
        \draw (-4.3,2) --  (-4.7,2);
        \draw (-5.3,2) --  (-5.7,2);
        \draw (-5,2.3) --  (-5,2.5);
        \draw (-5,1.7) --  (-5,1.5);
        \node[below] at (-5,1.5) {$0$};
        \filldraw[fill = white] (-4.3,1.7) rectangle node {$\mu$} (-3.7,2.3);
        \draw (-3.3,2) --  (-3.7,2);
        \draw (-4.3,2) --  (-4.7,2);
        \draw (-4,2.3) --  (-4,2.5);
        \draw (-4,1.7) --  (-4,1.5);
        \node[below] at (-4,1.5) {$1$};
        \filldraw[fill = white] (-3.3,1.7) rectangle node {$\mu$} (-2.7,2.3);
        \draw (-2.3,2) --  (-2.7,2);
        \draw (-3.3,2) --  (-3.7,2);
        \draw (-3,2.3) --  (-3,2.5);
        \draw (-3,1.7) --  (-3,1.5);
        \node[below] at (-3,1.5) {$2$};
        \filldraw[fill = white] (-2.3,1.7) rectangle node {$\mu$} (-1.7,2.3);
        \draw (-1.3,2) --  (-1.7,2);
        \draw (-2.3,2) --  (-2.7,2);
        \draw (-2,2.3) --  (-2,2.5);
        \draw (-2,1.7) --  (-2,1.5);
        \node[below] at (-2,1.5) {$3$};
\end{tikzpicture}\\
    =&\sum_{\{h_i\}} \Tr\biggl[\cdots D^{\mu}(h_0) D^{\mu}(h_1) \cdots\biggr]\ket{\{h_i\}}\bra{\{h_i\}},
    \label{eq:A_mu symmetry}
\eeq
where $\mu\in \text{Rep}(G)$ is a representation, and $D^{\mu}(h)$ is the representation matrix of the element $h\in G$. Moreover, given any group element $g\in G$ we can define
\beq
    U_{\rm TP}^{(g)} = \prod_i R_g^{i+\frac{1}{2}},\text{ where }R_g^{i+\frac{1}{2}}\equiv \sum_{g'}(\ket{g' \overline{g}}\bra{g'})_{i+\frac{1}{2}}.
\eeq
The above is symmetric under $\text{Rep}(G)$ and commutes with the quantum double Hamiltonian. A corresponding adiabatic cycle can be given by a conjugation of the Hamiltonian by $U(\theta)=\prod_i u_{i+1/2}$, such that $u_{i+1/2}(0)=I$ and $u_{i+1/2}(2\pi)=R_g^{i+1/2}$. The operators $U_{\rm TP}$ are also called the ribbon operators in quantum double models, which creates anyonic excitations at the end points~\cite{kitaev2003fault}. In Fig.~\ref{fig:symTFT_RepG}, we show that $U_{\rm TP}$ can be understood as the pumping of a flux anyon. The multiplication of Thouless pump operators $\{U_{\rm TP}^{(g)}\}$ form the group $G$, which is the group $\Gamma$ for the canonical fiber functor of the $\text{Rep}(G)$ category.

Note that the bulk has $D(G)$ topological order where, a flux anyon $g$ will be conjugated into $hgh^{-1}$ by a gauge transformation (or by the action of $A^{\text{QD}}_{i,h}$ on the lattice). Hence, the gauge invariant flux anyons in the bulk are labeled by conjugacy classes $[g]$, with an internal Hilbert space with dimension given by the order of $[g]$. However, on the physical boundary the gauge symmetry is absent. As a result, the topological invariant for the flux anyons on the boundary reduce to $G$ group elements.

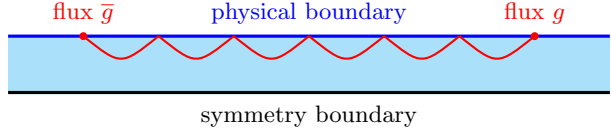
\begin{figure}[t!]
  \centering
  \begin{tikzpicture}[thick,scale=.5, every node/.style={scale=1.0}
  ]

  \def\n{5}       
  \def\gap{2}     
  \def\h{1.5}     
  \def\arc{\h-0.8}     
  \fill[cyan!30] (-1.5*\gap,0) rectangle (6.5*\gap,\h);
  \draw[line width=1.2pt] (-1.5*\gap,0) -- (6.5*\gap,0);
  \node[below] at (2.5*\gap,-.1) {symmetry boundary};
  \draw[blue,line width=1.2pt] (-1.5*\gap,\h) -- (6.5*\gap,\h);
  \node[blue,above] at (2.5*\gap,\h+.1) {physical boundary};
  \fill[red] (-\gap/2,\h) circle (3pt);
  \node[above] at (-\gap/2,\h+.1) {\textcolor{red}{flux $\overline{g}$}};
  \fill[red] (5.5*\gap,\h) circle (3pt);
  \node[above] at (5.5*\gap,\h+.1) {\textcolor{red}{flux $g$}};
  \foreach \i in {0,1,2,3,4,5} {
    \pgfmathsetmacro\x{\i*\gap}
    \draw[red] (\x-0.5*\gap,\h) .. controls (\x,\arc) .. (\x+0.5*\gap,\h);
    }
\end{tikzpicture}
  \caption{The $\text{Rep}(G)$ anyonic chain model can be interpreted as a quasi-1d system defined on the sandwich. A topological order $D(G)$ lives in the bulk, and a representation $\rho$ corresponds to an anyon $e_{\rho}$. Truncating the $U_{\rm TP}$ operator creates fluxes at the two endpoints, which cannot be condensed on the physical boundary for an SPT phase. Thus, the adiabatic cycle pumps a flux $g\in G$ along the quasi-1d system.}
  \label{fig:symTFT_RepG}
\end{figure}

\subsection{Minimal model for $\text{Rep}(G)$ symmetry}
\label{sec:minimal RepG}
To further simplify the quantum double model, we conjugate the system by a unitary $\prod_i CR_i CL_i$, where
\beq
    CR_i&\equiv \sum_{g, h}\left(\ket{g}\bra{g}\right)_{i-\frac{1}{2}}\otimes \left(\ket{h g}\bra{h}\right)_{i-1},\\
    CL_i&\equiv \sum_{g, h}\left(\ket{g}\bra{g}\right)_{i-\frac{1}{2}}\otimes \left(\ket{\overline{g} h}\bra{h}\right)_i.
\eeq
Then the vertex terms become $A_i'=\frac{1}{|G|}\sum_{k,g\in G}(\ket{k g}\bra{g})_{i-1/2}$, which fix all degrees of freedom on vertical edges to 
\beq
    \ket{+}\equiv \frac{1}{\sqrt{|G|}}\sum_{k\in G}\ket{k}.
\eeq
The rest of the system is effectively a $G$ spin chain. Since the vertex terms act as the 
identity on this low-energy subspace, the effective Hamiltonian for the canonical SPT phase 
is given by the conjugated plaquette terms alone,
\beq
    H_{\rm SPT} = -\sum_i B'_i, \qquad B'_i = \bigl(\ket{e}\bra{e}\bigr)_i .
\eeq
The ground state of this Hamiltonian is simply the product state $\ket{\phi}=\ket{\cdots,e,e,e,\cdots}$. An adiabatic cycle on this effective spin chain can be given by a conjugation of
\beq
    U^{(g)}(\theta) &= e^{\sum_j i\theta H^{(g)}_{j,j+1}}, \text{where} e^{i2\pi H^{(g)}_{j,j+1}} = R_g^{j}L_g^{j+1}.
    \label{eq:cycle unitary RL}
\eeq
The Thouless pump operator $U^{(g)}_{\rm TP}$ conjugates every $G$ spin by element $g\in G$. Applying the truncated unitary to the SPT state, we obtain 
\beq
    U_{\rm tr,TP}\ket{\phi} = \ket{\cdots, e, e, \overline{g}, e, e, \cdots, e, e, g, e, e, \cdots}.
\eeq
Namely, local defects $\overline{g}$ and $g$ are created at the end points by the truncated 
unitary. These are the localized invertible defects of Sec.~\ref{sec:qa evolution}, and 
since $U^{(g_1)}_{\rm TP}U^{(g_2)}_{\rm TP}=U^{(g_1g_2)}_{\rm TP}$, they realize the 
autoequivalence group $\Gamma=G$ for the canonical fiber functor.

We now turn to the remaining $\text{Rep}(G)$ SPT phases. It is shown in 
Appendix~\ref{app:QD} that a generic fiber functor for $\text{Rep}(G)$ corresponds to a pair 
$(K,\alpha)$, where $K\subset G$ is a subgroup and $\alpha\in Z^2(K,U(1))$ is a 2-cocycle of 
$K$; the canonical fiber functor above is the case $K=\{e\}$. Appendix~\ref{app:QD} showed that the SPT Hamiltonian for the 
pair $(K,\alpha)$ on the $G$ spin chain is composed of a sum of scalar multiples of mutually commuting projectors
\beq
    H_{\rm SPT} = -\sum_i \sum_{k\in K}  K'_{i,k} \Pi^{(K)}_{i-1} \Pi^{(K)}_i,
    \label{eq:hamiltonian K alpha}
\eeq
where 
\beq
   K'_{i,k}&= \sum_{k',k''\in K}\frac{\alpha(k,k'')}{\alpha(k'\overline{k},k)}\nonumber\\ 
   &\times \left(\ket{k'\overline{k}}\bra{k'}\right)_{i-1}\otimes \left(\ket{kk''}\bra{k''}\right)_{i}.
\eeq
The Rep($G$) symmetry operators are given by MPOs as in Eq.~\eqref{eq:A_mu symmetry}. 

For such generic $\text{Rep}(G)$ SPT phases, the group formed by Thouless pump operators is not necessarily identical to $G$. We now show that when $K$ is a normal subgroup it is instead an extension of the quotient $Q\equiv G/K$ by $H^1(K,U(1))$, built from two types of pump operator: one that acts by a phase on the $K$-spins, and one that shifts them.

We write the element $g\in G$ as a pair $g=(k,q)$, with multiplication rule
\beq
    (k_1, q_1) \cdot (k_2, q_2) = (k_1\cdot {}^{q_1}k_2\cdot n(q_1, q_2), q_1  q_2),
\eeq
where ${}^q k$ denotes the conjugation of $k\in K$ by $q\in Q$, and $n(q_1,q_2)\in K$ respects the cocycle condition with the conjugation action. When the conjugation is given by group homomorphism, $n(q_1, q_2)$ is a 2-cocycle representative in $H^2(Q,Z(K))$ associated with the conjugation action, with $Z(K)$ the center of $K$. We assume that the 2-cocycle $\alpha$ of $K$ is invariant in cohomology under the conjugation of $Q$, i.e., the conjugated $\alpha$ differs from $\alpha$ only by a coboundary of $\eta_q$,
\beq
    \frac{\alpha({}^q k_1, {}^q k_2)}{\alpha(k_1, k_2)} = \frac{\eta_q(k_1)\eta_q(k_2)}{\eta_q(k_1 k_2)}.
    \label{eq:Q-invariant cocycle}
\eeq

There are two types of Thouless pump operators. The first type, $U^{(\rho)}_{\rm TP_c}$, is diagonal: it acts on the total grading of a configuration by the corresponding character $\rho$,
\beq
    U_{\rm TP_c}^{(\rho)} = \prod_i Z_{\rho}^{(i)} =\sum_{\{k_i\}} \rho\Big(\prod_i k_i\Big)\ket{\{k_i\}}\bra{\{k_i\}},
\eeq
for a one-dimensional representation $\rho$ of $K$. Truncating it creates a $\rho$-charge at the cut. The operators $\{U^{(\rho)}_{\rm TP_c}\}$ are symmetric and commute with the Hamiltonian in Eq.~\eqref{eq:hamiltonian K alpha}, and they form the abelian group $H^1(K,U(1))$.

The second type shifts the spins and is defined for an element $q\in Q$ as
\beq
    U_{\rm TP_f}^{(q)} = \prod_i  R^{i}_{q} L^{i+1}_q \prod_i \left(Z_{\eta_q}^{(i)}\right)^\dagger,
    \\
    \text{where}\quad  Z_{\eta_q}^{(i)} = \sum_k\eta_q(k)(\ket{k}\bra{k})_{i}.
\eeq
The factor $\prod_i R^i_q L^{i+1}_q$ conjugates every spin by $q$, so that truncation creates a $q$-flux at the cut. The accompanying phase gate $\prod_i Z^{(i)}_{\eta_q}$ compensates the change of the cocycle $\alpha$ under conjugation by $q$, so that $U_{\rm TP_f}^{(q)}$ commutes with the Hamiltonian (see Appendix \ref{app:multiplication} for the proof).

The multiplication of two flux-type pump operators is (see Appendix~\ref{app:multiplication} for the derivation)
\beq
    U_{\rm TP_f}^{(q_1)}U_{\rm TP_f}^{(q_2)} = U_{\rm TP_f}^{(q_1 q_2)}U_{\rm TP_c}^{(\overline{\chi}\cdot\gamma)(q_1,q_2)},
    \label{eq:product of U_TP^q}
\eeq
where $(\overline{\chi}\cdot\gamma)(q_1,q_2)$ is a one-dimensional representation of $K$ for any $q_1,q_2\in Q$, defined as
\beq
    &(\overline{\chi}\cdot \gamma)(q_1,q_2) (x)\equiv \overline{\chi}(q_1,q_2)(x)\gamma_{k_{12}}(q_1,q_2)(k_{12}x\overline{k_{12}}),\\
    &\qquad \text{where}\ k_{12}={}^{\overline{q_1 q_2}}n(q_1,q_2).
\eeq
Moreover, $\chi(q_1, q_2)$ and $\gamma_{^{\overline{q_1 q_2}}n(q_1,q_2)}$ are functions of $K$ defined as
\beq
    \chi(q_1, q_2)(x)&\equiv\frac{\eta_{q_1}({}^{q_2}x)\eta_{q_2}(x)}{\eta_{q_1 q_2}(k_{12}x\overline{k_{12}})},\\
    \gamma_{k_{12}} (x)&\equiv \frac{\alpha(x,k_{12})}{\alpha(k_{12}, \overline{k_{12}}xk_{12})}.
\eeq
Eq.~\eqref{eq:product of U_TP^q} states that composing two flux-type pumps gives the flux-type pump for $q_1q_2$, but only up to a charge-type pump. The flux pumps therefore do not close into a group by themselves. This is precisely the statement that the full group of Thouless pumps is an extension of $Q$ by $H^1(K,U(1))$,
\beq
    1\rightarrow H^1(K, U(1)) \rightarrow G' \rightarrow Q \rightarrow 1.\label{eq:ext}
\eeq
Above, conjugation in $G$ induces the action of $Q$ on $H^1(K,U(1))$. The factor set $(\overline{\chi}\cdot \gamma)(q_1,q_2)$ appearing in the pump multiplication rule represents the extension class in $H^2(Q,H^1(K,U(1)))$.

To conclude, from the construction of Thouless pump operators we obtain that, when $K$ is a normal subgroup, the group of autoequivalences $\Gamma=G'$ for the module category $\mathcal{M}(K,\alpha)$ over $\text{Rep}(G)$, is given by the extension Eq.~\eqref{eq:ext}. The same result can be obtained mathematically following the derivation in Refs.~\cite{gelaki2007propertiesgrouptheoreticalcategories,naidu2007categorical,uribe2017classification}. When $K$ is abelian, this group extension is also discussed in Refs.~\cite{DAVYDOV2001273,nikshych2014categorical}. We obtained this result assuming that $\alpha$ is invariant under the action of $Q$. In general, when $\alpha$ is not invariant, $U^{(q)}_{\rm TP_f}$ is a Thouless pump operator only for $q\in Q'$, where $Q'$ is the subgroup of $Q$ that leaves $\alpha$ invariant in cohomology. The group of Thouless pumps is then given by the same extension with $Q$ replaced by $Q'$. For non-normal $K$, the quotient entering the extension is the subgroup of $N_G(K)/K$ that leaves $\alpha$ invariant in cohomology. We note that although in what follows our examples are Thouless pumps of SPT phases, the extension Eq.~\eqref{eq:ext} classifies the Thouless pumps of all gapped phases with $\text{Rep}(G)$ symmetry, not only the SPT ones.

In the rest of this section, we discuss some concrete examples. 
We first consider Rep($D_8$) SPT phases, presenting the SPT lattice models and the Thouless pumps. It is worth making explicit in what sense symmetries are non-invertible in the examples we study. A simple object $\mu$ of a fusion category is invertible precisely when it has an inverse under fusion, i.e. when $\mu\cdot\bar\mu=\mathds{1}$. For $\mathcal{D}=\text{Rep}(G)$ this happens if and only if $\dim\mu=1$. The category $\text{Rep}(G)$ is therefore group-like exactly when $G$ is abelian, in which case it reduces to $\text{Vec}_{\hat{G}}$ with $\hat{G}$ the group of characters. Non-invertible symmetry requires a non-abelian $G$, which is why we take $G=D_8$ below. The simple objects of $\text{Rep}(D_8)$ are the four one-dimensional representations $\{\mathds{1},a,b,ab\}$ and a single two-dimensional representation $\sigma$. The one-dimensional objects fuse as the group $\mathbb{Z}_2\times\mathbb{Z}_2$ and are invertible, whereas $\sigma\cdot\sigma = \mathds{1}+a+b+ab$ so that no object $\nu$ satisfies $\sigma\cdot\nu=\mathds{1}$.  Equivalently, $\sigma$ has quantum dimension $\dim\sigma=2\neq 1$, whereas invertibility requires unit quantum dimension. The symmetry operator implementing $\sigma$ therefore has no inverse, and $\text{Rep}(D_8)$ is a genuinely non-invertible symmetry.

The canonical SPT phase of Rep($D_8$) corresponds to the subgroup $K=\{e\}$, for which the lattice model have been discussed above, with its pumps classified by group $\Gamma=D_8$. Hence below, we focus on the other two SPT phases that correspond to two different $\mathbb{Z}_2^2$ subgroups of $D_8 = <a,x|a^4=x^2=axax=e>$, along with some 2-cocycle $\alpha$.

The second SPT phase corresponds to the normal subgroup \(K=\langle a^2,x\rangle\cong\mathbb Z_2\times\mathbb Z_2\).
If we denote the group elements as $(i, j)\equiv x^i a^{2j}$ with $i,j=0,1$, we can take the corresponding 2-cocycle to be
\beq
    \alpha ((i_1, j_1),(i_2, j_2)) = (-1)^{i_1 j_2}.
    \label{eq:subgroup cocycle 1}
\eeq
The Hamiltonian in Eq.~\eqref{eq:hamiltonian K alpha} projects every local degree of freedom onto a $\mathbb{Z}_2^2$ spin. If we decompose this spin into two qubits, and define the basis states as 
\beq
    \ket{0,\tilde{0}} = \ket{e},\quad \ket{1,\tilde{0}} = \ket{x},\\
    \ket{0,\tilde{1}} = \ket{a^2},\quad \ket{1,\tilde{1}} = \ket{a^2 x},
\eeq
then the ground state is stabilized by operators
\beq
    X_{i-1}\tilde{Z}_{i} X_i,\quad \tilde{X}_{i-1}Z_{i-1}\tilde{X}_i.
\eeq
Applying the Hadamard gates $\prod_i H_i \equiv \prod_i \frac{X_i+Z_i}{\sqrt{2}}$ to this state results in the well-known cluster state, i.e. the ground state of $H'_{\text{cluster}}$ in Eq.~\eqref{eq:hamiltonian cluster}. Since this $\mathbb{Z}_2^2$ subgroup is normal in $D_8$, the conjugation by $a\in D_8$ is an automorphism of $\mathbb{Z}_2^2$. We denote the conjugation action by  $a$ as follows,
\beq
    ^a(i,j) = (i, j+i).
\eeq
The conjugation acts on the 2-cocycle $\alpha$ in Eq.~\eqref{eq:subgroup cocycle 1} only by a coboundary
\beq
    &\alpha({}^a(i_1,j_1),{}^a(i_2,j_2)) \\
    &= \alpha((i_1,j_1),(i_2,j_2))\cdot \delta\eta((i_1,j_1),(i_2,j_2)),
\eeq
where the 1-cochain $\eta((i,j))=\text{i}^{i}$ and its coboundary is $\delta \eta = (-1)^{i_1i_2}$. The Thouless pump operator $U_{\rm TP_f}^{(a)}$ is given by
\beq
    U_{\rm TP_f}^{(a)} = \prod_i R^i_a L^{i+1}_a \prod_i \biggl(Z_{\eta}^{(i)}\biggr)^{\dagger}.
\eeq
Since the ground state $\ket{\phi}$ is stabilized by each term of the Hamiltonian in Eq.~\eqref{eq:hamiltonian K alpha}, for $k=a^2$ and a particular site $i$, we have
\beq
     \ket{\phi}
     =R^i_{a^2} L^{i+1}_{a^2} \cdot \left(\sum_{k} \eta^2(k)(\ket{k}\bra{k})_i \right)\ket{\phi}.
\eeq
Thus, the square of operator $U_{\rm TP_f}^{(a)}$ acts trivially on the ground state even with the truncation. The other types of Thouless pump operators are given by the representations of $\mathbb{Z}_2^2$ denoted as $\rho^{(r)}$. They are generated by the following operators,
\beq
    U_{\rm TP_c}^{(1)} &= \prod_i \rho^{(1)}(k_i)\ket{k_i}\bra{k_i},\\
    U_{\rm TP_c}^{(2)} &= \prod_i \rho^{(2)}(k_i)\ket{k_i}\bra{k_i},
\eeq
where $\rho^{(1)}((i,j))=(-1)^i$ and $\rho^{(2)}((i,j))=(-1)^j$.  In addition, one can show
\beq
    U_{\rm TP_f}^{(a)}U_{\rm TP_c}^{(2)} = U_{\rm TP_c}^{(1)}U_{\rm TP_c}^{(2)}U_{\rm TP_f}^{(a)}.
\eeq
The pumps generate a $D_8$ group since we can identify $U_{\rm TP_c}^{(1)}$ as $a^2$, $U_{\rm TP_c}^{(2)}$ as $x$, and $U_{\rm TP_f}^{(a)}$ as $xa$. These pumps classify the Thouless pumps for this Rep($D_8$) SPT.

Having discussed the canonical SPT phase, corresponding to \(K=\{e\}\), and the phase associated with \(K=\langle a^2,x\rangle\), we now consider the remaining \(\operatorname{Rep}(D_8)\) SPT phase. This phase corresponds to the other normal subgroup \(K=\langle a^2,xa\rangle\cong\mathbb Z_2\times\mathbb Z_2\), together with the corresponding cocycle. As in the preceding case, an appropriate cocycle representative gives a ground state that is equivalent to a cluster state up to on-site Hadamard gates. The Thouless pump operators can be constructed analogously and again form a group isomorphic to \(D_8\). Thus, although the pump group \(G'\) in the general extension need not be isomorphic to \(G\), all three \(\operatorname{Rep}(D_8)\) SPT phases considered here have pump group \(D_8\).

We can use our result in Eq.~\eqref{eq:ext} to determine the classification of Thouless pumps in Rep($G$) SPT phases without constructing lattice models. We consider Rep($D_8\times \mathbb{Z}_2$) symmetry as an example. Here, the canonical SPT phase that corresponds to $K=\{e\}$ has Thouless pumps classified by $D_8\times \mathbb{Z}_2$ as expected. There is a second SPT phase corresponding to $K = D_8\times \mathbb{Z}_2$~\cite{schnabel2016simple}. Since the quotient group is trivial for this example, the Thouless pumps of this SPT are classified by $H^1(K,U(1))=\mathbb{Z}_2^3$, the first cohomology group of $K$.

As a further example, consider $G=D_{2n}\times\mathbb Z_n$, with $n>1$ odd. We focus on the family of $\operatorname{Rep}(G)$-symmetric gapped phases specified by the normal subgroup $K=\mathbb Z_n\times\mathbb Z_n$ and a cocycle class $p\in H^2(K,U(1))\simeq\mathbb Z_n$. Within this family, the SPT phases are those for which $\gcd(p,n)=1$.

The available flux-type pumps are determined by the action of the quotient $Q=G/K\simeq\mathbb Z_2$ on the cocycle class. The nontrivial quotient element is represented by a dihedral reflection, which inverts the first $\mathbb Z_n$ factor of $K$ and leaves the second unchanged. Its action therefore sends $p\mapsto-p$. For the trivial class $p=0$, this action preserves the class, so both flux-type pumps $U^{(q)}_{\mathrm{TP}_f}$ and charge-type pumps $U^{(\rho)}_{\mathrm{TP}_c}$ are present. Together, they generate
\beq
\Gamma\simeq
(\mathbb Z_n\times\mathbb Z_n)\rtimes\mathbb Z_2
\simeq D_{2n}\times\mathbb Z_n.
\eeq
In contrast, for the SPT phases with \(\gcd(p,n)=1\), oddness of \(n\) implies \(p\not\equiv-p\pmod n\). The nontrivial quotient element therefore does not preserve the cocycle class and cannot support a flux-type pump. Only the charge-type pumps remain, giving
\beq
\Gamma=H^1(K,U(1))
\simeq\mathbb Z_n\times\mathbb Z_n.
\eeq
These results match the discussion in Ref.~\cite{disctor}, which gives
\beq\label{disctor-eq}
    &\operatorname{End}_{\operatorname{Vec}_{D_{2n}\times\mathbb{Z}_n}}(\mathcal{N}(\mathbb{Z}_n\times \mathbb{Z}_n,p))\\
    &=\begin{cases}
         \operatorname{Vec}_{D_{2n}\times\mathbb{Z}_n}, \,\,\,  p=0\\
         \operatorname{TY}(\mathbb{Z}_n\times \mathbb{Z}_n,p), \,\,\, \operatorname{gcd}(p,n)=1\\
         ``\operatorname{Vec}_{\mathbb{Z}_{q}\times\mathbb{Z}_q}\boxtimes\operatorname{TY}(\mathbb{Z}_{n/q}\times \mathbb{Z}_{n/q},p/q)",\\
         \operatorname{gcd}(p,n)=q
    \end{cases}
\eeq
where the quotation marks indicate a $\mathbb{Z}_q\times\mathbb{Z}_q$ graded extension of $\operatorname{TY}(\mathbb{Z}_{n/q}\times\mathbb{Z}_{n/q},p/q)$. The invertible objects of these categories form the group $\Gamma$ obtained above.

Finally, for a generic pair $(K,\alpha)$ the total number of distinct Thouless pumps of the corresponding $\text{Rep}(G)$ SPT state, i.e. the order of $\Gamma$, is~\cite{ostrik2006modulecategoriesdrinfelddouble,gelaki2007propertiesgrouptheoreticalcategories}
\beq
    |\Gamma| = \bigl|H^1(K,U(1))\bigr|\, \bigl|N_G(K,[\alpha])/K\bigr|,
    \label{eq:ostrik-count}
\eeq
where
\beq
    N_G(K,[\alpha]) = \{g\in N_G(K):[\alpha^g]=[\alpha]\},
\eeq
and $\alpha^g(k_1,k_2)\equiv\alpha(gk_1\bar g,\,gk_2\bar g)$ is the conjugated cocycle. The Thouless pump operators are still given by $U^{(\rho)}_{\rm TP_c}$ and $U^{(q)}_{\rm TP_f}$, where the former correspond to $\rho \in H^1(K,U(1))$ and the latter correspond to the embedding of $q\in N_G(K,[\alpha])/K$ into group the $G$.

\subsection{Minimal model for Rep($H$) symmetry}

The group-based models discussed above admit a natural generalization in which the finite group $G$ is replaced by a finite-dimensional
semisimple Hopf $C^*$-algebra $H$.  This replacement preserves the basic ingredients needed for a lattice construction. As will be defined later, the
multiplication in $H$ generalizes group multiplication, the comultiplication specifies how an algebra element acts consistently on a tensor product of local Hilbert spaces, and the antipode generalizes group inversion.  The Haar integrals of $H$ and its dual $H^*$ provide, respectively, the commuting-projector terms and the inner product needed to turn the underlying vector space of $H$ into a physical Hilbert space.

The corresponding categorical symmetry is $\operatorname{Rep}(H)$. Although this symmetry is generally non-invertible, it can still be
realized on an ordinary tensor-product spin chain by matrix-product operators. In the group-algebra limit, $H=\mathbb{C}[G]$, this construction reduces to the $G$-spin chain and the quantum-double models considered previously. The Hopf-algebra formulation should therefore be viewed not as a separate construction, but as the minimal extension of the group-based lattice model to genuinely non-invertible symmetries.

By the \emph{minimal model} for $\operatorname{Rep}(H)$ symmetry, we mean the fixed-point spin chain obtained by placing one copy of the
regular Hilbert space $H$ at each site.  The dual Haar integral selects a canonical product-state ground state, while the representations of
$H$ generate the $\operatorname{Rep}(H)$ symmetry MPOs. The Thouless pumps of this canonical SPT phase are then associated with
the group-like elements of $H$, namely the elements $k\in H$ satisfying $\Delta(k)=k\otimes k$. This condition is the Hopf-algebra analogue of the statement that a group element acts multiplicatively on a tensor product.

To illustrate the construction in the simplest genuinely non-group-like setting, we will specialize to the eight-dimensional Kac-Paljutkin Hopf algebra $H_8$. It is the smallest semisimple Hopf algebra that is neither commutative nor cocommutative, and $\operatorname{Rep}(H_8)$ contains a non-invertible simple object. Below, we define a general semisimple Hopf algebra, summarize the specific algebraic data of $H_8$, then construct the corresponding local Hilbert space, commuting-projector Hamiltonian, and symmetry MPOs, and finally identify the Thouless pumps of its canonical $\operatorname{Rep}(H_8)$ SPT state.

A fusion category that admits SPT phases is always a representation category of a finite dimensional semisimple Hopf algebra~\cite{etingof2015tensor,inamura2022lattice}. We use $H=(H;\mu,\eta;\Delta,\epsilon;S)$ to denote a Hopf algebra with underlying $\mathbb{C}$-vector space $H$, multiplication $\mu$, unit $\eta$, comultiplication $\Delta$, counit $\epsilon$, and antipode $S$. Given a finite group $G$, a finite dimensional  Hopf $C^*$-algebra can be defined on vector space $\mathbb{C}[G]$ with data
\beq
    \mu(g,h) = gh,\quad \eta = e,\quad \Delta(g) = g\otimes g,\\
    \epsilon(g) = 1,\quad S(g) = g^{-1},\quad g^*=g^{-1},
\eeq
where an element $g\in G$ denotes a basis vector in $\mathbb{C}[G]$. Further equipped with the inner product $(g,h)=\delta_{g,h}$, this Hopf algebra gives rise to a Hilbert space of a $G$ spin, which is the local degree of freedom on each edge of the quantum double model we discussed above. The generalized quantum double model from group algebra $\mathbb{C}[G]$ to Hopf $C^*$-algebra have been studied, as well as the mapping from this model to the Levin-Wen's string-net model~\cite{buerschaper2013hierarchy,buerschaper2013electric,balsam2012kitaev}. Ref.~\cite{jia2023boundary} also discusses the mapping between a class of boundaries for generalized quantum double model and that of the extended string-net model. Therefore, when the symmetry is the representation category of a finite dimensional Hopf $C^*$-algebra Rep($H$), using this map to the anyonic chain model defined in Eq.~\eqref{eq:anyonic chain}, we obtain a lattice model defined on a tensor product Hilbert space, which is called the Hopf ladder model or the generalized cluster model in Ref.~\cite{jia2024generalized}. In Appendix \ref{app:generalized qd}, we review the generalized quantum double model on a strip. In this section, we take the Kac-Paljutkin Hopf algebra $H_8$ as an example. We will demonstrate the minimal lattice model for the Rep($H_8$) SPT along with the Thouless pump operators on the SPT state.

Following Ref.~\cite{Masuoka1995SemisimpleHA,buerschaper2013hierarchy,meng2024non}, the Kac-Paljutkin Hopf algebra $H_8$ has generators $x,y,z$ and the following relations (the multiplication symbol $\mu$ is omitted here)
\beq
    &x^2=y^2=1,\quad z^2 = \frac{1}{2}(1+x+y-xy),\\
    &xy=yx,\quad zx=yz,\quad zy=xz.
\eeq
Hence, the set $\{1,x,y,xy,z,zx,zy,zxy\}$ forms the basis of the vector space $H_8$. The comultiplication relations are given by
\beq
    &\Delta(x)=x\otimes x,\quad \Delta(y)=y\otimes y,\\ &\Delta(z)=\frac{1}{2}(1\otimes 1+y\otimes 1+1\otimes x-y\otimes x)(z\otimes z).
\eeq
When an element $a\in H$ satisfies $\Delta(a)=a\otimes a$, it is called a group-like element. For $H_8$, they form the following group,
\beq
    \mathcal{G}(H_8) = \mathbb{Z}_2\times \mathbb{Z}_2 = <x,y>.
\eeq
The counit is defined as $\epsilon(x)=\epsilon(y)=\epsilon(z)=1$, and the antipod acts trivially, i.e., $S( h ) = h$. Lastly, the $C^*$-structure gives a conjugation involution
\beq
    x^*&=x,\quad y^*=y,\\ z^*&=z^{-1}=\frac{1}{2}(z+zx+zy-zxy).
\eeq
The Haar integral $h\in H_8$ is
\beq
    h=\frac{1}{8}(1+x+y+xy+z+zx+zy+zxy).
\eeq
Given a Hopf algebra $(H;\mu,\eta;\Delta,\epsilon;S)$, a dual Hopf algebra can be defined as $H^*=(H^*; \Delta^T,\epsilon^T;\mu^T,\eta^T;S^T)$. The Haar integral of the dual $H^*_8$ algebra is the dual of the unit vector,
\beq
    \phi = \delta_1.
\eeq

It has been shown in Ref.~\cite{buerschaper2013hierarchy} that the existence of Haar integral $\phi\in H^*$ allows for defining the inner product in $H$, 
\beq
    (a,b)_H := \phi(a^* b),
\eeq
which gives rise to a Hilbert space structure on $H$. Furthermore, with the Haar integral $h\in H$, a commuting projector Hamiltonian that generalizes the quantum double model from finite groups to finite dimensional $C^*$ Hopf algebra can be defined. The corresponding minimal model for the Rep($H_8$) SPT is defined on a spin chain, where each site hosts an eight dimensional Hilbert space given by $H_8$, and the Hamiltonian is
\beq
    H = -\sum_i \phi^{(i)},
\eeq
where $\phi^{(i)} = \sum_h \phi(h) \ket{h}_i \bra{h}$, and $\phi$ is the dual Haar integral. The ground state is therefore a product state $\ket{\cdots, 1, 1, 1, \cdots}$. The Rep($H_8$) symmetry operators are given by the following MPO,
\beq
    \mathcal{A}_{\mu} = &\begin{tikzpicture}[baseline=(current bounding box),scale=1.2]
        \node[text width=3cm] at (-5,2){$\cdots$};
        \node[text width=3cm] at (.1,2){$\cdots$};
        \filldraw[fill = white] (-5.3,1.7) rectangle node {$\mu$} (-4.7,2.3);
        \draw (-4.3,2) --  (-4.7,2);
        \draw (-5.3,2) --  (-5.7,2);
        \draw (-5,2.3) --  (-5,2.5);
        \draw (-5,1.7) --  (-5,1.5);
        \node[below] at (-5,1.5) {$0$};
        \filldraw[fill = white] (-4.3,1.7) rectangle node {$\mu$} (-3.7,2.3);
        \draw (-3.3,2) --  (-3.7,2);
        \draw (-4.3,2) --  (-4.7,2);
        \draw (-4,2.3) --  (-4,2.5);
        \draw (-4,1.7) --  (-4,1.5);
        \node[below] at (-4,1.5) {$1$};
        \filldraw[fill = white] (-3.3,1.7) rectangle node {$\mu$} (-2.7,2.3);
        \draw (-2.3,2) --  (-2.7,2);
        \draw (-3.3,2) --  (-3.7,2);
        \draw (-3,2.3) --  (-3,2.5);
        \draw (-3,1.7) --  (-3,1.5);
        \node[below] at (-3,1.5) {$2$};
        \filldraw[fill = white] (-2.3,1.7) rectangle node {$\mu$} (-1.7,2.3);
        \draw (-1.3,2) --  (-1.7,2);
        \draw (-2.3,2) --  (-2.7,2);
        \draw (-2,2.3) --  (-2,2.5);
        \draw (-2,1.7) --  (-2,1.5);
        \node[below] at (-2,1.5) {$3$};
\end{tikzpicture}\\
    =&\sum_{\{h_i\}} \Tr\biggl[D^{\mu}(h_0^{(2)}) D^{\mu}(h_1^{(2)}) \cdots\biggr]\ket{\{h_i^{(1)}\}}\bra{\{h_i\}},
\eeq
where $\mu\in \text{Rep}(H_8)$ labels representation, $D^{\mu}(h)$ is the representation matrix of element $h\in H_8$, and the notations $h^{(1)}, h^{(2)}$ come from the comultiplication of $h$,
\beq
    \Delta(h) = \sum h^{(1)}\otimes h^{(2)}.
\eeq
See Appendix~\ref{app:generalized qd} and Refs.~\cite{jia2023boundary,jia2024generalized} for more details. 

From our general result, the classification of Thouless pumps is given by group $\Gamma$. For the canonical SPT state of Rep($H$), $\Gamma$ is formed by the group-like elements in $H$~\cite{jia2023boundary,inamura20241+}. Particularly for $H_8$, different classes of Thouless pumps form group $\Gamma = \mathcal{G}(H_8)$, with the Thouless pumps operators
\beq
    U_{\rm TP}^{(k)} = \prod_j R_k^{j-1}L_k^{j},
\eeq
where $R_k^{j} = \sum_{a}\ket{a S(k)}_j \bra{a}$, $L_k^{j} = \sum_{a}\ket{k a}_j \bra{a}$, and $k\in \{1,x,y,xy\}$.

\subsection{Pump of non-invertible SPT states}

As mentioned in the beginning of this work, Kitaev used a pumping picture to propose that the space of SRE states $\{M^G_d\}$ with ordinary symmetries form an $\Omega$-spectrum. Namely, the homotopy classes of adiabatic cycles in $d$ dimension correspond to the pumps of $d-1$ dimensional SRE states. In this section, we will show its parallel in 1d non-invertible SPT states. 

As we have seen, the Rep($H$) symmetry can be realized on the spin chain as an MPO, which is also known as the on-site (strictly locality preserving) realization~\cite{meng2024non,evans2025operator}. Furthermore, it is known that for any fiber functor $F$ of $\mathcal{D}$, the pair $(\mathcal{D},F)$ is equivalent to the pair $(\text{Rep}(H), F')$, where $\text{Rep}(H)$ is the representation category of a Hopf algebra given from the Tannakian reconstruction, and $F'$ is the forgetful fiber functor~\cite{etingof2015tensor}. Hence, we can realize any non-invertible SPT on the spin chain as the canonical SPT state for some Rep($H$) symmetry. From the results of the last section, the classification of Thouless pumps is given by the group $\mathcal{G}(H)$.

To see the pump of such 0d SPT states on the lattice, we define the 0d symmetry operators by taking the on-site realization of $\text{Rep}(H)$ for a single site, which are the characters $\{\chi_{\mu}\}$. The action of symmetries operators on the basis of the single site Hilbert space $\{\ket{a}| a\in H\}$ is
\beq
    \chi_{\mu}\ket{a} =\Tr(D^{\mu}(a))\ket{a}.
\eeq
The fusion of symmetry operators is
\beq
    \chi_{\mu}\cdot \chi_{\nu}\ket{a} &= \Tr(D^{\mu}(a))\Tr(D^{\nu}(a))\ket{a}\\
    &=\Tr(D^{\mu}\otimes D^{\nu}(a\otimes a))\ket{a}.
\eeq
Now that we require the fusion rule of the non-invertible symmetries to align with that of $\text{Rep}(H)$, we have the following relation,
\beq
    \chi_{\mu}\cdot \chi_{\nu}\ket{a} &=\sum_{\rho}N^{\mu\nu}_{\rho}\Tr(D^{\rho}(a))\ket{a}\\
    &=\Tr(D^{\mu}\otimes D^{\nu}(\Delta(a)))\ket{a}.
\eeq
Combining the above two equations we have
\beq
    \Tr (D^{\mu}\otimes D^{\nu} (\Delta(a) - a\otimes a)) = 0,
\eeq
for every $\mu,\nu\in \text{Rep}(H)$. Therefore, the 0d state $\ket{a}$ is a non-invertible SPT when $a\in \mathcal{G}(H)$. 

We recall from the last section that the canonical SPT state is $\ket{\cdots, 1, 1, 1, \cdots}$, and the Thouless pump operator is $U^{(k)}_{\rm TP} = \prod_j R_k^{j-1} L_k^j$ for a group-like element $k\in\mathcal{G}(H)$. The truncated Thouless pump operator thus creates a state with invertible defects on site $i$ and $j$,
\beq
    \ket{\cdots,1,1,\overline{k}, 1, 1, \cdots, 1, 1, k,1,1, \cdots}.
\eeq
Therefore, the adiabatic cycles in 1d non-invertible SPT indeed correspond to the pump of 0d SPT states.

Generally speaking, the non-invertible symmetries in 0d form a ring $R$. The corresponding non-invertible symmetry for Rep($H$) is the fusion ring obtained by ignoring the $F$-symbols, i.e., for $\mu,\nu\in R$,
\beq
    \mu \cdot \nu = \sum_{\rho} N^{\mu\nu}_{\rho} \rho.
    \label{eq:0d fusion}
\eeq
An SPT state $\ket{\psi}$ with this non-invertible symmetry should satisfy 
\beq
    \mu\ket{\psi}\propto\ket{\psi}.
\eeq
Thus, inequivalent classes of such SPT states are classified by $\text{Hom}(R, \mathbb{C})$. As shown above, given a group-like element $a\in H$, its character qualifies as a homomorphism from $R$ to $\mathbb{C}$. However, there exist homomorphisms beyond the evaluation of characters on group-like elements, and vice versa not all such morphisms give inequivalent SPTs. For example, the fusion rings of $\text{Rep}(D_8)$ and $\text{Rep}(H_8)$ are the same, which we denote as $R=\{1,a,b,ab,\sigma\}$. There are five homomorphisms given by
\beq
    \phi(a) &=  \pm 1,\quad  \phi(b) = \pm 1,\quad \phi(\sigma) = 0,\\
    \text{or }
    \phi(a) &=  1,\quad \phi(b) = 1,\quad \phi(\sigma) =\pm 2.
\eeq
Only part of the above homomorphisms are obtained by evaluating the character of a representation of $H_8$ on one of its group elements (we recall that $\mathcal{G}(H_8)=\mathbb{Z}_2\times \mathbb{Z}_2$).
On the other hand, although all the above homomorphisms correspond to evaluating the characters on some elements in $D_8$, elements in the same conjugacy class give the same homomorphism (in contrast with the fact that $\mathcal{G}(\mathbb{C}[D_8])=D_8)$. Therefore, the map from $\mathcal{G}(H)$ to $\text{Hom}(R, \mathbb{C})$ is neither injective nor surjective. All Thouless pumps of 1d non-invertible SPT phases can be associated with 0d non-invertible SPTs, but this may not include all possible 0d non-invertible SPTs. Finally, while the set of group-like elements of a Hopf algebra is indeed a group, we do not expect a multiplicative structure on the space of 0d SPT protected by a non-invertible symmetry as there is no ``diagonal" ring homomorphism $R\to R\times R$.\footnote{The Hochschild cohomology $H^2(\mathcal{C},\Z)$ is discussed in Ref.~\cite{inamura20241+}, which also reduces to the 0d SPT classification  $H^1(G,U(1))$ in the invertible case $\mathcal{C}=\operatorname{Vec}_G$. However, it also suffers from the same drawbacks: the canonical map $\Gamma\to H^2(\mathcal{C},\Z)$ is neither injective nor surjective.}

\section{Floquet models from Thouless pumps}
\label{sec:floquet}

Having classified the quantized defects transported by Thouless pumps in the preceding sections, we now use these pumps as building blocks for periodically driven systems. On a closed chain, a complete pumping cycle acts trivially on the bulk up to an exact symmetry of the fixed-point Hamiltonian. When the same evolution is spatially truncated to an open chain, however, it leaves localized defects at the two boundaries. In a Floquet system, these boundary defects manifest themselves as robust pairings in the quasienergy spectrum and therefore provide dynamical signatures that distinguish different Floquet phases, a phenomenon well established for $\mathbb{Z}_2$ pumps. This connection allows the classification of Thouless pumps developed above to be translated directly into a classification of Floquet dynamics. In Refs.~\cite{motamarri2024symtftequilibriumtimecrystals,mclauchlan2025classifyingonedimensionalfloquetphases}, a classification of Floquet phases with non-invertible symmetry is proposed using symTFT under the assumption of many-body localization. Here, we instead take a bottom-up approach, constructing exactly solvable Floquet models with both invertible and non-invertible symmetries and study their phase diagrams. We begin with finite group symmetries and then extend the discussion to non-invertible Rep$(G)$ symmetries, presenting promising models for generalized Floquet SPT phases and Floquet symmetry-breaking phases~\cite{khemani2016phase,von2016phase2}.

In the previous sections we have shown that the Thouless pump operators $U_{\rm TP}$ in our fixed point model are manifestly exact symmetries of the Hamiltonian. This enables us to define periodically driven (Floquet) systems from 
\begin{align}
    U_{\rm TP} = e^{i 2\pi H_{\rm TP}}.
\end{align}
Non-trivial Floquet drives are those that have non-trivial pairings in the whole quasi-spectra. In this section, we will discuss Floquet binary drives, i.e, dynamics arising from piecewise time independent Hamiltonians,
\beq
H(t) =
\begin{cases}
H_{\rm SPT}, & 0 \leq t < T_1, \\
-H_{\rm TP}, & T_1 \leq t < T_1 + T_2,
\end{cases}
\eeq
with total period $T = T_1 + T_2$. The corresponding Floquet unitary (time evolution over one period) is then
\beq
U_F(T_1, T_2) = e^{i H_{\rm TP} T_2} \, e^{-i H_{\rm SPT} T_1}. \label{eq:floquet-binary-unitary}
\eeq

By definition, the Floquet unitary $U_F(T_1, T_2+2\pi)$ differs from $U_F(T_1, T_2)$ by a Thouless pump operator $U_{\rm TP}$. Therefore, by spatially truncating the evolution to an open chain, 
the quasi-spectra of $U_{F}(T_1, T_2)$ and $U_{F}(T_1, T_2+2\pi)$ have different pairings as the latter has additional edge-defects than the former. Spectral pairings on the open chain are the characteristics for different Floquet SPT (FSPT) phases. For group-like symmetries, FSPT phases are inherent dynamical phases defined by eigenstate order in the quasi-spectrum; their classification being given by $H^1(G,U(1))$~\cite{von2016phase,else2016classification,potter2016classification,roy2016abelian,harper2020topology}, which is exactly the classification of $G$ Thouless pumps. This correspondence is natural from the perspective of Ref.~\cite{else2016classification}. In a localized Floquet system, evolution through one driving period generates a discrete time-translation symmetry \(\mathbb{Z}\), enlarging the symmetry group from \(G\) to \(G\times\mathbb{Z}\). In the decomposition
$$
H^2(G\times\mathbb{Z},U(1))
\simeq H^2(G,U(1))\times H^1(G,U(1)),
$$
the \(H^1(G,U(1))\) factor arises from the projective commutation relation between \(G\) and one-period time evolution and labels the \(G\) charge transported during each driving period. It therefore reproduces precisely the classification of Thouless pumps obtained in the preceding sections. Eigenstate SPT order with non-invertible $\mathrm{Rep}(D_8)$
symmetry was studied for disordered Hamiltonians in
Ref.~\cite{Yabo-Mitra26}, which also investigated Floquet
edge modes with period-doubled dynamics. The corresponding
study of eigenstate SPT order in disordered Floquet systems
was left open in that work. In the following, we will present exactly solvable Floquet problems with various symmetries, along with the phase diagram in the parameter space $(T_1, T_2)$, and study how the notion of FSPT phases and Floquet symmetry breaking phases~\cite{khemani2016phase,von2016phase2} can be extended to non-invertible symmetries.

\subsection{$G$ Floquet problems}

For the minimal model with a finite abelian group-like symmetry $G$, the Thouless pump Hamiltonians are defined as
\beq
H_{\rm TP}^{(\rho)} = \sum_j H^{(\rho)}_{j-1,j},
\eeq
where each local term satisfies  $e^{i2\pi H^{(\rho)}_{j-1,j}} = Z^{(j-1)}_{\rho}(Z^{(j)}_{\rho})^{\dagger}$ for $\rho\in H^1(G,U(1))$. The SPT Hamiltonians are defined by a 2-cocycle $\omega$ as described in Sec.~\ref{sec:minimal VecG}. They are related to each other by SPT entanglers, which are diagonal phase gates. Since the Thouless pump Hamiltonians we use are also diagonal, they commute with each other. Hence, we will focus on the trivial SPT Hamiltonian with a product of $\ket{+}=\frac{1}{\sqrt{|G|}}\sum_g\ket{g}$ at each site as its ground state, whereas the Floquet unitary for other SPT Hamiltonians is given by conjugations of SPT entanglers. We will start by introducing the Floquet Ising problem for $\mathbb{Z}_2$, moving on to abelian symmetries, and finally discussing general finite groups. Using results from Onsager algebra~\cite{onsager1944crystal,dolan1982conserved,vladimir2017integrable,vernier2019onsager,jones2025pivoting,jones2025charge}, we will identify a family of integrable Floquet problems, making it easier to identify the phase structures. The subtle issue of localization effects with disorder is left for future studies.

The $\mathbb{Z}_2$ Floquet unitary corresponds to the well-known Floquet Ising model~\cite{prosen1998new,prosen2000exact}, 
\beq
    U_F(T_1, T_2) = e^{i \frac{T_2}{4} \sum_j Z_{j-1}Z_{j}}e^{i \frac{T_1}{4} \sum_j X_j},
\eeq
Here $H_{\rm SPT}=-\frac14\sum_jX_j$ and $H_{\rm TP}=+\frac14\sum_jZ_{j-1}Z_j$, so that $H_{\rm SPT}$ has the product ground state $\otimes_j|+\rangle_j$ and $e^{+i\pi Z_{j-1}Z_j/2}=iZ_{j-1}Z_j$ implements the charge pump. The convention with negative signs in both Floquet exponents is obtained by $(T_1,T_2)\mapsto(-T_1,-T_2)$. This Floquet unitary corresponds to the row-to-row transfer matrix in the classical Ising model on a 2d square lattice through analytic continuation~\cite{onsager1944crystal,baxter1982exactly}. It shares the same phase diagram and phase transitions as shown in Fig.~\ref{fig:phase} (and discussed in more detail below). This phase diagram not only includes the standard ordered/disordered phases in the classical Ising model with a self-dual critical point, it also shows additional phases. In particular, there are a total of four distinct Floquet phases due to different stroboscopic dynamics. In fact, this Floquet unitary, after a Jordan-Wigner transformation, can be mapped to an evolution under a time-independent fermion bilinear Floquet Hamiltonian, with the Floquet phases distinguished by the winding numbers of the single-particle eigenstates in momentum space~\cite{Asboth14,khemani2016phase,von2016phase,harper2020topology,yates2017entanglement,Yates18,yeh2023decay}. 

Now we present an alternate approach to obtain the phase structure,  by employing the Onsager algebra generated by the driving Hamiltonians. This algebra characterizes the phase structure of not only the Floquet Ising model, but various other Floquet problems discussed below. 

The Onsager algebra generated by operators $\{G_n\}$ and $\{A_n\}$ is defined as~\cite{dolan1982conserved,vladimir2017integrable}\footnote{The exact expressions for the generators in the Ising model are discussed in Ref.~\cite{prosen2000exact}.},  
\beq
    [A_m, A_n] &= 4 G_{m-n},\\
    [G_m, A_n] &= 2(A_{n+m}-A_{n-m}),\\
    [G_m, G_n] &= 0.
    \label{eq:onsager}
\eeq
For the Onsager realizations considered below, we use the convention
\begin{equation}
    A_0=-4H_{\rm SPT},\qquad A_1=-4H_{\rm TP}.
    \label{eq:floquet-onsager-sign-convention}
\end{equation}
In the Ising example this gives $A_0=\sum_jX_j$ and $A_1=-\sum_jZ_{j-1}Z_j$. Equation~\eqref{eq:floquet-binary-unitary} then takes the form
\begin{align}
U_F = e^{-\frac{i T_2}{4}A_1}e^{\frac{i T_1}{4}A_0},
\end{align}
a complete set of conserved charges in the Onsager algebra that commute with $U_F$ can be obtained from combinations of the elements above~\cite{prosen2000exact}, 
\beq
    Q_m = &c_1 s_2 (A_{m+1} + A_{-m+1}) - s_1 c_2(A_m + A_{-m}) \\
    &+ i s_1 s_2 (G_{m+1} -G_{m-1}),
    \label{eq:conserved charges}
\eeq
where $s_i=\sin(T_i/2)$ and $c_i=\cos(T_i/2)$. In principle, these conserved charges and their eigenvalues fully characterize this Floquet model, with the Floquet Hamiltonian being related to these conserved charges. To classify phases in the parameter space $(T_1, T_2)$ of a Floquet model, we usually rely on the existence of localized modes on the edge, which reveals stroboscopic motion within each period. However, they are not straightforwardly attainable even in integrable models because the charges are not localized and are by definition preserved in each period. In addition, the localized edge modes that appear with open boundaries are not related to the standard charges of integrable models. In Ref.~\cite{eric2024strong}, the localized modes on the edge of the Floquet XXZ model are constructed using transfer matrices. Here, we use a different but relevant quantity to classify the Floquet phases by performing a formal Fourier transform on the generators,
\beq
    A_m &= 2\int d\theta \big(e^{-im\theta}E_\theta^+ + e^{im\theta}E_\theta^-\big),\\
    G_m &= 2\int d\theta \big(e^{-im\theta} - e^{im\theta}\big)H_\theta,
\eeq
for rapidity $\theta\in [0,2\pi]$. The operators $\{E_\theta^{\pm}\}$ and $\{H_\theta\}$ satisfy $[E_\theta^+, E_{\theta'}^-] = 2\delta(\theta-\theta')H_\theta$, $[E_\theta^{\pm}, H_{\theta'}]=\mp \delta(\theta-\theta')E_\theta^{\pm}$. This means that the Onsager algebra is a fixed-point subalgebra of the $\mathfrak{sl}_2$ loop algebra. With suitable regularization this reduces to direct product of copies of $\mathfrak{sl}_2$ Lie algebra~\cite{ahn1991onsager}. The conserved charges in Eq.~\eqref{eq:conserved charges} is block diagonal in rapidity sectors. If we denote $J^{(1)}_\theta = \frac{1}{2}(E_\theta^+ + E_\theta^-)$, $J^{(2)}_\theta = \frac{1}{2i} (E_\theta^+ - E_\theta^-)$, $J^{(3)}_\theta = H_\theta$,
\beq
    Q_m &= 8\int d\theta \cos(m\theta) \Vec{n}_\theta \cdot \Vec{J}_\theta,\\
    \Vec{n}_\theta &=  (c_1 s_2 \cos{\theta} - s_1 c_2,  c_1 s_2 \sin{\theta},  s_1 s_2 \sin{\theta})^T.\label{eq:onsager-conserved-charge-vector}
\eeq
We have ignored the scalar charge in the above as they do not contribute in the winding number calculation.

For a finite system in a fixed Onsager representation sector,
the Fourier spin operators can be written as
$\vec J_\theta=\sum_a\delta(\theta-\theta_a)\vec S_a$,
where $\vec S_a$ acts in a spin-$j_a$ representation.
The rapidities $\theta_a$ and spins $j_a$ are fixed
within this sector.

At each rapidity, a rotation of the spin axis maps
$\vec n_{\theta_a}\cdot\vec S_a$ to
$|\vec n_{\theta_a}|S_a^z$.
Its eigenvalues are therefore $|\vec n_{\theta_a}|\mu_a$,
where $\mu_a=-j_a,-j_a+1,\ldots,j_a$ are integer or half-integer spin-projection eigenvalues.
Since the spin operators belonging to distinct factors commute,
the same spin-projection quantum numbers $\{\mu_a\}$
label a common eigenstate of all the conserved charges.
We encode these quantum numbers in the distribution
\begin{equation}
    \mu(\theta)
    = \sum_a \mu_a\,\delta(\theta-\theta_a).
    \label{eq:onsager-charge-projection-density}
\end{equation}

Substituting the spin-projection eigenvalues into
Eq.~\eqref{eq:onsager-conserved-charge-vector}, and using
$|\vec n_\theta|^2
=1-(c_1c_2+s_1s_2\cos\theta)^2$,
gives the corresponding conserved-charge eigenvalues:
\begin{equation}
    \begin{aligned}
        q_m
        &= 8\sum_a \mu_a\cos(m\theta_a)
           |\vec n_{\theta_a}|\\
        &= 8\int d\theta\,\mu(\theta)\cos(m\theta)
           |\vec n_\theta|\\
        &= 8\int d\theta\,\mu(\theta)\cos(m\theta)
           \sqrt{1-(c_1c_2+s_1s_2\cos\theta)^2}.
    \end{aligned}
    \label{eq:onsager-conserved-charge-eigenvalues}
\end{equation}

The quantum numbers $\mu_a$ specify the chosen eigenstate,
whereas the vector $\vec n_\theta$ is determined by
the drive parameters and the rapidity.
Its magnitude sets the mode-resolved charge amplitude.
In particular, this amplitude vanishes when $T_1=T_2$
at $\theta=0$, and when $T_1=2\pi-T_2$ at $\theta=\pi$.
Wherever $|\vec n_\theta|\neq 0$, the direction of
the spin-projection axis is described by the normalized vector
\begin{equation}
    \vec n^{\rm norm}_\theta
    = \frac{\vec n_\theta}{|\vec n_\theta|}
    = \frac{1}{|\vec n_\theta|}
      \begin{pmatrix}
          c_1s_2\cos\theta-s_1c_2\\
          c_1s_2\sin\theta\\
          s_1s_2\sin\theta
      \end{pmatrix}.
    \label{eq:onsager-normalized-projection-axis}
\end{equation}

The above vector lives on a plane as can be seen by introducing an orthonormal basis in the tilted plane,
\begin{equation}
\vec e_1 = (1,0,0), \qquad
\vec e_2 = (0,c_1,s_1),
\end{equation}
and decomposing the normal vector as
\begin{equation}
\Vec{n}^{\rm norm}
=\frac{1}{|\Vec{n}_\theta|}
\biggl[X(\theta)\,\vec e_1
+
Y(\theta)\,\vec e_2\biggr] .
\end{equation}
with $X(\theta)=c_1 s_2 \cos{\theta} - s_1 c_2$, $Y(\theta)=s_2\sin\theta$.
As $\theta$ varies from $0$ to $2\pi$, this vector traces a closed curve whose winding number around the origin takes the integer values shown in Fig.~\ref{fig:phase}.

To summarize, the Onsager structure organizes the conserved
charges into commuting $\mathfrak{sl}_2$ factors within each representation
sector. Their common eigenstates are labeled by the
spin-projection quantum numbers $\{\mu_a\}$, which are encoded
in the distribution $\mu(\theta)$. The vector $\vec n_\theta$
determines the projection axis and the mode-resolved charge
amplitude. As the rapidity $\theta$ varies, its normalized
direction traces a closed planar curve. The winding of this
curve is invariant under smooth deformations that preserve
the planar structure and keep $|\vec n_\theta|\neq 0$
for all $\theta$, providing a topological diagnostic for
the binary Floquet family considered here. 

For the translation-invariant Floquet Ising model, the
Jordan--Wigner transformation gives a free-fermion realization
of this mode decomposition. In this case, the rapidity can
be related to crystal momentum, and the spin factors describe
the momentum-resolved fermionic pseudospins. The associated
single-particle Floquet bands and their winding invariants
have been extensively studied
~\cite{Asboth14,yates2017entanglement,yeh2023decay,verga2023entanglement}.

We note that a more careful treatment can enable us to in fact define two different winding numbers, which can be used to calculate the number of $0$ and $\pi$ spectral pairings in Floquet systems with open boundaries, and completely characterize the four Floquet phases~\cite{yates2017entanglement}. A detailed explication using the Onsager algebra is left for future studies.

\begin{figure}[h]
  \centering
  \begin{tikzpicture}[scale=0.5]
\draw (0,0) rectangle (4*pi,4*pi);
\foreach \x/\label in {0/$-2\pi$, 2*pi/$0$, 4*pi/$2\pi$}
  \draw (\x,0) -- (\x,-0.1) node[below] {\label};
\foreach \y/\label in {0/$-2\pi$, 2*pi/$0$, 4*pi/$2\pi$}
  \draw (0,\y) -- (-0.1,\y) node[left] {\label};
  \node[below] at (2*pi,-1) {$T_1$};
  \node[left] at (-1,2*pi) {$T_2$};
\fill[cyan!20] (0,0) -- (pi,pi) -- (0,2*pi) -- cycle; 
\fill[cyan!20] (0,2*pi) -- (pi,3*pi) -- (0,4*pi) -- cycle; 
\fill[cyan!20] (4*pi,0) -- (3*pi,pi) -- (4*pi,2*pi) -- cycle; 
\fill[cyan!20] (4*pi,2*pi) -- (3*pi,3*pi) -- (4*pi,4*pi) -- cycle; 
\fill[red!20]  (2*pi,2*pi) -- (pi,pi) -- (2*pi,0) -- (3*pi,pi) -- cycle; 
\fill[red!20]  (2*pi,4*pi) -- (pi,3*pi) -- (2*pi,2*pi) -- (3*pi,3*pi) -- cycle; 
\fill[gray!20] (0,0) -- (pi,pi) -- (2*pi,0) -- cycle; 
\fill[gray!20]  (2*pi,2*pi) -- (pi,pi) -- (0,2*pi) -- (pi,3*pi) -- cycle;
\fill[gray!20]  (0,4*pi) -- (pi,3*pi) -- (2*pi,4*pi) -- cycle;
\fill[gray!20] (2*pi,0) -- (3*pi,pi) -- (4*pi,0) -- cycle; 
\fill[gray!20]  (4*pi,2*pi) -- (3*pi,pi) -- (2*pi,2*pi) -- (3*pi,3*pi) -- cycle;
\fill[gray!20]  (2*pi,4*pi) -- (3*pi,3*pi) -- (4*pi,4*pi) -- cycle;
\draw (0,0) -- (4*pi,4*pi);
\draw (0,2*pi) -- (2*pi,0);
\draw (4*pi,2*pi) -- (2*pi,4*pi);
\draw (2*pi,0) -- (4*pi,2*pi);
\draw (0,2*pi) -- (2*pi,4*pi);
\draw (4*pi,0) -- (0,4*pi);
\node[align=center,text width=2cm] at (pi/2-0.5,pi) {$-1$};
\node[align=center,text width=2cm] at (pi/2-0.5,3*pi) {$-1$};
\node[align=center,text width=2cm] at (7*pi/2+0.5,pi) {$-1$};
\node[align=center,text width=2cm] at (7*pi/2+0.5,3*pi) {$-1$};
\node[align=center,text width=2cm] at (2*pi,pi) {$+1$};
\node[align=center,text width=2cm] at (2*pi,3*pi) {$+1$};
\node[below, align=center,text width=2cm] at (pi,pi/2) {$0$};
\node[above, align=center,text width=2cm] at (pi,7*pi/2) {$0$};
\node[align=center,text width=2cm] at (pi,2*pi) {$0$};
\node[below, align=center,text width=2cm] at (3*pi,pi/2) {$0$};
\node[above, align=center,text width=2cm] at (3*pi,7*pi/2) {$0$};
\node[align=center,text width=2cm] at (3*pi,2*pi) {$0$};
\end{tikzpicture}
  \caption{The winding number of the single particle band in the parameter space of the $\mathbb{Z}_2$ Floquet problem. On the boundaries of each colored regions, the mode-resolved charge amplitude $|\vec n_\theta|$ vanishes for at least one rapidity $\theta$, rendering the winding number undefined. The generalization of this phase structure to binary Floquet systems with $\mathbb{Z}_N$ symmetry can be obtained by extending the parameter ranges to $[-N\pi,N\pi)$.}
  \label{fig:phase}
\end{figure}

The extension to other finite abelian symmetries is also straightforward by virtue of the Onsager algebra discovered in the superintegrable chiral Potts model~\cite{ALBERTINI1989excitation,fendley2012prafermionic}. We take cyclic $\mathbb{Z}_N$ symmetry as a demonstration, and the constructions for other abelian symmetries (which are isomorphic to products of cyclic symmetries)  follow directly. We take the trivial $\mathbb{Z}_N$ SPT Hamiltonian and the Thouless pump Hamiltonian to be~\cite{ALBERTINI1989excitation,fendley2012prafermionic,vladimir2017integrable,jones2025pivoting,jones2025charge}, 
\beq
    H_{\rm SPT} &= -\frac{1}{N}\sum_j \sum_{m=1}^{N-1}\alpha_m X_j^{m},\\
    H_{\rm TP} &= +\frac{1}{N}\sum_j \sum_{m=1}^{N-1}\alpha_m Z_{j-1}^{-m}Z_{j}^m,
    \label{eq:potts}
\eeq
where 
\beq
\alpha_m = \frac{1}{1-\omega^m}, \quad \omega = e^{2\pi i /N}.
\eeq
The local terms in the Thouless pump Hamiltonian satisfy $e^{\frac{i2\pi}{N}\sum_m \alpha_m Z_{j-1}^{-m}Z_j^m} = \omega^{\frac{N-1}{2}}Z_{j-1}^{-1}Z_j$~\cite{jones2025charge}. On
an open chain with the two edges indexed $L,R$, the product of the bond operators lead to 
$Z_{\rm L}^{-1}Z_{\rm R}$, corresponding to a generator of
$ H^1(\mathbb{Z}_N,U(1))\cong\mathbb{Z}_N$. Consequently, evolution under $H_{\rm TP}$ for a time $2\pi$
implements, up to an overall phase, the elementary charge pump.
The pump of charge $r\in\mathbb{Z}_N$ is obtained by replacing $Z_j \rightarrow Z_j^r$ in all terms in  
$H_{\rm TP}$. For the displayed pair in Eq.~\eqref{eq:potts}, the normalization
in Eq.~\eqref{eq:floquet-onsager-sign-convention} gives an Onsager
representation on the periodic clock chain
\cite{ALBERTINI1989excitation,fendley2012prafermionic,vladimir2017integrable,jones2025pivoting,jones2025charge}. Therefore, any
part of the preceding Floquet analysis that depends only on the
Onsager relations carries over, including the conserved-charge and winding construction, and consequently, the
repeated
phase-boundary structure shown in Fig.~\ref{fig:phase}. For $N>2$, however, the representation is
interacting and the physical phases need not be identical to those
of the free-fermion Ising chain.

Regarding Fig.~\ref{fig:phase} applied to $\mathbb{Z}_N$, note that since the Floquet unitary is strictly periodic in the parameter space, i.e, $U_F(T_1+c_1\cdot 2N\pi, T_2+c_2\cdot 2N\pi) = U_F(T_1, T_2)$ for any integer $c_1, c_2$, we need to extend the parameter range in Fig.~\ref{fig:phase} from $[-2\pi,2\pi]$ to $[-N\pi, N\pi]$. Thus, the binary Floquet problem for the $\mathbb{Z}_N$ symmetry has a richer phase structure than for $N=2$ because not only are there more than one charge pump, but even for a specific charge pump there are more critical lines in parameter space. In a later work, we will expand further on the  discussion of winding numbers and spectral pairings to an open chain with the goal being to elucidate the physical meanings of the windings for $N>2$.

Since there are usually more than one non-trivial Thouless pumps for abelian symmetries, we can define Floquet problems beyond binary drives,
\beq
    U_F(T_1, T_2, T_3) = e^{iT_3 H_{\rm TP}^{(\rho')}}e^{iT_2 H_{\rm TP}^{(\rho)}}e^{-iT_1 H_{\rm SPT}},
\eeq
with a richer phase structure. For example, if we include both the non-trivial Thouless pump Hamiltonians for $\mathbb{Z}_3$ symmetry, they combine into the hopping terms $\sum_j (Z^{-1}_{j-1}Z_j + h.c.)$ of the $\mathbb{Z}_3$ clock model, for the Floquet unitary $U_F(T_1, T_2, T_2)$, i.e, $T_2=T_3$. The parameter space $(T_1, T_2)$ does not contain an inherently dynamical FSPT phase, because the product of the two non-trivial Thouless pumps with $\mathbb{Z}_3$ symmetry is a trivial Thouless pump. However, the Floquet phase structure in the $(T_1, T_2, T_3)$ space is more interesting and awaits further studies. Our observation that  $T_2=T_3$ is a trivial Floquet SPT is consistent with numerical studies that indicate no stable edge modes for the $Z_3$ model with real Ising couplings \cite{Alicea14,Yeh24}.

In general, for a finite group $G$ which could be non-abelian, the elements $gh\overline{g}\overline{h}$ for any $g,h\in G$ generate a normal subgroup of $G$, denoted as $[G,G]$. Any 1d representation $\rho$ of $G$ is trivial in $[G,G]$, i.e., $\rho(gh\overline{g}\overline{h})\equiv 1$. Thus $\rho$ factors through the abelianization $Q=G/[G,G]$:
$\rho=\rho_Q\circ\pi$, where $\pi:G\to Q$ is the quotient map. The charge operator labeling the Thouless pump therefore depends only on the quotient degrees of freedom. We now exhibit a quotient-clock sector explicitly. The Onsager and winding statements below concern this sector. Set,
\begin{align}
    K=[G,G],
\end{align}
and denote the quotient map by
\begin{align}
    \pi:G\longrightarrow G/K.
\end{align}
Suppose that $q_1\in G/K$ generates a cyclic subgroup of order $n>1$,
\begin{align}
    C:=\langle q_1\rangle=\{e,q_1,q_1^2,\ldots,q_1^{n-1}\}\cong\mathbb{Z}_n.
\end{align}
Choose an element $s\in G$ satisfying
\begin{align}
    \pi(s)=q_1.
\end{align}
Thus \(s\) is a coset representative of the quotient-group element $q_1$. We define the corresponding normalized
$K$-invariant state by
\begin{align}
    \ket{\ell}_K
    =
    \frac{1}{\sqrt{|K|}}
    \sum_{a\in K}\ket{as^\ell},
    \qquad
    \ell=0,\ldots,n-1.
    \label{eq:K-invariant-coset-state}
\end{align}
These states form an orthonormal basis of the $n$-dimensional subspace
\begin{equation}
    \mathcal V_C
    :=\operatorname{span}\{\ket{\ell}_K:\ell=0,\ldots,n-1\}
    \subset\mathbb C[G].
    \label{eq:floquet-quotient-clock-sector}
\end{equation}
The state $\ket{\ell}_K$ is invariant under left multiplication by $K$. Indeed, for any $k\in K$,
\begin{align}
    L_k\ket{\ell}_K=
    \frac{1}{\sqrt{|K|}}
    \sum_{a\in K}\ket{kas^\ell}=\ket{\ell}_K,
\end{align}
because the map $a\mapsto ka$ merely permutes the elements of $K$. Furthermore, using the
normality of \(K\) we find
\begin{align}
    L_{ks}\ket{\ell}_K&=\frac{1}{\sqrt{|K|}}
    \sum_{a\in K}\ket{ksas^\ell}
    \nonumber\\
    &=\frac{1}{\sqrt{|K|}}
    \sum_{a\in K}
    \ket{k(sas^{-1})s^{\ell+1}}
    \nonumber\\
    &=\ket{\ell+1}_K.
    \label{eq:Lks-clock-action}
\end{align}
The last equality follows because $a\mapsto k(sas^{-1})$ permutes $K$.
Thus every $L_{ks}$ preserves $\mathcal V_C$ and restricts to the same
cyclic shift $X\ket{\ell}_K=\ket{\ell+1}_K$. In particular,
\begin{align}
    \left.\frac{1}{|K|}\sum_{k\in K}(L_{ks})^m
    \right|_{\mathcal V_C}=X^m.
    \label{eq:averaged-L-clock}
\end{align}
The clock operator is defined as
\begin{align}
    Z:=\sum_{\ell=0}^{n-1}\sum_{a\in K}
       \omega^\ell\ket{as^\ell}\bra{as^\ell}.
       \label{eq:floquet-quotient-clock-diagonal}
\end{align}

For a chain of $L$ sites, define the following Hamiltonians on $\mathcal V_C^{\otimes L}$:
\begin{align}
    H_{\rm SPT}
    &=-\frac{1}{n}\sum_j\sum_{m=1}^{n-1}\alpha_m
      \left.\left[\frac{1}{|K|}\sum_{k\in K}(L_{ks}^j)^m\right]
      \right|_{\mathcal V_C^{\otimes L}}
      \nonumber\\
    &=-\frac{1}{n}\sum_j\sum_{m=1}^{n-1}\alpha_m X_j^m,
      \label{eq:H-SPT-clock}\\
    H_{\rm TP}
    &=\frac{1}{n}\sum_j\sum_{m=1}^{n-1}
      \alpha_m Z_{j-1}^{-m}Z_j^m.
      \label{eq:floquet-quotient-clock-pump}
\end{align}
With periodic boundary conditions, the restricted generators
$A_0^{(C)}=-4H_{\rm SPT}|_{\mathcal V_C^{\otimes L}}$ and
$A_1^{(C)}=-4H_{\rm TP}|_{\mathcal V_C^{\otimes L}}$ obey the Dolan--Grady relations,
\begin{align}
 [A_0^{(C)},[A_0^{(C)},[A_0^{(C)},A_1^{(C)}]]]
 &=16[A_0^{(C)},A_1^{(C)}],\nonumber\\
 [A_1^{(C)},[A_1^{(C)},[A_1^{(C)},A_0^{(C)}]]]
 &=16[A_1^{(C)},A_0^{(C)}].
 \label{eq:floquet-quotient-sector-dolan-grady}
\end{align}
Consequently, the binary drive
\begin{equation}
 U_F^{(C)}(T_1,T_2)
 =e^{iT_2H_{\rm TP}}e^{-iT_1H_{\rm SPT}}
 \label{eq:floquet-quotient-sector-unitary}
\end{equation}
admits the conserved-charge and winding construction of the cyclic model.

To summarize, the preceding constructions yield conserved charges and winding diagnostics in the specified Onsager representations, which is characterized in Fig.~\ref{fig:phase}. Establishing this connection as a Floquet-phase invariant requires demonstrating that it characterizes the dynamics throughout the relevant quasienergy spectrum and is stable under symmetry-preserving perturbations. Identifying the resulting regions as Floquet symmetry-protected topological (FSPT) or Floquet symmetry-breaking phases therefore requires an analysis of the appropriate dynamical invariants and bulk eigenstate properties, together with a mechanism supporting their stability, such as localization~\cite{khemani2016phase}.

For the Floquet Ising realization, the connection between bulk winding invariants and strong zero or $\pi$ edge modes accounts for the corresponding quasienergy pairings on open chains~\cite{Fendley_2016,Yates2019}. Such spectral pairing should nevertheless be distinguished from bulk eigenstate SPT or symmetry-breaking order, which must be diagnosed independently
using suitable string or correlation functions. Establishing analogous relations between the winding and boundary modes for the other Onsager realizations, and determining the stability of any associated eigenstate order beyond the integrable limit, remain directions for future work.

\subsection{Rep$(G)$ Floquet problems}

In this section, we construct Rep$(G)$-symmetric Floquet drives
and identify invariant cyclic sectors admitting an Onsager
description. As shown in Sec.~\ref{sec:minimal RepG}, the minimal model
for the canonical Rep$(G)$ SPT phase is defined on a
$G$ spin chain and has the unique product ground state
$\ket{\phi}=\otimes_j\ket{e}_j$. This state is invariant
under the Rep$(G)$ symmetry, implemented by the MPO in
Eq.~\eqref{eq:A_mu symmetry}. For a nonidentity element $k\in G$,
the associated Thouless pump operator in the canonical SPT phase is $U_{\rm TP}^{(k)} = \prod_j R^{j-1}_k L^j_k$ as shown in Eq.~\eqref{eq:cycle unitary RL}. 
We denote the cyclic subgroup generated by $k$ as
$K=\langle k\rangle$ in the following. We consider the following Hamiltonians constituting the Floquet drive,
\beq
    H^{(k)}_{\rm SPT} &= -\frac{1}{|K|}\sum_j \sum_{m=1}^{|K|-1}\alpha_m (Z^{(j)}_{(k)})^{m},\\
    H^{(k)}_{\rm TP} &= +\frac{1}{|K|}\sum_j \sum_{m=1}^{|K|-1}\alpha_m R^{j-1}_{k^m}L^{j}_{k^m},
\eeq
where the local operator $Z^{(j)}_{(k)}$ is defined as
\beq
    Z^{(j)}_{(k)} := \sum_{m=1}^{|K|} e^{\frac{2\pi i m}{|K|}}\left(\ket{k^m}\bra{k^m}\right)_j. \label{eq: Zk-def}
\eeq

To make the clock structure explicit, let $r=|K|$ and
$\omega=e^{2\pi i/r}$. We use the coefficients $\alpha_m=(1-\omega^m)^{-1}$.
The local Hilbert space is
$\mathbb C[G]$, with orthonormal basis
$\{|g\rangle:g\in G\}$. We consider the invariant cyclic
sector $\mathcal H_K=\mathbb C[K]$, with basis
\begin{equation}
    |\ell\rangle:=|k^\ell\rangle,
    \qquad \ell=0,\ldots,r-1.
    \label{eq:repG-floquet-cyclic-basis}
\end{equation}
Since $k^r=e$, the label $\ell$ is understood modulo $r$. The clock and shift operators on this sector are given by
\begin{equation}
    Z|\ell\rangle=\omega^\ell|\ell\rangle,
    \qquad
    X|\ell\rangle=|\ell+1\rangle.
    \label{eq:repG-floquet-clock-actions}
\end{equation}
Using the conventions $L_k|g\rangle=|kg\rangle$ and
$R_k|g\rangle=|gk^{-1}\rangle$, their restrictions to
$\mathcal H_K$ are
\begin{equation}
    L_k\big|_{\mathcal H_K}=X,
    \qquad
    R_k\big|_{\mathcal H_K}=X^{-1},
    \qquad
    Z_{(k)}\big|_{\mathcal H_K}=Z.
    \label{eq:repG-floquet-clock-identification}
\end{equation}

Both driving Hamiltonians preserve
$\mathcal H_K^{\otimes N}$, with the clock reduction and
associated Onsager description understood to hold in this
invariant sector. On the full local Hilbert space
$\mathbb C[G]$, the definition of $Z_{(k)}$ in Eq.~\eqref{eq: Zk-def}
vanishes on basis states outside $K$ and therefore obeys
\begin{equation}
    (Z_{(k)})^r=P_K,
    \qquad
    P_K=\sum_{\ell=0}^{r-1}
    |k^\ell\rangle\langle k^\ell|.
    \label{eq:repG-floquet-cyclic-projector}
\end{equation}
The projector $P_K$ acts as the identity on
$\mathcal H_K$.

Thus, on the invariant sector $\mathcal H_K^{\otimes N}$
with periodic boundary conditions, these Hamiltonians
form an Onsager pair with $A_0=-4H^{(k)}_{\rm SPT}$ and
$A_1=-4H^{(k)}_{\rm TP}$ obeying the Dolan--Grady relations.
The on-site Hamiltonian $H^{(k)}_{\rm SPT}$ is gapped and has the
same unique product ground state $\ket{\phi}$ on the full space.
The mutually commuting bond operators also give the full-space identity
\begin{align}
 e^{i2\pi H^{(k)}_{\rm TP}}
 &=e^{i\pi(r-1)N_b/r}U^{(k)}_{\rm TP},
 \label{eq:repG-floquet-full-space-pump-phase}
\end{align}
where $N_b$ is the number of included bonds and the same bonds define
$U^{(k)}_{\rm TP}$. Thus the Hamiltonians and pump evolution are defined
on the full Hilbert space, whereas the clock identification and Onsager
description established here apply to the invariant cyclic sector.
 
 Within this invariant cyclic sector, the Onsager relations
yield an analogous winding construction and phase-boundary
structure to the group-like cyclic models in
Fig.~\ref{fig:phase}. Establishing whether the resulting
regions realize distinct Rep$(G)$-protected Floquet
eigenstate phases requires a separate analysis of their
symmetry order and stability, including the sectors outside
$\mathcal H_K^{\otimes N}$. Possible realizations through many-body localization or
long-lived prethermal dynamics remain to be investigated.
In the latter case, the relevant order concerns prethermal
states and dynamics rather than exact Floquet eigenstates.

\section{Conclusions and outlook}\label{Conc}

In this work, we have developed a classification of Thouless pumps in one-dimensional gapped phases with non-invertible symmetries. Using quasiadiabatic evolution, we established a correspondence between homotopy classes of adiabatic cycles and invertible defects generated by truncated Thouless pump operators. With the classification of gapped phases by module categories, we showed that these adiabatic cycles are classified by the group of $\mathcal{D}$-autoequivalences.

We constructed explicit lattice realizations of these cycles for SPT phases, and demonstrated how the pumping picture naturally reproduces known mathematical results about autoequivalence groups. Furthermore, we showed how the Thouless pump operators can be used for the constructions of Floquet drives. We then studied binary Floquet drives that obey the Onsager algebra, and derived the phase diagram, with distinct phases characterized by distinct winding  numbers and quasi-spectral structures.

We can always think of a more general family of gapped Hamiltonians beyond just an adiabatic cycle. Namely, we can consider a parameter space $X$ other than $S^1$. For some three-dimensional manifold $X$, it is shown that invertible phases (without any symmetry) over parameter space $X$ are classified by an integer (higher Berry curvature), which can be understood as the pump of a Chern number in the state~\cite{kapustin2020higher,wen2023flow}. In addition, we can construct an adiabatic cycle for every closed loop in $X$, which gives rise to a Thouless pump operator $U_{\rm TP}$. As we showed earlier, different classes of Thouless pumps correspond to different elements in the group of $\mathcal{D}$-autoequivalences $\Gamma$. Thus, the gapped phases over $X$ should also be classified by different assignments of elements in $\Gamma$ on the loops of $X$. For example when $X=T^k$,  a family of gapped Hamiltonians from assigning $U_{{\rm TP}_i}$ on the $i$-th loop of $X$ can be constructed
\begin{align}
&    H(\theta_1,\cdots, \theta_k)=\biggl[e^{i\theta_k H_{{\rm TP}_k}}\cdots e^{i\theta_1 H_{{\rm TP}_1}}\biggr]\nonumber\\
&H \biggl[e^{-i\theta_1 H_{{\rm TP}_1}}\cdots e^{-i\theta_k H_{{\rm TP}_k}}\biggr].
\end{align}
For SPT phases, this gives a classification of gapped phases over $X$ as 
\beq
H^1(X,\Gamma)\oplus H^3(X,\mathbb{Z}),
\eeq
where $H^1(X,\Gamma)$ comes from different assignments of loops and $H^3(X,\mathbb{Z})$ comes from the higher Berry curvature. This result agrees with the conjectured classification in Ref.~\cite{inamura20241+}.  We expect this classification to extend to other gapped phases, subject to possible constraints arising from mixed anomalies between Thouless pump operators associated with distinct adiabatic loops.

Looking forward, 
it would be interesting to extend our classification and lattice construction of adiabatic cycles to more general systems (fermionic systems, abstract spin chains~\cite{evans2025operator}, etc), with more general symmetries (time-reversal symmetry, spatially modulated symmetries, etc).  Secondly, in Ref.~\cite{aasen2022adiabatic} a conjecture is made about the homotopy groups of the space of two-dimensional gapped Hamiltonians  with topological order, along with the description of adiabatic cycles in the extended string-net model. It would be interesting to investigate other higher-dimensional phases with (higher) non-invertible symmetries, developing explicit lattice constructions. Finally, the Floquet constructions presented here point to a broader interplay between non-invertible symmetries and dynamical phases of matter, but the characterisation of their edge dynamics and stability away from exactly solvable limits, requires further development. We will leave this for future work.

\section{Acknowledgments}
We would like to thank Ruochen Ma and Yifan Wang for the feedback and comments about the draft. YL thanks Mikhail Litvinov, Meng Cheng, and Ruochen Ma for useful discussions. This work was supported by the U.S. National Science Foundation under Grant No. NSF DMR-2316598 (YL, AM). The work of MDA has received funding from the MacCracken Fellowship. GPT-6 Astra was used to proofread and revise the manuscript.

\begin{widetext}
\appendix

\section{Quasiadiabatic evolution for the conjugation cycle}
\label{app:berry}
In this appendix we show that conjugation unitaries reproduce the quasiadiabatic evolution on the ground subspace up to a Berry phase. Consider the following conjugation cycle of a gapped Hamiltonian,
\beq
    H(\theta) = V(\theta) H V(\theta)^{\dagger},
\eeq
where $A(\theta):=-i \left[\partial_{\theta}V(\theta)\right] V(\theta)^{\dagger}$ such that the conjugation unitary can be written as
\beq
    V(\theta) &= \mathcal{S}\exp{i \int_0^{\theta} ds A(s)}.
\eeq
The corresponding quasiadiabatic evolution is~\cite{bachmann2012automorphic} 
\beq
    U(\theta) &= \mathcal{S}\exp{i \int_0^{\theta} ds K(s)},
    \label{eq:qa evolution}
\eeq
where
\beq
    K(\theta) &= \int_{-\infty}^{\infty}dt  w_{\gamma}(t) \int_0^t du e^{iuH(\theta)}(\partial_{\theta} H(\theta))e^{-iuH(\theta)}\\
    &=\int_{-\infty}^{\infty}dt w_{\gamma}(t)\int_0^t du e^{iuH(\theta)}\big(i [A(\theta),H(\theta)]\big)e^{-iuH(\theta)}\\
    &=\int_{-\infty}^{\infty}dt w_{\gamma}(t)\int_0^t du \partial_u \big(-e^{iuH(\theta)}A(\theta)e^{-iuH(\theta)} \big)\\
    &=\int_{-\infty}^{\infty}dt w_{\gamma}(t)\big( A(\theta) - e^{itH(\theta)}A(\theta)e^{-itH(\theta)} \big).
\eeq
Using the fact that $\int dt w_{\gamma}(t)=1$, and that the Fourier transform $\hat{w}_{\gamma}(\omega)=0$ when $|\omega|\geq \gamma$ with the cutoff parameter $\gamma$ being smaller than the spectral gap $\Delta$, we conclude that
\beq
    P(\theta)K(\theta)P(\theta)=0,\qquad P(\theta)K(\theta)(1-P(\theta))=P(\theta)A(\theta)(1-P(\theta)),\label{eq:A5}
\eeq
where $P(\theta)$ is the projection into the ground subspace of $H(\theta)$.

To relate the quasiadiabatic evolution $U(\theta)$ to the conjugation unitary $V(\theta)$, we write these two evolution operators as
\beq
    U(\theta) = V(\theta) W(\theta).
\eeq 
Differentiating $U=VW$ gives $\partial_\theta W=-iV^\dagger(A-K)VW$. Thus, we have
\beq
    W(\theta) =& \mathcal{S}\exp{-i \int_0^{\theta} ds V(s)^{\dagger} \Big(A(s)-K(s)\Big)V(s) }\\
    =& \mathcal{S}\exp{-i \int_0^{\theta} ds  P(0) V(s)^{\dagger} A(s)V(s) P(0)} \\
     &\cdot\mathcal{S}\exp{-i \int_0^{\theta} ds (1-P(0))V(s)^{\dagger} \Big(A(s)-K(s)\Big)V(s)(1-P(0))}\\
     =& W_P\cdot W_{1-P}.
\eeq
Here $W_P$ and $W_{1-P}$ denote the evolutions within the two fixed subspaces. This unitary is exactly the Wilczek-Zee holonomy in the degenerate ground subspace associated with the conjugation cycle~\cite{wilczek1984appearance}.
For the cases we are concerned with, the generator $A(\theta)$ is a sum of local symmetric operators. Since we assume there is no accidental ground state degeneracy, there is no local symmetric operator that connects orthogonal ground states. Therefore, $A(\theta)$ is diagonal in the subspace $P$, i.e., the non-abelian holonomy reduces to the Berry phase for each ground state~\cite{Berry:1984jv}. When defining the invertible defect by $U(\theta)$ for an adiabatic conjugation cycle given by $V(\theta)$, we ignore these Berry phases. We claim that it is safe to ignore these phases for our purposes because, first they are continuous and do not depend on the topology of cycles, and second they are uniform since $W(\theta)$ is by definition symmetric. 

Let us review our example in Sec.~\ref{sec:qa evolution} on the $\mathbb{Z}_2$ spin chain,
\beq
    H(\theta) = -\sum_j \left(Je^{\frac{i \theta (X_j+X_{j+1})}{4}}  Z_j Z_{j+1} e^{-\frac{i \theta (X_j+X_{j+1})}{4}} +g X_j\right),
\eeq
The generator of $V(\theta)$ is a constant operator $A=\sum_j X_j/4$. When $J>g$, the Hamiltonian $H(0)$ is in the $\mathbb{Z}_2$ SSB phase, and the ground subspace $P(0)$ is spanned by two ferromagnetic ground states $\ket{\rm FM_{\uparrow}}$ and $\ket{\rm FM_{\downarrow}}$. The Wilczek-Zee unitary in the ground subspace $W_P$ is generated by a $2\times 2$ matrix,
\beq
    \bra{\rm FM_{i}}V(\theta)^{\dagger}A(\theta)V(\theta)\ket{\rm FM_{i'}}= \frac{1}{4}\sum_j \bra{\rm FM_{i}}X_j\ket{\rm FM_{i'}},
\eeq
with matrix indices $i,i' = \uparrow, \downarrow$. The off-diagonal entries vanish because the two SRE ground states are not related by local $\mathbb{Z}_2$ symmetric operators. Therefore, the unitary $W_P$ is diagonal,
\beq
    W_P = \begin{pmatrix}
        e^{i \delta_{\uparrow}} & 0\\
        0 & e^{i \delta_{\downarrow}}
    \end{pmatrix}.
    \label{eq:diagonal WP}
\eeq
Furthermore, since $W(\theta)$ commutes with the $\mathbb{Z}_2$ symmetry operator, which exchanges the SRE ground states, we conclude that $\delta_{\uparrow}\equiv \delta_{\downarrow}$. Hence, the Thouless pump operator $U_{\rm TP}$ acts on the ground subspace in the same way as $V(2\pi) = \prod_j X_j$ with a phase factor.

As a final remark, we note that although the Berry phases are continuous and not important for our purpose, for ordinary symmetries, the difference of the Berry phases between the ground states in the twisted and untwisted sectors is quantized and works as a homotopy invariant of adiabatic cycles~\cite{shiozaki2022adiabatic}. Similar phenomenon is used for the study of higher Thouless pumps for continuous symmetry~\cite{artymowicz2024quantization}.

\section{Module categories over $\text{Vec}_G$}
\label{app:vecG}
In this section, we discuss the module categories over $\text{Vec}_G$ with only one simple object, this corresponds to fiber functors. It is known that these module categories are classified by the second cohomology group $H^2(G,U(1))$. A 2-cocycle representative $[\omega]\in H^2(G,U(1))$ satisfies the cocycle condition
\beq
    \omega(h,k)\omega(g,hk)=\omega(gh,k)\omega(g,h).
\eeq

Denote the simple object in $\mathcal{M}(\omega)$ as $m$, then the fusion action is
\beq
    \delta_g \rhd m = m,
\eeq
for any simple object $\delta_g\in\text{Vec}_G$. The fusion in $\text{Vec}_G$ and fusion action obey the following associativity,
\beq
    \delta_g \rhd (\delta_h \rhd m) =  \omega(g,h) \delta_{gh} \rhd m,
    \label{eq:F move}
\eeq
where the 2-cocycle $\omega$ is the $^{\triangledown}F$ symbol for the group-like category. Now we are able to derive the plaquette terms in the anyonic chain model in Eq.~\eqref{eq:B_i string-net}, 
\beq
&B_i \ket{\begin{tikzpicture}[baseline={([yshift=-.4ex]current bounding box.center)},thick,scale=0.8, every node/.style={scale=1.0},
  arrow/.style={
    line width=1.2pt, 
    postaction={decorate},
    decoration={markings, mark=at position 0.6 with {\arrow{latex}}}
  },
  modulearrow/.style={
    dashed,
    blue,
    postaction={decorate},
    decoration={markings, mark=at position 0.6 with {\arrow{latex}}}
  }
  ]
  \def\n{5}       
  \def\gap{2}     
  \def\h{1.5}     
    \fill[black] (0,0) circle (3pt); 
    \draw[arrow] (0,0) -- (0,\h);  
    \node[left] at (0-0.1,0.5*\h) {$b_{i-\frac{1}{2}}$};
    \fill[blue] (0,\h) circle (3pt);
    \fill[black] (\gap,0) circle (3pt); 
    \draw[arrow] (\gap,0) -- (\gap,\h);  
    \node[right] at (\gap+0.1,0.5*\h) {$b_{i+\frac{1}{2}}$};
    \fill[blue] (\gap,\h) circle (3pt);
  \draw[arrow] (0,0) -- (\gap,0); 
    \node[below] at (\gap/2,-0.1) {$a_i$};
  \draw[arrow] (-1,0) -- (0,0);  
  \draw[arrow] (\gap,0) -- (\gap+1,0);  
    \draw[modulearrow] (\gap,\h) -- (0,\h); 
  \draw[modulearrow] (0,\h) -- (-1,\h);  
  \draw[modulearrow] (\gap+1,\h) -- (\gap,\h); 
\end{tikzpicture}  }
=\sum_{g\in G}
\ket{\begin{tikzpicture}[baseline={([yshift=-.4ex]current bounding box.center)},thick,scale=0.8, every node/.style={scale=1.0},
  arrow/.style={
    line width=1.2pt, 
    postaction={decorate},
    decoration={markings, mark=at position 0.6 with {\arrow{latex}}}
  },
  modulearrow/.style={
    dashed,
    blue,
    postaction={decorate},
    decoration={markings, mark=at position 0.6 with {\arrow{latex}}}
  }
  ]
  \def\n{5}       
  \def\gap{2}     
  \def\h{1.5}     
    \fill[black] (0,0) circle (3pt); 
    \draw[arrow] (0,0) -- (0,\h);  
    \node[left] at (0-0.1,0.5*\h) {$b_{i-\frac{1}{2}}$};
    \fill[blue] (0,\h) circle (3pt);
    \fill[black] (\gap,0) circle (3pt); 
    \draw[arrow] (\gap,0) -- (\gap,\h);  
    \node[right] at (\gap+0.1,0.5*\h) {$b_{i+\frac{1}{2}}$};
    \fill[blue] (\gap,\h) circle (3pt);
  \draw[arrow] (0,0) -- (\gap,0); 
    \node[below] at (\gap/2,-0.1) {$a_i$};
  \draw[arrow] (-1,0) -- (0,0);  
  \draw[arrow] (\gap,0) -- (\gap+1,0);  
    \draw[modulearrow] (\gap,\h) -- (0,\h); 
  \draw[modulearrow] (0,\h) -- (-1,\h);  
  \draw[modulearrow] (\gap+1,\h) -- (\gap,\h); 
  \draw[thick,line width=1.2pt] (0.5*\gap, 0.5*\h) circle (0.4*\h);
  \draw[thick, -{Latex}] ([shift={(10:0.4*\h)}] 0.5*\gap,.5*\h) arc[start angle=-1, end angle=1,radius=0.4*\h];
  \node[right] at (0.5*\gap-0.1,.5*\h) {$g$};
\end{tikzpicture}  }\\
&=\sum_{g\in G}
\ket{\begin{tikzpicture}[baseline={([yshift=-.4ex]current bounding box.center)},thick,scale=0.8, every node/.style={scale=1.0},
  arrow/.style={
    line width=1.2pt, 
    postaction={decorate},
    decoration={markings, mark=at position 0.6 with {\arrow{latex}}}
  },
  arrown/.style={
    line width=1.2pt, 
    postaction={decorate},
    decoration={markings, mark=at position 0.4 with {\arrow{latex}}, mark=at position 0.9 with {\arrow{latex}}}
  },
  arrownn/.style={
    line width=1.2pt, 
    postaction={decorate},
    decoration={markings, mark=at position 0.8 with {\arrow{latex}}}
  },
  modulearrow/.style={
    dashed,
    blue,
    postaction={decorate},
    decoration={markings, mark=at position 0.6 with {\arrow{latex}}}
  }
  ]
  \def\n{5}       
  \def\gap{2}     
  \def\h{1.5}     
    \fill[black] (0,0) circle (3pt); 
    \draw[arrown] (0,0) -- (0,\h);  
    \node[left] at (0-0.1,0.8*\h) {$b_{i-\frac{1}{2}}$};
    \node[left] at (0-0.1,0.3*\h) {$b_{i-\frac{1}{2}}\overline{g}$};
    \fill[blue] (0,\h) circle (3pt);
    \fill[black] (\gap,0) circle (3pt); 
    \draw[arrown] (\gap,0) -- (\gap,\h);  
    \node[right] at (\gap+0.1,0.8*\h) {$b_{i+\frac{1}{2}}$};
    \node[right] at (\gap+0.1,0.3*\h) {$gb_{i+\frac{1}{2}}$};
    \fill[blue] (\gap,\h) circle (3pt);
    \fill[blue] (0.4*\gap,\h) circle (3pt);
    \fill[blue] (0.6*\gap,\h) circle (3pt);
  \draw[arrow] (0,0) -- (\gap,0); 
    \node[below] at (\gap/2,-0.1) {$g a_i$};
  \draw[arrow] (-1,0) -- (0,0);  
  \draw[arrow] (\gap,0) -- (\gap+1,0);  
    \draw[modulearrow] (\gap,\h) -- (0,\h); 
  \draw[modulearrow] (0,\h) -- (-1,\h);  
  \draw[modulearrow] (\gap+1,\h) -- (\gap,\h); 
  \draw[arrownn] (0,\h/2) -- (0.4*\gap,\h);
  \draw[arrownn] (\gap,\h/2) -- (0.6*\gap,\h);
  \node[left] at (0.5*\gap-0.1,.5*\h) {$\overline{g}$};
  \node[right] at (0.5*\gap+0.1,.5*\h-0.05) {$g$};
\end{tikzpicture}  }\\
&=\sum_{g\in G}\omega(b_{i-\frac{1}{2}}, \overline{g})\omega(g,b_{i+\frac{1}{2}})
\ket{\begin{tikzpicture}[baseline={([yshift=-.4ex]current bounding box.center)},thick,scale=0.8, every node/.style={scale=1.0},
  arrow/.style={
    line width=1.2pt, 
    postaction={decorate},
    decoration={markings, mark=at position 0.6 with {\arrow{latex}}}
  },
  modulearrow/.style={
    dashed,
    blue,
    postaction={decorate},
    decoration={markings, mark=at position 0.6 with {\arrow{latex}}}
  }
  ]
  \def\n{5}       
  \def\gap{2}     
  \def\h{1.5}     
    \fill[black] (0,0) circle (3pt); 
    \draw[arrow] (0,0) -- (0,\h);  
    \node[left] at (0-0.1,0.5*\h) {$b_{i-\frac{1}{2}} \overline{g}$};
    \fill[blue] (0,\h) circle (3pt);
    \fill[black] (\gap,0) circle (3pt); 
    \draw[arrow] (\gap,0) -- (\gap,\h);  
    \node[right] at (\gap+0.1,0.5*\h) {$g b_{i+\frac{1}{2}}$};
    \fill[blue] (\gap,\h) circle (3pt);
  \draw[arrow] (0,0) -- (\gap,0); 
    \node[below] at (\gap/2,-0.1) {$g a_i$};
  \draw[arrow] (-1,0) -- (0,0);  
  \draw[arrow] (\gap,0) -- (\gap+1,0);  
    \draw[modulearrow] (\gap,\h) -- (0,\h); 
  \draw[modulearrow] (0,\h) -- (-1,\h);  
  \draw[modulearrow] (\gap+1,\h) -- (\gap,\h); 
\end{tikzpicture}  },
\eeq
where we use the $^{\triangledown}F$ move in Eq.~\eqref{eq:F move} to obtain the last equality.

An autoequivalence of this module category $\mathcal{M}(\omega)$ over $\text{Vec}_G$ is an equivalence $\gamma$ of $\mathcal{M}(\omega)$ together with a natural isomorphism $\beta$ compatible with the fusion actions,
\beq
    \beta: \gamma(\delta_g \rhd m)\rightarrow \delta_g\rhd \gamma(m).
\eeq 
Because objects in $\text{Vec}_G$ are invertible, and there is only one simple object, 
\beq
    \delta_g \rhd m =  m,\quad \gamma(m) = m,
\eeq
the isomorphism only gives a complex phase $\beta_g$ for $g\in G$. Since the isomorphism should be compatible with the associativity we have,
\beq
    &(\delta_g\cdot \delta_h)\rhd \gamma(m) \xrightarrow{\beta_{gh}^{-1}}\gamma((\delta_g\cdot \delta_h)\rhd m) \xrightarrow{\omega(g,h)^{-1}} \gamma(\delta_g\rhd( \delta_h\rhd m)) \\
    \xrightarrow{\beta_{g}}&\delta_g\rhd\gamma(\delta_h \rhd m)\xrightarrow{\beta_h} \delta_g\rhd(\delta_h\rhd \gamma(m))\xrightarrow{\omega(g,h)}(\delta_g\cdot \delta_h)\rhd \gamma(m),
\eeq
for any $g,h\in G$. The above implies the one-cocycle condition $\beta^{-1}_{gh}\beta_g \beta_h=1$.
Therefore, $\beta$ is a one-dimensional representation of $G$, i.e., $\beta\in H^1(G,U(1))$.

\section{Classification of lifted module categories}
\label{app:proof}
The goal of this section is to explain when the structure of a module category $\mathcal{M}$ over a fusion category $\mathcal{D}$ can be extended to a module for the graded extension $\mathcal{C}=\mathcal{D}\boxtimes \operatorname{Vec}_{\mathbb{Z}_n}^\omega$ without changing the underlying linear category. We will show first by a direct argument and then by applying a general construction that, this is related to embeddings of $\operatorname{Vec}_{\mathbb{Z}_n}^\omega$ into $\mathcal{D}_\mathcal{M}^*$, or equivalently to picking elements of $\Gamma=\operatorname{Aut}_\mathcal{D}(\mathcal{M})$ of order $n$ (with extra properties).

\subsection{(Un)folding trick}
In this subsection we will show that since the two factors of $\mathcal C=\mathcal{D}\boxtimes \operatorname{Vec}_{\Z_n}^\omega$ commute, extending the given
$\mathcal D$-action is equivalent to specifying an action of
$\operatorname{Vec}_{\mathbb Z_n}^{\omega}$ on $\mathcal M$ by $\mathcal D$-module
autoequivalences. 

From a module structure over $\mathcal C$:
\begin{equation}
    \blacktriangleright : \mathcal{C}\times \mathcal{M}=\mathcal{D}\boxtimes \operatorname{Vec}_{\mathbb Z_n}^{\omega}\times \mathcal{M}\to \mathcal{M},
\end{equation}
we can read off two module structures for $\mathcal{M}$ over the two factors: the original one over $\mathcal{D}$ by considering element of the form $(x\boxtimes \mathds{1})$
\begin{equation}
    \mathcal{D}\times \mathcal{M}\to \mathcal{M}, \quad x\rhd_\mathcal{D} m:= (x\boxtimes \mathds{1})\blacktriangleright  m,
\end{equation}and the new one $\operatorname{Vec}_{\Z_n}^\omega$ by considering element of the form $(
 \mathds{1}\boxtimes y)$
\begin{equation}
    \operatorname{Vec}_{\Z_n}^\omega\times \mathcal{M}\to \mathcal{M}, \quad y\rhd_{ \operatorname{Vec}_{\Z_n}^\omega}  m:= ( \mathds{1}\boxtimes y)\blacktriangleright  m.
\end{equation}
Now, those two structure are not unrelated but they are compatible in the following sense. The $\mathcal C$ module associator gives the following equivalences:
\begin{equation}
\begin{gathered}
    x\rhd_\mathcal{D}(y\rhd_{ \operatorname{Vec}_{\Z_n}^\omega} m)=(x\boxtimes \mathds{1})\blacktriangleright((\mathds{1}\boxtimes y)\blacktriangleright m)\simeq ((x\boxtimes \mathds{1})\boxtimes (\mathds{1}\boxtimes y))\blacktriangleright m\simeq (x\boxtimes y)\blacktriangleright m,\\
    y\rhd_{ \operatorname{Vec}_{\Z_n}^\omega}(x\rhd_\mathcal{D} m)\simeq (\mathds{1}\boxtimes y)\blacktriangleright((x\boxtimes \mathds{1})\blacktriangleright m)\simeq ((\mathds{1}\boxtimes y)\boxtimes (x\boxtimes \mathds{1}))\blacktriangleright m\simeq (x\boxtimes y)\blacktriangleright m.
    \end{gathered}
\end{equation}
Since the right hand side is the same, we can put them together and obtain a coherence datum:
\begin{equation}\label{eq: compatibility}
    x\rhd_\mathcal{D}(y\rhd_{ \operatorname{Vec}_{\Z_n}^\omega} m)\simeq y\rhd_{ \operatorname{Vec}_{\Z_n}^\omega}(x\rhd_\mathcal{D} m).
\end{equation}

From the module structure, $\operatorname{Vec}_{\Z_n}^\omega$ acts by endofunctors
\begin{equation}
    F:\ \operatorname{Vec}_{\Z_n}^\omega\to \operatorname{End}(\mathcal{M}), \quad y\mapsto F_y(m)=y\rhd_{\operatorname{Vec}_{\Z_n}^\omega} m,
\end{equation}
and moreover, from the above compatibility condition, it acts by elements of $\mathcal{D}_\mathcal M^*:=\operatorname{Fun}_\mathcal{D}(\mathcal{M}, \mathcal{M})$. In other words, $F$ describe monoidal functor $\operatorname{Vec}_{\mathbb{Z}_n}^{\omega}\to \mathcal{D}_\mathcal{M}^*$ (which is an embedding if the action is faithful) .

At the level of objects, such a functor is determined by the image
$U_{\mathrm{TP}}$ of the generator $1\in\mathbb Z_n$, which is invertible hence an element of $\Gamma =
    \operatorname{Aut}_{\mathcal D}(\mathcal M)$ and of order dividing $n$ (faithfulness would require the order to be exactly $n$). 

It remains to impose the monoidal coherence condition. 
Denoting by $[\widetilde F]$ the associator class of the pointed
subcategory generated by $U_{\mathrm{TP}}$, a functor exists only when $[\widetilde F]
  =
    [\omega]
    \text{ in }
    H^3(\mathbb Z_n,U(1))$.
    
We have thus concluded that faithful lifts are specified by invertible objects
$U_{\mathrm{TP}}\in\mathcal D_{\mathcal M}^{*}$ of exact order $n$ whose
generated pointed fusion subcategory has associator class $[\omega]$.

\subsection{Relation with the obstruction-theoretic description}
\label{subsec:obstruction-description}

Let us now relate this simple description to the general classification of module categories over graded fusion categories discussed in ~\cite{meir2012module}\footnote{See e.g. \cite{thorngren2024fusion,antinucci2023anomaliesnoninvertibleselfdualitysymmetries} for a more physics-oriented review.}. 

Consider a $G$-graded fusion category
\begin{equation}
    \mathcal C
    =
    \bigoplus_{g\in G}\mathcal C_g,
    \qquad
    \mathcal C_e=\mathcal D,
\end{equation}
and let $\mathcal L$ be an indecomposable $\mathcal C$-module category.
The restriction of $\mathcal L$ to $\mathcal D$ need not remain
indecomposable. Rather, the $G$-action permutes its indecomposable
$\mathcal D$-module components.

Choosing one such component $\mathcal N$, let $
    H\leq G$
be its stabiliser. The full $\mathcal C$-module category is then obtained
by induction from an $\mathcal C_H=\bigoplus_{h\in H}\mathcal{C}_{h}$-module structure on $\mathcal N$:
\begin{equation}
    \mathcal L
    \simeq
    \operatorname{Ind}_{\mathcal C_H}^{\mathcal C}(\mathcal N)
    =
    \mathcal C\boxtimes_{\mathcal C_H}\mathcal N.
    \label{eq:induced-module-category}
\end{equation}

The remaining problem is therefore to determine when the
$\mathcal D$-module category $\mathcal N$ can be promoted to a
$\mathcal C_H$-module category. Let $\Gamma
    =
    \operatorname{Aut}_{\mathcal D}(\mathcal N)$
be the group of equivalence classes of invertible $\mathcal D$-module
autoequivalences of $\mathcal N$. Following \cite{meir2012module}, the data defining such a lift can be organised into
an action of $H$ on $\Gamma$, together with two successive choices of
coherence data. More precisely, one first obtains an outer action $
    \overline{\Phi} :
    H
    \to
    \operatorname{Out}(\Gamma)$.
The first obstruction asks whether this outer action can be lifted to an
actual action $\Phi :
    H
    \to
    \operatorname{Aut}(\Gamma)$.
    
Once such a lift has been chosen, there is a second obstruction valued in $
    H^2(H,Z(\Gamma))$. 
When this obstruction vanishes, its trivializations form a torsor over $
    H^1(H,Z(\Gamma))$.
We denote such a choice by $v$.

Finally, there is an associativity obstruction valued in
$
    H^3(H,U(1))$.
When it vanishes, the possible choices of associativity data form a torsor
over $ H^2(H,U(1))$,
which we denote schematically by $\beta$.

Thus, the classification can be summarized by the data
\begin{equation}
    (\mathcal N,H,\Phi,v,\beta),
\end{equation}
subject to the vanishing of the corresponding obstruction classes.
We now specialize this general construction to the case relevant for us.

In our problem we require the lift to have the same underlying linear
category as the original $\mathcal D$-module category $\mathcal M$.
Consequently there is only one $\mathcal D$-module component, and hence
the symmetry is unbroken:
\begin{equation}
    H=\mathbb Z_n,
    \qquad
    \mathcal N=\mathcal M.
\end{equation}

 The first ingredient needed to equip $\mathcal{M}$ with a module category over $\mathcal{C}$ is an action by fusion. Moreover, the action needs to be compatible with the grading of $\mathcal C$. The first two obstructions combined ensure exactly this! 
 
 Let us fix a family of equivalences $\psi_g:\mathcal{C}_g\boxtimes_\mathcal{D}\mathcal{N}\xrightarrow[]{\sim}\mathcal{N}$. Compatibility with the fusion rules is equivalent to asking that the following composition:
\begin{equation}\label{first-two-obstructions}
    Y_{a,b}: \mathcal{N}\xrightarrow{\psi_a^{-1}}\mathcal{C}_a\boxtimes_\mathcal{D}\mathcal{N}\xrightarrow{\psi_b^{-1}}\mathcal{C}_a\boxtimes_\mathcal{D}\mathcal{C}_b\boxtimes_\mathcal{D}\mathcal{N}\xrightarrow{M_{a+b}}\mathcal{C}_{a+b}\boxtimes_\mathcal{D}\mathcal{N}\xrightarrow{\psi_{a+b}}\mathcal{N},
\end{equation}
 is the identity. This constraint can be split into two obstructions by defining an action $ \Phi : H \to \operatorname{Aut}(\Gamma)$ as follows: for $a\in H$ and $f\in \Gamma$:
 \begin{equation}
     a\cdot f: \mathcal{N}\xrightarrow[]{\psi_a^{-1}}\mathcal{C}_a\boxtimes_{\mathcal{D}}\mathcal{N}\xrightarrow[]{\mathds{1}_{\mathcal{C}_a}\boxtimes f}\mathcal{C}_a\boxtimes_{\mathcal{D}}\mathcal{N} \xrightarrow[]{\psi_a}\mathcal{N}.
 \end{equation}
 A direct computation gives:
 \begin{equation}\label{eq: lift}
     \Phi(a)\Phi(b)=\Phi(a+b)c_{Y_{a,b}},
 \end{equation}
 i.e. $\Phi$ fails to be a homomorphism due to a conjugation $c_{Y_{a,b}}\in \operatorname{Inn}(\Gamma)$ by $Y_{a,b}$. Moreover, a redefinition of the maps $\psi_k\to \gamma_k\psi_k$ by an element $\gamma_k\in \Gamma$ changes $\Phi(a)\to \Phi(a)c_{\gamma_a}$. We therefore conclude that $\Phi$ is a homomorphism when we consider it valued in the quotient $\operatorname{Out}(\Gamma)$ and it is now independent of $\psi_k$.

 With these definitions, the first obstruction is the (in)ability of lifting $\Psi$ to $\operatorname{Aut}(\Gamma)$. Once such a lift is chosen, Eq.~\eqref{eq: lift} implies  $Y_{a,b}\in Z(\Gamma)$.
 
 The second obstruction asks whether the remaining central elements $Y_{a,b}$ can be removed by a redefinition of the $\psi_a$\footnote{One can actually prove that $Y_{a,b}\in Z^2(H,Z(\Gamma))$ and recast the obstruction to the cohomological problem of finding a cocycle $v\in C^1(H, Z(\Gamma))$ trivializing $Y$
 \begin{equation}
     Y=\dd v.
 \end{equation}}.
 
Finally, the third obstruction consists in the compatibility of the aforementioned identifications with the $F$-symbols measuring the (lack of) associativity. 

Let us now focus on our case,  where $\mathcal{C}_k=\mathcal{D}$ as a $\mathcal{D}-\mathcal{D}$ bimodule category for all $k\in \mathbb{Z}_n$, the fusion rules $M_{g,h}:\mathcal{C}_g\boxtimes\mathcal{C}_h\to \mathcal{C}_{gh}$ follows from that of $\mathcal{D}$, and the $F$ symbol factorizes as that of $\mathcal{D}$ times $\omega$\footnote{In the language of \cite{etingof2009fusioncategorieshomotopytheory}, the tuple $(c, M, \alpha)$ defining the graded extension is given by the constant identity map $c:G\xrightarrow{\mathcal{D}}\operatorname{BrPic}(\mathcal{D})$, the tensor product of elements of $\mathcal{D}$ as multiplication $M_{g,h}: \mathcal{D}\boxtimes \mathcal{D}\xrightarrow{\otimes}\mathcal{D}$ and $\alpha=\omega$.}. Indeed, the objects of $\mathcal C_k$ can be written
as $A\cdot U_{\mathrm{TP}}^k$, and $U_{\mathrm{TP}}^k$ commutes with all
$\mathcal D$ defects\footnote{Even without the commutativity condition, we can conclude $\mathcal{C}_k\simeq \mathcal{D}$ as left $\mathcal{D}$ module or, in the language of \cite{etingof2009fusioncategorieshomotopytheory} $\mathcal{C}_k$ is quasi-trivial. These types of bimodules are classified by an autoequivalence $f_k\in \operatorname{Aut}(\mathcal{D})$ twisting the right action:
\begin{equation}
     (X\boxtimes Y)\rhd_{f_k}(A\cdot U^k_{TP})=XAf_k(Y)\cdot U_{\rm TP}^k
\end{equation}}. Hence
\begin{equation}
    X\triangleright
    (A\cdot U_{\mathrm{TP}}^k) \triangleleft Y
    =
    XAY\cdot U_{\mathrm{TP}}^k.
    \label{eq:Ck-bimodule-action}
\end{equation}
Therefore an equivalence 
\begin{equation}
    \psi_k:
    \mathcal C_k\boxtimes_{\mathcal D}\mathcal M\simeq \mathcal{D}\boxtimes_{\mathcal D} \mathcal{M}\simeq \mathcal{M}
    \xrightarrow{\sim}
    \mathcal M,
\end{equation}
is simply a $\mathcal D$-module autoequivalence of $\mathcal M$, $\psi_k\in\Gamma$. More explicitly, given $\psi_k\in\Gamma$, the action of
$A\cdot U_{TP}^k\in \mathcal C_k$ is
\begin{equation}
     m
    \longmapsto
    A\triangleright\psi_k(m).
    \label{eq:Ck-action-special}
\end{equation}
It follows immediately that $    \Phi(k)=c_{\psi_k}$. Hence, the induced outer action
is trivial, and the first
obstruction vanishes automatically.

Furthermore,
\begin{equation}
    Y_{a,b}
    =
    \psi_{a+b}\psi_b^{-1}\psi_a^{-1}.
    \label{eq:Yab-special}
\end{equation}
Consequently the first two stages of the obstruction theory reduce to
requiring that the equivalence classes of the $\psi_k$ define a group
homomorphism from $\mathbb Z_n$ to $\Gamma$. 
Such a homomorphism is determined by the image
$U_{\mathrm{TP}}\in\Gamma$ of the generator. For a faithful action,
$U_{\mathrm{TP}}$ must have exact order $n$.

The third obstruction is more subtle\footnote{As explained above, the fact that $\mathcal{D}$ and $\operatorname{Vec}_{\mathbb{Z}_n}^\omega$ factorise allows us to treat them independently. Moreover, the fact that $\mathcal{M}$ was a module category over $\mathcal{D}$ to begin with means that this obstruction is already solved for the $\mathcal{D}$ part.}. $U_{TP}^{a}(U_{TP}^{b}(m))$ might differ from $(U_{TP}^{a}U_{TP}^{b})(m)$ by a phase $\eta(a, b,m)$ that needs to be compatible with the anomaly $\omega$ of $U_{TP}^{a}$ itself. The third obstruction is the existence of $\eta$ such that the cocycle $O_3$ in $Z^3(\mathbb{Z}_n,U(1))$ vanishes\footnote{To be more precise, here and in \eqref{eq: last-obstruction} the obstruction classes vanish in cohomology.}:
\begin{equation}\label{eq: thir-obstruction}
    O_3(a,b,c):=\frac{\omega(a,b,c)\eta(a, b+c;m)\eta(b,c;m)}{\eta(a+b,c;m)\eta(a,b;\psi_c(m))}=1
\end{equation}
As noted in \cite{meir2012module}, the above quantity is independent of $m$. 

Let us now make another remark. The autoequivalence $\psi_k$ can be viewed as an (invertible) object of the category $\mathcal{D}_\mathcal{M}^*$. Denoting $\tilde{F}$ the associator of $(\mathcal{D}_\mathcal{M}^*)^{op}$, we can say, with a slight abuse of notation, that $\psi_k$ spans a subcategory $\operatorname{Vec}_{\mathbb{Z}_n}^{\tilde{F}}$.  Moreover, $\mathcal{M}$ enjoys a canonical module structure over $(\mathcal{D}_\mathcal{M}^*)^{op}$ whose defining data $\gamma:=\prescript{\lhd}{}{F}_{((\mathcal{D}_\mathcal{M}^*)^{op}, \mathcal{M})}$ automatically satisfies:
\begin{equation}\label{eq: last-obstruction}
    \frac{\tilde{F}(a,b,c)\gamma(a, b+c;m)\gamma(b,c;m)}{\gamma(a+b,c;m)\gamma(a,b;\psi_c(m))}=1
\end{equation}
Thus, choosing $\eta=\gamma$, we have $O_3=\omega/\tilde{F}$. We can conclude that the third obstruction vanishes exactly when $\omega=\tilde{F}$, and since $H^2(\mathbb{Z}_n;U(1))=0$ there is no freedom in choosing $\beta$.

Thus, in the
present factorized case, the general sequence of obstructions reduces
precisely to the monoidal-functor description
of the previous subsection.

\subsection{Composition of adiabatic cycles}
\label{subsec:composition-adiabatic-cycles}
Now, while the set of module categories over a fixed category does not have a canonical group structure, $\Gamma$ does. This group law has a direct physical interpretation in terms of the composition of the adiabatic cycles.

Let $\gamma$ and $\eta$ be two cycles, with corresponding pumped defects $U_{\rm TP}^{\gamma,\eta}\in \Gamma$. Since the defects correspond to the evaluation of the adiabatic unitary $U^{\gamma,\eta}(t)$ at $t=2\pi$, it is now evident that the defect pumped by the composite cycle
 \begin{equation}
     \gamma*\eta(t)=\begin{cases}\gamma(2t)\ \text{ for } 0\leq t<\pi\\
 \eta(2t-2\pi)\ \text{ for } \pi\leq t<2\pi \end{cases}, \quad
 U^{\gamma*\eta}(t)=\begin{cases}U^\gamma(2t)\ \text{ for } 0\leq t<\pi\\U^\eta(2t-2\pi)U^\gamma(2\pi)\ \text{ for } \pi\leq t<2\pi \end{cases}.\end{equation} 
is given by the composition of the defects of the two composed cycles:
 \begin{equation}
     U_{\rm TP}^{\gamma*\eta}=U_{\rm TP}^{\eta} U_{\rm TP}^{\gamma},
 \end{equation}
Thus composition of adiabatic cycles is mapped directly to the group operation in $\Gamma$.

\subsection{The $\mathbb Z$ description and the effective cyclic symmetry}
\label{subsec:Z-description}

There is also a useful formulation in which one does not fix the order of
$U_{\mathrm{TP}}$ from the outset. To be more precise, given $U_{\rm TP}\in \Gamma$, its powers defines an homomorphism from $\Z$ to $\Gamma$. Since $\Z$ is free, there is no $H^3$-obstruction to realizing it as a $\operatorname{Vec_\Z}$ action. If $U_{\rm TP}$ has finite order $n$, this action is nonfaithful and the effective symmetry is $\Z_n$. However, in order to pass to the quotient, we need to identify $U^a_{\rm TP}U_{\rm TP}^b$ with $U_{\rm TP}^{[a+b]_n}$ up to a phase $\eta(a,b)$. The failure of this identification to be associative leads to:
\begin{equation}
    \omega(a,b,c)=\frac{\eta(a,b+c)\eta(b,c)}{\eta(a+b,c)\eta(a,b)},
\end{equation}
which is in the trivial class of $H^3(\Z,U(1))$ but may become nontrivial in $H^3(\Z_n,U(1))$. The effective $\operatorname{Vec}_{\Z_n}^\omega$ picture is considered to be more physical; however, the unfaithful $\operatorname{Vec}_\Z$ extended symmetry gives a clearer connection to the classifying space interpretation of the adiabatic cycles.

To compare with the results in Ref.~\cite{inamura20241+}, since $S^1\simeq B\Z$, the lifting data defines an element of $\check{H}^1(B\Z,B\underline{\operatorname{Fun}}_\mathcal{C}(\mathcal{M},\mathcal{M})^{\text{Inv}})$ as they conjecture. The same perspective extends to a generic anomalous group $G$. By running an analogous argument in this section for $X=BG^\sigma$\footnote{With $X=BG^\sigma$ we denote the extension $B^2U(1)\to X\to G$ with $k$-invariant given by $\sigma$.}, we can see that an element of $\check{H}^1(BG^\sigma,B\underline{\operatorname{Fun}}_\mathcal{C}(\mathcal{M},\mathcal{M})^{\text{Inv}})$ exactly encodes the trivializations needed to lift a module category $\mathcal{M}$ over the fusion category $\mathcal{D}$ to a module category over the graded extension $\mathcal{D}\boxtimes\operatorname{Vec}_G^\sigma$.

\section{The quantum double model on a strip}
\label{app:QD}

We start by writing a quantum double model~\cite{kitaev2003fault} on a strip. To do that, we need to specify the upper and lower boundary conditions. It is known that the complete set of gapped boundaries is given by the set of pairs $(K\subset G,\alpha\in Z^2(K,U(1)))$~\cite{beigi2011quantum,bullivant2017twisted}. The corresponding quantum double ground state with a $(K,\alpha)$ boundary can be obtained by applying the gauging map~\cite{yoshida2017gapped} on a 2d $G$ trivial SPT state with a $K$-SPT state on the boundary. 

For our current purpose, we always choose $(G, 1)$ for the bottom boundary and choose a specific pair $(K,\alpha)$ for the top boundary, such that there is only one irreducible projective representation of $K$, i.e., there is only one simple object in $\text{Rep}^{
\alpha}(K)$. Given such a pair $(K,\alpha)$, we can define a tensor product Hilbert space $\mathcal{H}=\mathbb{C}[K]^{\otimes \text{upper edges}}\bigotimes \mathbb{C}[G]^{\otimes \text{other edges}}$ on the following strip. The basis states are given by assigning group elements $g_i, h_j\in G$ and $k_i\in K$ to label corresponding edges,
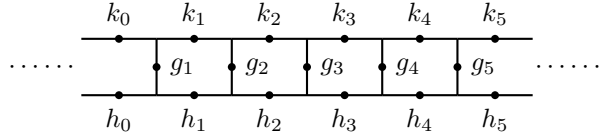
\begin{align*}
    \begin{tikzpicture}[thick,scale=0.5, every node/.style={scale=1.0}
  ]
  \def\gap{2}     
  \def\h{1.5}     
  \foreach \i in {1,2,3,4,5} {
    \pgfmathsetmacro\x{\i*\gap}
    \draw[thick] (\x,0) -- (\x,\h);
    \node[right] at (\x+0.1,0.5*\h) {$g_{\i}$};
    \fill[black] (\x,0.5*\h) circle (3pt);
  }
  \foreach \i in {0,1,2,3,4} {
    \pgfmathsetmacro\xA{\i*\gap}
    \pgfmathsetmacro\xB{\xA + \gap}
    \draw[thick] (\xA,0) -- (\xB,0);
    \fill[black] ({(\xA+\xB)/2},0) circle (3pt);
    \node[below] at ({(\xA+\xB)/2},-0.1) {$h_{\i}$};
  }
  \fill[black] (5.5*\gap,0) circle (3pt);
  \node[below] at (5.5*\gap,-0.1) {$h_5$};
  \draw[thick] (5*\gap,0) -- (6*\gap,0);
  \foreach \i in {0,1,2,3,4} {
    \pgfmathsetmacro\xA{\i*\gap}
    \pgfmathsetmacro\xB{\xA + \gap}
    \draw[thick] (\xA,\h) -- (\xB,\h);
    \fill[black] ({(\xA+\xB)/2},\h) circle (3pt);
    \node[above] at ({(\xA+\xB)/2},\h+0.1) {$k_{\i}$};
  }
  \fill[black] (5.5*\gap,\h) circle (3pt);
  \node[above] at (5.5*\gap,\h+0.1) {$k_5$};
  \draw[thick] (5*\gap,\h) -- (6*\gap,\h);
  \node at (6.5*\gap,0.5*\h) {$\cdots\cdots$};
  \node at (-0.5*\gap,0.5*\h) {$\cdots\cdots$};
\end{tikzpicture}
\end{align*}

The commuting projector Hamiltonian of this quantum double model is as follows,
\beq
    H^{\text{QD}}=-\sum_i A_i^{\text{QD}} - \sum_{i} B_i^{\text{QD}} - \sum_i Q_i,
\eeq
where the vertex terms and the plaquette terms are the conventional terms in the quantum double model, 
\begin{align}
A^{\text{QD}}_i&=\frac{1}{|G|}\sum_{s\in G}A^{\text{QD}}_{i,s},\nonumber\\ 
A^{\text{QD}}_{i,s}
&\ket{\begin{tikzpicture}[baseline={([yshift=-.5ex]current bounding box.center)},thick,scale=0.6, every node/.style={scale=1.0},
  arrow/.style={
    line width=1.2pt, 
    postaction={decorate},
    decoration={markings, mark=at position 0.6 with {\arrow{latex}}}
  }
  ]
  \def\gap{2}     
  \def\h{1.5}     
    \fill[black] (0.5*\h,0-\h/2) circle (3pt); 
    \fill[black] (-0.5*\h,0-\h/2) circle (3pt); 
    \fill[black] (0.5*\h,\h-\h/2) circle (3pt); 
    \fill[black] (-0.5*\h,\h-\h/2) circle (3pt); 
    \fill[black] (0,0.5*\h-\h/2) circle (3pt); 
    \draw[thick] (0,0-\h/2) -- (0,\h-\h/2);  
    \node[right] at (0+0.1,0.5*\h-\h/2) {$g_i$};  
  \draw[thick] (0,0-\h/2) -- (\h,0-\h/2); 
  \draw[thick] (-\h,\h-\h/2) -- (\h,\h-\h/2); 
    \node[above] at (\h/2,0.1+\h/2) {$k_i$};
    \node[above] at (-\h/2,0.1+\h/2) {$k_{i-1}$};
    \node[below] at (\h/2,-0.1-\h/2) {$h_i$};
    \draw[thick] (-\h,0-\h/2) -- (0,0-\h/2);  
    \node[below] at (-\h/2,-0.1-\h/2) {$h_{i-1}$};
\end{tikzpicture} }
=\ket{\begin{tikzpicture}[baseline={([yshift=-.5ex]current bounding box.center)},thick,scale=0.6, every node/.style={scale=1.0},
  arrow/.style={
    line width=1.2pt, 
    postaction={decorate},
    decoration={markings, mark=at position 0.6 with {\arrow{latex}}}
  }
  ]
  \def\gap{2}     
  \def\h{1.5}     
    \fill[black] (0.5*\h,0-\h/2) circle (3pt); 
    \fill[black] (-0.5*\h,0-\h/2) circle (3pt); 
    \fill[black] (0.5*\h,\h-\h/2) circle (3pt); 
    \fill[black] (-0.5*\h,\h-\h/2) circle (3pt); 
    \fill[black] (0,0.5*\h-\h/2) circle (3pt); 
    \draw[thick] (0,0-\h/2) -- (0,\h-\h/2);  
    \node[right] at (0+0.1,0.5*\h-\h/2) {$s g_i$};  
  \draw[thick] (0,0-\h/2) -- (\h,0-\h/2); 
  \draw[thick] (-\h,\h-\h/2) -- (\h,\h-\h/2); 
    \node[above] at (\h/2,0.1+\h/2) {$k_i$};
    \node[above] at (-\h/2,0.1+\h/2) {$k_{i-1}$};
    \node[below] at (\h/2,-0.1-\h/2) {$s h_i$};
    \draw[thick] (-\h,0-\h/2) -- (0,0-\h/2);  
    \node[below] at (-\h/2,-0.1-\h/2) {$h_{i-1}\overline{s}$};
\end{tikzpicture} },\nonumber\\
B_i^{\text{QD}}&\ket{\begin{tikzpicture}[baseline={([yshift=-.4ex]current bounding box.center)},thick,scale=0.6, every node/.style={scale=1.0}]
  \def\h{1.5}     
    \fill[black] (0.5*\h,0) circle (3pt);
    \fill[black] (0.5*\h,\h) circle (3pt);
    \fill[black] (0,0.5*\h) circle (3pt); 
    \fill[black] (\h,0.5*\h) circle (3pt); 
    \draw[thick] (0,0) -- (0,\h);  
    \node[left] at (0-0.1,0.5*\h) {$g_i$};
    \draw[thick] (\h,0) -- (\h,\h);  
    \node[right] at (\h+0.1,0.5*\h) {$g_{i+1}$};
  \draw[thick] (0,0) -- (\h,0);
  \draw[thick] (-\h/2,\h) -- (\h+\h/2,\h);
  \node[above] at (\h/2,\h+0.1) {$k_i$};
  \node[below] at (\h/2,-0.1) {$h_i$};
  \draw[thick] (-\h/2,0) -- (0,0);  
  \draw[thick] (\h,0) -- (\h+\h/2,0);  
\end{tikzpicture} } = \delta_{k_i, \overline{g_i}h_i g_{i+1}}\ket{\begin{tikzpicture}[baseline={([yshift=-.4ex]current bounding box.center)},thick,scale=0.6, every node/.style={scale=1.0}]
  \def\h{1.5}     
    \fill[black] (0.5*\h,0) circle (3pt); 
    \fill[black] (0.5*\h,\h) circle (3pt);
    \fill[black] (0,0.5*\h) circle (3pt); 
    \fill[black] (\h,0.5*\h) circle (3pt); 
    \draw[thick] (0,0) -- (0,\h);  
    \node[left] at (0-0.1,0.5*\h) {$g_i$};
    \draw[thick] (\h,0) -- (\h,\h);  
    \node[right] at (\h+0.1,0.5*\h) {$g_{i+1}$};
  \draw[thick] (0,0) -- (\h,0);
  \draw[thick] (-\h/2,\h) -- (\h+\h/2,\h);
  \node[above] at (\h/2,\h+0.1) {$k_i$};
  \node[below] at (\h/2,-0.1) {$h_i$};
  \draw[thick] (-\h/2,0) -- (0,0);  
  \draw[thick] (\h,0) -- (\h+\h/2,0);  
\end{tikzpicture} },
\end{align}
where we have embedded the $K$ subgroup into $G$. The vertex terms on the upper boundary are given by
\begin{align}
   Q_i &=  \frac{1}{|K|}\sum_{k\in K}Q_{i,k},\nonumber \\  Q_{i,k}
&\ket{\begin{tikzpicture}[baseline={([yshift=-.5ex]current bounding box.center)},thick,scale=0.6, every node/.style={scale=1.0},
  arrow/.style={
    line width=1.2pt, 
    postaction={decorate},
    decoration={markings, mark=at position 0.6 with {\arrow{latex}}}
  }
  ]
  \def\gap{2}     
  \def\h{1.5}     
    \fill[black] (0.5*\h,0-\h/2) circle (3pt); 
    \fill[black] (-0.5*\h,0-\h/2) circle (3pt); 
    \fill[black] (0.5*\h,\h-\h/2) circle (3pt); 
    \fill[black] (-0.5*\h,\h-\h/2) circle (3pt); 
    \fill[black] (0,0.5*\h-\h/2) circle (3pt); 
    \draw[thick] (0,0-\h/2) -- (0,\h-\h/2);  
    \node[right] at (0+0.1,0.5*\h-\h/2) {$g_i$};  
  \draw[thick] (0,0-\h/2) -- (\h,0-\h/2); 
  \draw[thick] (-\h,\h-\h/2) -- (\h,\h-\h/2); 
    \node[above] at (\h/2,0.1+\h/2) {$k_i$};
    \draw[thick] (-\h,0-\h/2) -- (0,0-\h/2);  
    \node[above] at (-\h/2,0.1+\h/2) {$k_{i-1}$};
    \node[below] at (\h/2,-0.1-\h/2) {$h_i$};
    \node[below] at (-\h/2,-0.1-\h/2) {$h_{i-1}$};
\end{tikzpicture} }
=\frac{\alpha(k,k_i)}{\alpha(k_{i-1}\overline{k},k)}\ket{\begin{tikzpicture}[baseline={([yshift=-.5ex]current bounding box.center)},thick,scale=0.6, every node/.style={scale=1.0},
  arrow/.style={
    line width=1.2pt, 
    postaction={decorate},
    decoration={markings, mark=at position 0.6 with {\arrow{latex}}}
  }
  ]
  \def\gap{2}     
  \def\h{1.5}     
    \fill[black] (0.5*\h,0-\h/2) circle (3pt); 
    \fill[black] (-0.5*\h,0-\h/2) circle (3pt); 
    \fill[black] (0.5*\h,\h-\h/2) circle (3pt); 
    \fill[black] (-0.5*\h,\h-\h/2) circle (3pt); 
    \fill[black] (0,0.5*\h-\h/2) circle (3pt); 
    \draw[thick] (0,0-\h/2) -- (0,\h-\h/2);  
    \node[right] at (0+0.1,0.5*\h-\h/2) {$g_i \overline{k}$};  
  \draw[thick] (0,0-\h/2) -- (\h,0-\h/2); 
  \draw[thick] (-\h,\h-\h/2) -- (\h,\h-\h/2); 
    \node[above] at (\h/2,0.1+\h/2) {$k k_i$};
    \draw[thick] (-\h,0-\h/2) -- (0,0-\h/2);  
    \node[above] at (-\h/2,0.1+\h/2) {$k_{i-1}\overline{k}$};
    \node[below] at (\h/2,-0.1-\h/2) {$h_i$};
    \node[below] at (-\h/2,-0.1-\h/2) {$h_{i-1}$};
\end{tikzpicture} }.
\end{align}

To check the ground state degeneracy, we can use a unitary $\prod_i CR^{(i)} CL^{(i)} CS^{(i)}$ to conjugate the system, where
\beq
    CR^{(i)}&\equiv \sum_{g_i, h_{i-1}}\ket{g_i}\bra{g_i}\otimes \ket{h_{i-1} g_i}\bra{h_{i-1}},\\
    CL^{(i)}&\equiv \sum_{g_i, h_{i}}\ket{g_i}\bra{g_i}\otimes \ket{\overline{g_i} h_{i}}\bra{h_{i}},\\
    CS^{(i)}&\equiv \sum_{h_i, k_i} \ket{k_i}\bra{k_i}\otimes \ket{h_i \overline{k_i}}\bra{h_i}.
\eeq
The conjugated system has decoupled edges $g_i, h_i$ that are fixed to 
\beq
    \ket{g_i}=\frac{1}{\sqrt{|G|}}\sum_{g\in G}\ket{g},\quad \ket{h_i}=\ket{e}
\eeq
in the low-energy subspace. The rest of the system is effectively a $K$ spin chain with the Hamiltonian being a sum of scalar multiples of mutually commuting projectors
\beq
    H' &= -\sum_i \sum_{k\in K} K'_{i,k} = -\sum_i  \sum_{k\in K}\sum_{k',k''}\frac{\alpha(k,k'')}{\alpha(k'\overline{k},k)}\left(\ket{k'\overline{k}}\bra{k'}\right)_{i-1}\otimes \left(\ket{kk''}\bra{k''}\right)_{i}.
    \label{eq:K hamiltonian}
\eeq
It can be checked that the operator $K'_{i,k}$ satisfies $K'_{i,k_1}K'_{i,k_2}=K'_{i,k_1 k_2}$, and when $i\neq j$, $\ K'_{i,k_1}K'_{j,k_2}=K'_{j,k_2}K'_{i,k_1}$.

Therefore, a ground state of this Hamiltonian should be invariant under $K'_{i,k}$ on any site $i$ for any element $k\in K$. In principle, we can create such a to-be-ground-state by applying the projector
\beq
    \ket{\phi_g} = \prod_i \left(\frac{1}{|K|}\sum_k K'_{i,k}\right)\ket{\cdots, e, g, e, \cdots},
\eeq
where $\ket{...,e,g,e,...}$ is the state when we have $\ket{g}$ on site 1, and $\ket{e}$ on all the other sites. These states $\{\ket{\phi_g}\}$ have different grading $\prod_i k_i$, and they are the only possible ground states, since any configuration $\ket{\{k_i\}}$ is contained in one of them. In other words, a state $\ket{\phi_g}$ is a (weighted) superposition of all configurations with grading $g$. Now we show that $\ket{\phi_g}$ vanishes when $g\neq e$. Notice that for any grading-$g$ configuration $\{k_i\}$ in $\ket{\phi_g}$,
\beq
    \ket{\phi_g} = \prod_i \left(\frac{1}{|K|}\sum_k K'_{i,k}\right)\ket{\{k_i\}}.
\eeq
We expand the product on the RHS above, and only keep track of the terms proportional to $\ket{\{k_i\}}$ itself, i.e,  take the inner-product $\langle \{k_i\} |\phi_g\rangle$. They are given by the action of $\prod_i K'_{i,g_i}$ for some specific set of elements $\{g_i\}$ that transform $ k_i \rightarrow g_i k_i \bar{g}_{i+1}$. The terms that return $\ket{\{k_i\}}$ to itself require $g_i k_i \bar{g}_{i+1}=k_i$ at every 
site, i.e.
\beq
    g_{i+1}=\overline{k_i}\, g_i \,k_i .
\eeq
The whole set $\{g_i\}$ is therefore fixed by the first element. Writing $g_1=k$,
\beq
    g_1 = k, \quad g_2 = \overline{k_1} k k_1, \quad 
    g_3 = \overline{k_1 k_2} \,k\, k_1 k_2, \quad \cdots,
    \quad g_{i+1} = \overline{k_1\cdots k_i}\, k\, k_1\cdots k_i .
\eeq
Closing the chain requires $g_{L+1}=g_1$, which gives $\bar{g} k g = k$ with 
$g=\prod_i k_i$ the grading. Hence the diagonal terms are labelled by elements of the 
centralizer, $k\in C_K(g)=\{h\in K \,|\, hg=gh\}$.

Each factor $K'_{i,g_i}$ contributes the phase 
$\alpha(g_i,k_i)/\alpha(g_{i-1}k_{i-1}\bar{g}_i,\,g_i)$. The denominator simplifies 
immediately from  the invariance condition, which for site $i-1$ reads 
$g_{i-1}k_{i-1}\bar{g}_i=k_{i-1}$, so the total phase accumulated by 
$\prod_i K'_{i,g_i}$ acting on $\ket{\{k_i\}}$ is
\beq
    \prod_{i=1}^{L}\frac{\alpha(g_i,k_i)}{\alpha(k_{i-1},g_i)},
\eeq
with the site labels understood periodically, $k_0\equiv k_L$ and $g_{L+1}\equiv g_1$. 
Reindexing the denominator around the ring, 
$\prod_i \alpha(k_{i-1},g_i)=\prod_i \alpha(k_i,g_{i+1})$, and using 
$g_{i+1}=\overline{k_i}\,g_i\,k_i$ this becomes
\beq
    \prod_{i=1}^{L}\frac{\alpha(g_i,k_i)}{\alpha(k_i,\,\overline{k_i}\,g_i\,k_i)}
    =\prod_{i=1}^{L}\theta_{g_i}(k_i),
\eeq
where the slant product is defined as
\beq
    \theta_k(k')\equiv \frac{\alpha(k,k')}{\alpha(k',\overline{k'}kk')}.
    \label{eq:slant}
\eeq
The slant product obeys $\theta_k(xy)=\theta_k(x)\,\theta_{\bar{x}kx}(y)$. Applying this 
repeatedly, with $g_1=k$ and $g_{i+1}=\overline{k_i}g_ik_i$, the product telescopes to a 
single phase,
\beq
    \prod_{i=1}^{L}\theta_{g_i}(k_i)=\theta_k(k_1k_2\cdots k_L)=\theta_k(g).
\eeq
Since $k$ and $g$ commute, the slant product reduces to 
$\theta_k(g)=\alpha(k,g)/\alpha(g,k)$, and the cocycle condition gives 
$\theta_k(g)\theta_{k'}(g)=\theta_{kk'}(g)$. Thus $\theta_{\cdot}(g)$ is a one-dimensional 
representation of the centralizer group $C_K(g)=\{h\in K\,|\,hg=gh\}$. Summing over the 
allowed labels $k\in C_K(g)$, the weight of the term $\ket{\{k_i\}}$ in $\ket{\phi_g}$ is
\beq
    \bra{\{k_i\}}\phi_g\rangle \propto \sum_{k\in C_K(g)}\theta_k(g).
\eeq
By orthogonality of characters this sum vanishes unless $\theta_{\cdot}(g)$ is the trivial 
representation. 
Write $P\equiv\prod_i\bigl(\frac{1}{|K|}\sum_{k}K'_{i,k}\bigr)$ and 
$\ket{\psi}\equiv\ket{\{k_i\}}$, so that $\ket{\phi_g}=P\ket{\psi}$. Each factor is 
idempotent by $K'_{i,k_1}K'_{i,k_2}=K'_{i,k_1k_2}$, and Hermitian since 
$K'^{\dagger}_{i,k}=K'_{i,\bar{k}}$ and the sum runs over all of $K$; the factors at 
different sites commute. Hence $P$ is a projector, $P^2=P$ and $P^{\dagger}=P$, so
\beq
    \bra{\psi}P\ket{\psi}=\bra{\psi}P^{\dagger}P\ket{\psi}=\bigl\|P\ket{\psi}\bigr\|^2 .
\eeq
The diagonal weight is therefore the squared norm of $\ket{\phi_g}$, and its vanishing 
implies $\ket{\phi_g}=0$. It therefore remains to 
determine for which $g$ one has $\theta_k(g)\equiv 1$ for all $k\in C_K(g)$.

We now show that this happens only for $g=e$. For that, we look at the unique $\alpha$-projective irreducible representation $U$ satisfying
\beq
    U_g U_h =\alpha(g,h)U_{gh}.
\eeq
For any element $g\in K$, we can define operator 
\beq
    A_g \equiv \sum_{h\in K} U_h^{-1} U_g U_h  = \sum_{h\in K}\frac{\alpha(g,h)}{\alpha(h,\bar{h}gh)}U_{\bar{h}gh} = \sum_{h\in K}\theta_g(h) U_{\bar{h}gh}
\eeq
defined for an element $g\in K$. This operator commutes with $U_k$ for any $k\in K$ because
\beq
    U_k^{-1} A_g U_k = \sum_{h\in K} U_k^{-1} U_h^{-1} U_g U_h U_k = \sum_{h\in K} U_{hk}^{-1}U_g U_{hk} = A_g.
\eeq
Due to Schur's lemma,
\beq
    A_g = \lambda I = \lambda U_e.
\eeq
On the other hand, 
\beq
    A_g = \sum_{h\in K}\theta_g(h) U_{\bar{h}gh} = \left(\sum_{h\in C_K(g)}\theta_g(h)\right)U_{g} + \sum_{x\neq g, x\in [g]}\left(\sum_{h: \bar{h}gh=x}\theta_{hx\bar{h}}(h)\right)U_{x}.
\eeq
Combining the above two equations, we conclude that when $g\neq e$, $\sum_{k\in C_K(g)}\theta_g(k)=\sum_{k\in C_K(g)}\overline{\theta_k(g)}$ must vanish. Therefore, according to our previous discussion $\ket{\phi_g}=0$ for every $g\neq e$, leaving a single ground state $\ket{\phi_e}$ with 
trivial grading.

To conclude, the quantum double model on the strip has a unique ground state for our choice 
of the pair $(K,\alpha)$. It therefore defines an SPT state with the matrix product operator 
symmetries $\text{Rep}(G)$.

\section{Mapping from quantum double to string-net models}
\label{app:map}

To map a quantum double model with gapped boundary as those discussed in the last section, we can first follow Ref.~\cite{buerschaper2009mapping} to make a basis change on the $g_i$ and $h_i$ edges,
\beq
    \ket{\mu_i,m'_i,m_i}_i&=\sqrt{\frac{|\mu_i|}{|G|}}\sum_{g\in G}[D^{\mu_i}(g)]_{m'_i, m_i}\ket{g}_i,\\
    \ket{\nu_i,n'_{i},n_i}_i&=\sqrt{\frac{|\nu_i|}{|G|}}\sum_{h\in G}[D^{\nu_i}(h)]_{n'_{i}, n_{i}}\ket{h}_i.
\eeq
where $\mu_i$ and $\nu_i$ run through all the irreducible representations of $G$. From the orthogonality relations, the new basis are orthonormal,
\beq
    \langle \mu,i,j| \nu, k,l\rangle &=\frac{\sqrt{|\mu|\cdot |\nu|}}{|G|}\sum_{g,h\in G}[D^{\mu}(h)]^*_{i,j}[D^{\nu}(g)]_{k,l}\langle h | g\rangle \\
    &= \delta_{\mu,\nu}\delta_{i,k}\delta_{j,l}.
\eeq
The inverse change of basis is given by
\beq
    \ket{g}_i &= \sum_{\mu_i\in \text{Rep}(G)}\sqrt{\frac{|\mu_i|}{|G|}}\sum_{m'_i,m_i}[D^{\mu_i}(g)]_{m'_i,m_i}^* \ket{\mu_i,m'_i,m_i}_i,\\
    \ket{h}_i &= \sum_{\nu_i\in \text{Rep}(G)}\sqrt{\frac{|\nu_i|}{|G|}}\sum_{n'_{i},n_{i}}[D^{\nu_i}(h)]_{n'_{i},n_{i}}^* \ket{\nu_i,n'_{i},n_{i}}_i.
\eeq
Meanwhile, there is a unique projective representation $M$ of the subgroup $K$ associated with the 2-cocycle $\alpha$. We make a basis change correspondingly on the $i$-th upper edge,
\beq
    \ket{r'_{i},r_{i}}_i=\frac{1}{|K|^{1/4}}\sum_{k\in K}[M(k)]_{r'_{i},r_{i}}\ket{k}_i.
\eeq
with the inverse transformation
\beq
    \ket{k}_i = \frac{1}{|K|^{1/4}}\sum_{r'_{i},r_{i}}[M(k)]^*_{r'_{i},r_{i}}\ket{r'_{i},r_{i}}_i.
\eeq
Following the orthogonality relations~\cite{kobayashi2025projective}, the new basis are also orthonormal,
\beq
    \langle r'_1, r_1 | r'_2, r_2\rangle &= \frac{1}{\sqrt{|K|}}\sum_{k,k'\in K}[M(k)]_{r'_2,r_2}[M(k')]^*_{r'_1,r_1}\langle k' | k\rangle \\
    &= \frac{1}{\sqrt{|K|}}\sum_{k\in K}[M(k)]_{r'_2,r_2}[M(k)]^*_{r'_1,r_1}\\
    &= \delta_{r'_1,r'_2}\delta_{r_1,r_2}.
\eeq

The state after the above basis change becomes
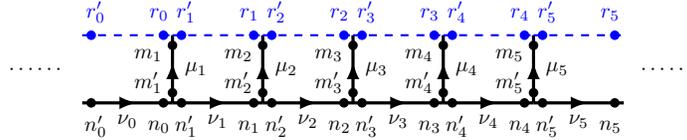
\begin{align*}
    \begin{tikzpicture}[thick,scale=0.6, every node/.style={scale=0.8}, arrow/.style={
    line width=1.2pt, 
    postaction={decorate},
    decoration={markings, mark=at position 0.6 with {\arrow{latex}}}
  }
  ]
  \def\gap{2}     
  \def\h{1.5}     
  \foreach \i in {1,2,3,4,5} {
    \pgfmathsetmacro\x{\i*\gap}
    \draw[arrow] (\x,0) -- (\x,\h);
    \node[right] at (\x+0.1,0.5*\h) {$\mu_{\i}$};
    \node[left] at (\x-0.05,0.5*\h-0.2*\h) {$m'_{\i}$};
    \fill[black] (\x,0.5*\h-0.35*\h) circle (3pt);
    \node[left] at (\x-0.05,0.5*\h+0.2*\h) {$m_{\i}$};
    \fill[black] (\x,0.5*\h+0.35*\h) circle (3pt);
  }
  \foreach \i in {0,1,2,3,4} {
    \pgfmathsetmacro\xA{\i*\gap}
    \pgfmathsetmacro\xB{\xA + \gap}
    \draw[arrow] (\xA,0) -- (\xB,0);
    \fill[black] ({(\xA+\xB)/2- 0.4*\gap},0) circle (3pt);
    \fill[black] ({(\xA+\xB)/2+ 0.4*\gap},0) circle (3pt);
    \node[below] at ({(\xA+\xB)/2},-0.1) {$\nu_{\i}$};
    \node[below] at ({(\xA+\xB)/2 - 0.35*\gap},-0.1) {$n'_{\i}$};
    \node[below] at ({(\xA+\xB)/2 + 0.35*\gap},-0.22) {$n_{\i}$};
  }
  \fill[black] ({5.5*\gap- 0.4*\gap},0) circle (3pt);
    \fill[black] ({5.5*\gap+ 0.4*\gap},0) circle (3pt);
  \node[below] at ({5.5*\gap - 0.35*\gap},-0.1) {$n'_{5}$};
    \node[below] at ({5.5*\gap + 0.35*\gap},-0.22) {$n_{5}$};
  \node[below] at (5.5*\gap,-0.1) {$\nu_5$};
  \draw[arrow] (5*\gap,0) -- (6*\gap,0);
  \foreach \i in {0,1,2,3,4} {
    \pgfmathsetmacro\xA{\i*\gap}
    \pgfmathsetmacro\xB{\xA + \gap}
    \draw[blue, thick, dashed] (\xA,\h) -- (\xB,\h);
    \fill[blue] ({(\xA+\xB)/2- 0.4*\gap},\h) circle (3pt);
    \fill[blue] ({(\xA+\xB)/2+ 0.4*\gap},\h) circle (3pt);
    \node[above] at ({(\xA+\xB)/2 - 0.35*\gap},\h+0.1) {\textcolor{blue}{$r'_{\i}$}};
    \node[above] at ({(\xA+\xB)/2 + 0.35*\gap},\h+0.15) {\textcolor{blue}{$r_{\i}$}};
  }
  \fill[blue] ({5.5*\gap- 0.4*\gap},\h) circle (3pt);
    \fill[blue] ({5.5*\gap+ 0.4*\gap},\h) circle (3pt);
  \node[above] at ({5.5*\gap - 0.35*\gap},\h+0.1) {\textcolor{blue}{$r'_{5}$}};
    \node[above] at ({5.5*\gap + 0.35*\gap},\h+0.15) {\textcolor{blue}{$r_{5}$}};
  \draw[blue, thick, dashed] (5*\gap,\h) -- (6*\gap,\h);
  \node at (6.5*\gap,0.5*\h) {$\cdots\cdots$};
  \node at (-0.5*\gap,0.5*\h) {$\cdots\cdots$};
\end{tikzpicture}
\end{align*}

As we discussed in the last section, the quantum double model is given by,
\beq
    H^{\text{QD}}=-\sum_i A_i^{\text{QD}} - \sum_{i} B_i^{\text{QD}} - \sum_i Q_i.
\eeq
The vertex terms in the new basis becomes
\beq
    A_i^{\text{QD}} &\ket{\nu_{i-1},n'_{i-1},n_{i-1}}\otimes\ket{\nu_{i},n'_{i},n_{i}}\otimes\ket{\mu_{i},m'_{i},m_{i}}\\
    &=\sum_{\tilde{n}_{i-1},\tilde{n}'_{i},\tilde{m}'_i} W^{(\nu_{i-1},\nu_i,\mu_i)}_{(\tilde{n}_{i-1},\tilde{n}'_{i},\tilde{m}'_i), (n_{i-1},n'_{i},m'_i)}\ket{\nu_{i-1},n'_{i-1},\tilde{n}_{i-1}}\otimes\ket{\nu_{i},\tilde{n}'_{i},n_{i}}\otimes\ket{\mu_{i},\tilde{m}'_{i},m_{i}},
\eeq
where 
\beq
    W^{(\nu_{i-1},\nu_i,\mu_i)}_{(\tilde{n}_{i-1},\tilde{n}'_{i},\tilde{m}'_i), (n_{i-1},n'_{i},m'_i)}=\frac{1}{|G|}\sum_{s\in G} [D^{\nu_{i-1}}(s)]_{\tilde{n}_{i-1},n_{i-1}}[D^{\nu_{i}}(s)]^{\dagger}_{n'_{i},\tilde{n}'_{i}}[D^{\mu_{i}}(s)]^{\dagger}_{m'_{i},\tilde{m}'_{i}}.
\eeq
When $N_{\nu_i,\mu_i}^{\nu_{i-1}}> 0$, following the orthogonality relation, there is a unitary matrix $w$ that decomposes this tensor product into a direct sum such that
\beq
    W^{(\nu_{i-1},\nu_i,\mu_i)}_{(\tilde{n}_{i-1},\tilde{n}'_{i},\tilde{m}'_i), (n_{i-1},n'_{i},m'_i)}=\sum_{A_i=1,\cdots,N_{\nu_i,\mu_i}^{\nu_{i-1}}}\left(w^{(\nu_{i-1},\nu_i,\mu_i)}_{\tilde{n}_{i-1},\tilde{n}'_{i},\tilde{m}'_{i}; A_i}\right)^* w^{(\nu_{i-1},\nu_i,\mu_i)}_{n_{i-1},n'_{i},m'_{i}; A_i},
\eeq
in which each summand is a one-dimensional projector in the $\{\ket{n_{i-1},n'_{i},m'_{i}}\}$ space. On the other hand, using the 2-cocycle conditions we can write the vertex terms on the upper boundary after the basis change as
\beq
    Q_i &\ket{r'_{i-1},r_{i-1}}\otimes\ket{r'_{i},r_{i}}\otimes \ket{\mu_{i},m'_{i},m_{i}}\\
    &=\sum_{\tilde{r}_{i-1},\tilde{r}'_{i},\tilde{m}_i} V^{(\mu_i)}_{(\tilde{r}_{i-1},\tilde{r}'_{i},\tilde{m}_i), (r_{i-1},r'_{i},m_i)}\ket{r'_{i-1},\tilde{r}_{i-1}}\otimes\ket{\tilde{r}'_{i},r_{i}}\otimes \ket{\mu_{i},m'_{i},\tilde{m}_{i}},
\eeq
where 
\beq
    V^{(\mu_i)}_{(\tilde{r}_{i-1},\tilde{r}'_{i},\tilde{m}_i), (r_{i-1},r'_{i},m_i)} &= \frac{1}{|K|}\sum_{k\in K}\frac{1}{\alpha(\overline{k},k)} [D^{\mu_{i}}(k)]_{\tilde{m}_{i},m_i}[M(k)]_{\tilde{r}_{i-1},r_{i-1}}[M(\overline{k})]_{r_{i},\tilde{r}_{i}},\\
    &= \frac{1}{|K|}\sum_{k\in K} [D^{\mu_{i}}(k)]_{\tilde{m}_{i},m_i}[M(k)]_{\tilde{r}_{i-1},r_{i-1}}[M(k)]^{\dagger}_{r_{i},\tilde{r}_{i}}.
\eeq
The tensor product of a $K$ linear representation $\mu$ and the projective representation $M$ is another projective representation of $K$ associated with the 2-cocycle $\alpha$. Since $M$ is the only irreducible projective representation, this tensor product representation can be decomposed into a direct sum of $|\mu|$ projective representations $M$ up to isomorphism. Therefore, following from the orthogonality relation, there is another unitary $v$ that split the above $V$ operator into a direct sum of orthogonal rank-one projectors,
\beq
    V^{(\mu_i)}_{(\tilde{r}_{i-1},\tilde{r}'_{i},\tilde{m}_i), (r_{i-1},r'_{i},m_i)} = \sum_{B_i = 1, \cdots, |\mu_i|} \left( v^{(\mu_i)}_{\tilde{r}_{i-1},\tilde{r}'_{i},\tilde{m}_i; B_i}\right)^* v^{(\mu_i)}_{r_{i-1},r'_{i},m_i; B_i}.
\eeq
Hence, in the $A_i^{\text{QD}} = Q_i = 1$ subspace the representation matrix indices at the same vertex are all contracted into the $A_i$ and $B_i$ indices, according to the projectors given by the matrices $w$ and $v$. The basis wavefunctions in this subspace are of the form:
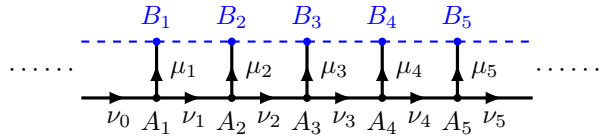
\begin{align*}
    \begin{tikzpicture}[thick,scale=0.5, every node/.style={scale=1.0}, arrow/.style={
    line width=1.2pt, 
    postaction={decorate},
    decoration={markings, mark=at position 0.6 with {\arrow{latex}}}
  }
  ]
  \def\gap{2}     
  \def\h{1.5}     
  \foreach \i in {1,2,3,4,5} {
    \pgfmathsetmacro\x{\i*\gap}
    \draw[arrow] (\x,0) -- (\x,\h);
    \node[right] at (\x+0.1,0.5*\h) {$\mu_{\i}$};
  }
  \foreach \i in {1,2,3,4,5} {
    \pgfmathsetmacro\xA{\i*\gap}
    \pgfmathsetmacro\xB{\xA + \gap}
    \draw[arrow] (\xA,0) -- (\xB,0);
    \fill[black] (\xA,0) circle (3pt);
    \node[below] at (\xA,-0.1) {$A_{\i}$};
    \node[below] at ({(\xA+\xB)/2},-0.1) {$\nu_{\i}$};
  }
  \node[below] at (0.5*\gap,-0.1) {$\nu_0$};
  \draw[arrow] (0*\gap,0) -- (1*\gap,0);
  \foreach \i in {1,2,3,4,5} {
    \pgfmathsetmacro\xA{\i*\gap}
    \pgfmathsetmacro\xB{\xA + \gap}
    \draw[blue, thick, dashed] (\xA,\h) -- (\xB,\h);
    \fill[blue] (\xA,\h) circle (3pt);
    \node[above] at ({\xA},\h+0.1) {\textcolor{blue}{$B_{\i}$}};
  }
  \draw[blue, thick, dashed] (0*\gap,\h) -- (1*\gap,\h);
  \node at (6.5*\gap,0.5*\h) {$\cdots\cdots$};
  \node at (-0.5*\gap,0.5*\h) {$\cdots\cdots$};
\end{tikzpicture}
\end{align*}

We note that the above basis states are exactly that of the anyonic chain model shown in Fig.~\ref{fig:anyonic-chain}. When the module category $\mathcal{M}$ is a fiber functor, the $A^t_i$ terms in the anyonic chain Hamiltonian in Eq.~\eqref{eq:anyonic chain} vanish. The constraints $N_{\nu_i,\mu_i}^{\nu_{i-1}}>0$ exactly correspond to the $A^d_i$ projectors. Following Ref.~\cite{buerschaper2009mapping} to use the $3j$ symbols of $G$ representations to write the $F$-symbols of $\text{Rep}(G)$, and similarly use the $3j$ symbols between linear and projective $K$ representations to write the $^{\triangledown}F$-symbols of the module category $\mathcal{M}(K,\alpha)$ over $\text{Rep}(G)$, it can be showed that the plaquette term $B_i^{\text{QD}}$ acts on the subspace of quantum double model in the same way as the plaquette term $B_i$ in the string-net model.

In the above, we showed the explicit map from quantum double with some specific $(K,\alpha)$ boundary to the string-net model with boundary associated with a fiber functor $\mathcal{M}$. This map can be made between all possible pairs of $(K,\alpha)$ and module category $\mathcal{M}$ with minor modifications. This correspondence between two descriptions of the boundaries comes naturally, since it is known that each possible choice of pairs $(K,\alpha)$ corresponds to a fiber functor $\mathcal{M}$ over $\text{Rep}(G)$. 

\section{Derivation of $U_{\mathrm{TP}_f}^{(q)}$ and multiplication rules for the Thouless pump operators}\label{app:multiplication}

Before deriving the multiplication rule, we first verify that $U_{\mathrm{TP}_f}^{(q)}$ commutes with the Hamiltonian in Eq.~\eqref{eq:hamiltonian K alpha}, since this is what fixes the phase gate $\prod_i Z_{\eta_q}^{(i)}$ appearing in its definition. Write
\beq
U_{\mathrm{TP}_f}^{(q)} = C_q Z_{\eta_q}^\dagger,
\qquad
C_q\equiv \prod_i R_q^i L_q^{i+1},
\qquad
Z_{\eta_q}\equiv\prod_i Z_{\eta_q}^{(i)} .
\label{eq:F1}
\eeq
Note that $C_q$ conjugates every physical $K$-spin, such that $C_q\ket{k}=\ket{{}^q k}$.

It is useful to write the $i$-th local term of Eq.~\eqref{eq:hamiltonian K alpha} as
\beq
B_i^{(\alpha)} = \sum_{k,k',k''\in K}
\frac{\alpha(k,k'')}{\alpha(k'\bar{k},k)} \left(\ket{k'\bar{k}}\bra{k'}\right)_{i-1} \otimes \left(\ket{kk''}\bra{k''}\right)_i ,
\label{eq:F2}
\eeq
so that
\beq
H_{\mathrm{SPT}}=-\sum_i B_i^{(\alpha)}.
\eeq
Here $k'$ and $k''$ label the physical spin states, while $k$ labels the term in the sum over $K$.

We first conjugate a local term by $C_q$,
\beq
C_q^\dagger B_i^{(\alpha)}C_q=\sum_{k,k',k''\in K} \frac{\alpha(k,k'')}{\alpha(k'\bar{k},k)}
\left(\ket{{}^{\bar q}(k'\bar{k})}\bra{{}^{\bar q}k'}\right)_{i-1}
\otimes
\left(\ket{{}^{\bar q}(kk'')}\bra{{}^{\bar q}k''}\right)_i.
\label{eq:F4}
\eeq
Now change variables according to
$\widetilde{k}'={} ^{\bar q}k',\widetilde{k}''={} ^{\bar q}k'',\widetilde{k}={}^{\bar q}k$. Dropping the tildes after this change of variables gives
\beq
C_q^\dagger B_i^{(\alpha)}C_q=\sum_{k,k',k''\in K}\frac{\alpha({}^qk,{}^qk'')}{\alpha({}^q(k'\bar{k}),{}^qk)}
\left(\ket{k'\bar{k}}\bra{k'}\right)_{i-1}
\otimes
\left(\ket{kk''}\bra{k''}\right)_i .
\label{eq:F5}
\eeq

We now use Eq.~\eqref{eq:Q-invariant cocycle},
\beq
\frac{\alpha({}^q k_1,{}^q k_2)}{\alpha(k_1,k_2)} = \frac{\eta_q(k_1)\eta_q(k_2)}{\eta_q(k_1k_2)}.
\label{eq:F6}
\eeq
For the numerator of Eq.~\eqref{eq:F5},
\beq
\alpha({}^qk,{}^qk'') = \alpha(k,k'') \frac{\eta_q(k)\eta_q(k'')}{\eta_q(kk'')},
\eeq
whereas for the denominator,
\beq
\alpha({}^q(k'\bar{k}),{}^qk)= \alpha(k'\bar{k},k) \frac{\eta_q(k'\bar{k})\eta_q(k)}{\eta_q(k')}.
\eeq
Consequently,
\beq
\frac{\alpha({}^qk,{}^qk'')}{\alpha({}^q(k'\bar{k}),{}^qk)}
=
\frac{\alpha(k,k'')}{\alpha(k'\bar{k},k)}
\frac{\eta_q(k')\eta_q(k'')}{\eta_q(k'\bar{k})\eta_q(kk'')}.
\label{eq:F9}
\eeq
In particular, the factors $\eta_q(k)$ cancel. The remaining factor depends only on the initial and final physical spin configurations. This remaining factor is precisely generated by the diagonal phase gate. Since
\beq
Z_{\eta_q}^{(j)} = \sum_{x\in K} \eta_q(x)\ket{x}\bra{x}_j ,
\eeq
we have
\beq
Z_{\eta_q}^\dagger \left[ \left(\ket{k'\bar{k}}\bra{k'}\right)_{i-1} \otimes \left(\ket{kk''}\bra{k''}\right)_i\right] Z_{\eta_q}
=
\frac{\eta_q(k')\eta_q(k'')}{\eta_q(k'\bar{k})\eta_q(kk'')} \left[\left(\ket{k'\bar{k}}\bra{k'}\right)_{i-1} \otimes \left(\ket{kk''}\bra{k''}\right)_i\right].
\label{eq:F11}
\eeq
Comparing Eqs.~\eqref{eq:F9} and \eqref{eq:F11}, we therefore find
\beq
C_q^\dagger B_i^{(\alpha)}C_q = Z_{\eta_q}^\dagger B_i^{(\alpha)}Z_{\eta_q}.
\label{eq:F12}
\eeq
Summing over $i$ gives
\beq
C_q^\dagger H_{\mathrm{SPT}}C_q = Z_{\eta_q}^\dagger H_{\mathrm{SPT}}Z_{\eta_q}.
\label{eq:F12a}
\eeq
Therefore, the Thouless pump $U_{\mathrm{TP}_f}^{(q)} = C_q Z_{\eta_q}^\dagger$ commutes with $H_{\mathrm{SPT}}$. 

We now derive the product of two flux-pump operators:
\begin{align}
    U_{\mathrm{TP}_f}^{(q_1)} U_{\mathrm{TP}_f}^{(q_2)}=
    C_{q_1}Z_{\eta_{q_1}}^\dagger C_{q_2}Z_{\eta_{q_2}}^\dagger=
    C_{q_1}C_{q_2}Z_{\eta_{q_1}\circ{}^{q_2}}^\dagger Z_{\eta_{q_2}}^\dagger,
    \label{eq:product-flux-step-one}
\end{align}
where
\begin{equation}
    \left(\eta_{q_1}\circ{}^{q_2}\right)(k):=\eta_{q_1}({}^{q_2}k).
\end{equation}
We define
\begin{equation}
    \chi(q_1,q_2)(k):=\frac{\eta_{q_1}({}^{q_2}k)\eta_{q_2}(k)}{\eta_{q_1q_2}(k_{12} k \overline{k_{12}})},
    \label{eq:chi-definition}
\end{equation}
where $k_{12}\equiv {}^{\overline{q_1 q_2}}n(q_1,q_2)$. Moreover, the group-action part of Eq.~\eqref{eq:product-flux-step-one} is simplified using the multiplication rule,
\begin{align}
    C_{q_1}C_{q_2}= C_{n(q_1,q_2)}C_{q_1 q_2}.
    \label{eq:C-factor-set}
\end{align}
Combining Eqs.~\eqref{eq:product-flux-step-one},
\eqref{eq:chi-definition}, and
\eqref{eq:C-factor-set}, we obtain
\begin{align}
    U_{\mathrm{TP}_f}^{(q_1)} U_{\mathrm{TP}_f}^{(q_2)}= C_{q_1q_2} Z_{\eta_{q_1 q_2}}^\dagger C_{k_{12}}Z_{\chi(q_1,q_2)}^\dagger =
    U_{\mathrm{TP}_f}^{(q_1q_2)} C_{k_{12}}Z_{\chi(q_1,q_2)}^\dagger.
    \label{eq:product-of-flux-pumps-with-residual-C}
\end{align}
Restoring the original notation, this is
\beq
    U_{\rm TP_f}^{(q_1)}U_{\rm TP_f}^{(q_2)} = U_{\rm TP_f}^{(q_1 q_2)}\cdot \prod_i \left(R^i_{^{\overline{q_1 q_2}}n(q_1,q_2)} L^{i+1}_{^{\overline{q_1 q_2}}n(q_1,q_2)}\right)\prod_i \left(Z_{\chi(q_1, q_2)}^{(i)}\right)^\dagger.
    \label{eq:product of U_F with RL}
\eeq
It can be shown using Eq.~\eqref{eq:Q-invariant cocycle} that $\chi(q_1, q_2)\in C^1(K,U(1))$ respects the following relation:
\beq
    \chi(q_1, q_2)(x_1 x_2) &= \frac{\eta_{q_1}({}^{q_2}(x_1 x_2))\eta_{q_2}(x_1 x_2)}{\eta_{q_1 q_2}(k_{12} x_1 x_2 \overline{k_{12}})}\\
    &= \frac{\eta_{q_1}({}^{q_2}x_1)\eta_{q_1}({}^{q_2}x_2)\big(\frac{\alpha({}^{q_2}x_1, {}^{q_2}x_2)}{\alpha(n(q_1,q_2){}^{q_1 q_2}x_1\overline{n(q_1,q_2)}, n(q_1,q_2){}^{q_1 q_2}x_2\overline{n(q_1,q_2)})}\big)\eta_{q_2}(x_1)\eta_{q_2}(x_2)\big(\frac{\alpha(x_1, x_2)}{\alpha({}^{q_2}x_1, {}^{q_2}x_2)}\big)}{\eta_{q_1 q_2}(k_{12} x_1 \overline{k_{12}})\eta_{q_1 q_2}(k_{12} x_2 \overline{k_{12}})\big(\frac{\alpha(k_{12} x_1 \overline{k_{12}}, k_{12} x_2 \overline{k_{12}})}{\alpha(n(q_1,q_2){}^{q_1 q_2}x_1\overline{n(q_1,q_2)}, n(q_1,q_2){}^{q_1 q_2}x_2\overline{n(q_1,q_2)})}\big)}\\
    &=\frac{\eta_{q_1}({}^{q_2}x_1)\eta_{q_2}(x_1)}{\eta_{q_1 q_2}(k_{12} x_1 \overline{k_{12}})}\cdot\frac{\eta_{q_1}({}^{q_2}x_2)\eta_{q_2}(x_2)}{\eta_{q_1 q_2}(k_{12} x_2 \overline{k_{12}})}\cdot \frac{\alpha(x_1,x_2)}{\alpha(k_{12} x_1 \overline{k_{12}},k_{12} x_2 \overline{k_{12}})}\\
    &= \chi(q_1, q_2)(x_1) \cdot \chi(q_1, q_2)(x_2)\cdot \frac{\alpha(x_1,x_2)}{\alpha(k_{12} x_1 \overline{k_{12}},k_{12} x_2 \overline{k_{12}})}.
    \label{eq: chi multiplication}
\eeq

From the Hamiltonian in Eq.~\eqref{eq:hamiltonian K alpha} one notes that the ground state $|\phi\rangle$ is locally stabilized by the product of a diagonal-phase gate and $R_k^iL_k^{i+1}$ such that,
\beq
    R^i_k L^{i+1}_k\ket{\phi} = \left(\sum_{k',k''}\frac{\alpha(k',k)}{\alpha(k,\overline{k}k'')}\left(\ket{k'}\bra{k'}\right)_{i}\otimes \left(\ket{k''}\bra{k''}\right)_{i+1}\right)\ket{\phi}.
\eeq
Hence,
\beq
    \prod_i R^i_k L^{i+1}_k\ket{\phi} = \cdots R^{i-1}_k L^{i}_k R^i_k L^{i+1}_k \cdots\ket{\phi} = \prod_i Z_{\gamma_k}^{(i)}\ket{\phi}, \gamma_k (x)\equiv \frac{\alpha(x,k)}{\alpha(k, k^{-1}xk)}. 
\eeq
In addition, it can be shown from the cocycle condition of $\alpha$ that
\beq
    \frac{\gamma_k(x_1 x_2)}{\gamma_k(x_1)\gamma_k (x_2)}
    &= \frac{\alpha(k, k^{-1}x_1 k)\alpha(k, k^{-1}x_2 k)\alpha(x_1 x_2,k)}{\alpha(x_1,k) \alpha(x_2,k)\alpha(k, k^{-1}x_1 x_2 k)}\\
    &=\frac{\cancel{\alpha(k, k^{-1}x_1 k)}\alpha(k, k^{-1}x_2 k)\alpha(x_1 x_2,k)}{\alpha(x_1,k) \alpha(x_2,k)\cancel{\alpha(k, k^{-1}x_1 x_2 k)}}\cdot\frac{\alpha(k^{-1}x_1 k,k^{-1}x_2 k)\cancel{\alpha(k, k^{-1}x_1 x_2 k)}}{\alpha(x_1 k,k^{-1}x_2 k)\cancel{\alpha(k,k^{-1}x_1 k)}}\\
    &=\frac{\alpha(k, k^{-1}x_2 k)\cancel{\alpha(x_1 x_2,k)}}{\alpha(x_1,k) \cancel{\alpha(x_2,k)}}\frac{\alpha(k^{-1}x_1 k,k^{-1}x_2 k)}{\alpha(x_1 k,k^{-1}x_2 k)}\cdot \frac{\cancel{\alpha(x_2, k)}\alpha(x_1, x_2 k)}{\cancel{\alpha(x_1 x_2, k)}\alpha(x_1, x_2)}\\
    &=\frac{\cancel{\alpha(k, k^{-1}x_2 k)}}{\cancel{\alpha(x_1,k)} }\frac{\alpha(k^{-1}x_1 k,k^{-1}x_2 k)}{\cancel{\alpha(x_1 k,k^{-1}x_2 k)}}\frac{\cancel{\alpha(x_1, x_2 k)}}{\alpha(x_1, x_2)}\cdot\frac{\cancel{\alpha(x_1 k,k^{-1}x_2 k)}\cancel{\alpha(x_1,k)}}{\cancel{\alpha(k,k^{-1}x_2 k)}\cancel{\alpha(x_1,x_2 k)}}\\
    & = \frac{\alpha(k^{-1}x_1 k, k^{-1}x_2 k)}{\alpha(x_1, x_2)}.
\eeq
Combining Eq.~\eqref{eq: chi multiplication} and the above equation with $k=k_{12}$, we conclude that
\beq
    (\overline{\chi}\cdot \gamma)(q_1,q_2) (x)\equiv \chi(q_1,q_2)^{-1}(x)\gamma_{k_{12}}(q_1,q_2)(k_{12}x\overline{k_{12}})
\eeq
is a one-dimensional representation of $K$ for any $q_1,q_2\in Q$. We note that in the particular case where $K$ is Abelian, both $\chi(q_1,q_2)$ and $\gamma_{k_{12}}$ are one-dimensional representations of $K$.

\section{Generalized quantum double for Hopf algebra}
\label{app:generalized qd}
In Ref.~\cite{jia2023boundary}, it is shown that the string-net model with Dirichlet/forgetful boundary can be mapped from a generalized quantum double with smooth/rough boundary, while the boundary that corresponds to the forgetful fiber functor of Rep($H$) in the string-net model can be mapped from the rough boundary of the generalized quantum double. Therefore, for the lattice model that maps to the Rep($H$) anyonic chain model in Eq.~\eqref{eq:anyonic chain} with the forgetful fiber functor on the top boundary, the basis states are given as follows, where we assign an element $a_j,b_j\in H$ to each edge,
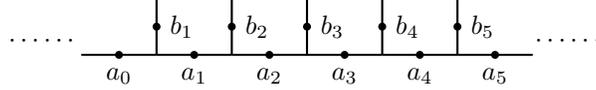
\begin{align*}
    \begin{tikzpicture}[thick,scale=0.5, every node/.style={scale=1.0}
  ]
  \def\gap{2}     
  \def\h{1.5}     
  \foreach \i in {1,2,3,4,5} {
    \pgfmathsetmacro\x{\i*\gap}
    \draw[thick] (\x,0) -- (\x,\h);
    \node[right] at (\x+0.1,0.5*\h) {$b_{\i}$};
    \fill[black] (\x,0.5*\h) circle (3pt);
  }
  \foreach \i in {0,1,2,3,4} {
    \pgfmathsetmacro\xA{\i*\gap}
    \pgfmathsetmacro\xB{\xA + \gap}
    \draw[thick] (\xA,0) -- (\xB,0);
    \fill[black] ({(\xA+\xB)/2},0) circle (3pt);
    \node[below] at ({(\xA+\xB)/2},-0.1) {$a_{\i}$};
  }
  \fill[black] (5.5*\gap,0) circle (3pt);
  \node[below] at (5.5*\gap,-0.1) {$a_5$};
  \draw[thick] (5*\gap,0) -- (6*\gap,0);
  \node at (6.5*\gap,0.25*\h) {$\cdots\cdots$};
  \node at (-0.5*\gap,0.25*\h) {$\cdots\cdots$};
\end{tikzpicture}
\end{align*}
The Hamiltonian is formed by the mutually commuting projector terms
\beq
    H^{\text{GQD}}=-\sum_i A_i^{\text{GQD}} - \sum_{i} B_i^{\text{GQD}}.
\eeq
To write the terms in the Hamiltonian, we introduce Sweedler's notation for iterated comultiplication as follows,
\beq
    \Delta^{(n)}(a):=\underbrace{\Delta(\cdots \Delta}_{n}(a)\cdots)=\sum_{(a)}a^{(1)}\otimes a^{(2)}\otimes \cdots\otimes a^{(n+1)}.
\eeq
The plaquette terms are given by
\begin{align}
&B_i^{\text{GQD}} \ket{\begin{tikzpicture}[baseline={([yshift=-.4ex]current bounding box.center)},thick,scale=0.6, every node/.style={scale=1.0}]
  \def\h{1.5}     
    \fill[black] (0.5*\h,0) circle (3pt); 
    \fill[black] (0,0.5*\h) circle (3pt); 
    \fill[black] (\h,0.5*\h) circle (3pt); 
    \draw[thick] (0,0) -- (0,\h);  
    \node[left] at (0-0.1,0.5*\h) {$b_i$};
    \draw[thick] (\h,0) -- (\h,\h);  
    \node[right] at (\h+0.1,0.5*\h) {$b_{i+1}$};
  \draw[thick] (0,0) -- (\h,0); 
    \node[below] at (\h/2,-0.1) {$a_i$};
  \draw[thick] (-\h/2,0) -- (0,0);  
  \draw[thick] (\h,0) -- (\h+\h/2,0);  
\end{tikzpicture} }
=\sum \phi(S(b_i^{(2)}) a_i^{(2)} b_{i+1}^{(2)})
\ket{\begin{tikzpicture}[baseline={([yshift=-.4ex]current bounding box.center)},thick,scale=0.6, every node/.style={scale=1.0}]
  \def\h{1.5}     
    \fill[black] (0.5*\h,0) circle (3pt); 
    \fill[black] (0,0.5*\h) circle (3pt); 
    \fill[black] (\h,0.5*\h) circle (3pt); 
    \draw[thick] (0,0) -- (0,\h);  
    \node[left] at (0-0.1,0.5*\h) {$b_i^{(1)}$};
    \draw[thick] (\h,0) -- (\h,\h);  
    \node[right] at (\h+0.1,0.5*\h) {$b_{i+1}^{(1)}$};
  \draw[thick] (0,0) -- (\h,0); 
    \node[below] at (\h/2,-0.1) {$a_i^{(1)}$};
  \draw[thick] (-\h/2,0) -- (0,0);  
  \draw[thick] (\h,0) -- (\h+\h/2,0);  
\end{tikzpicture} }.
\end{align}
The vertex terms are given by
\begin{align}
A^{\text{GQD}}_{i}
\ket{\begin{tikzpicture}[baseline={([yshift=-.5ex]current bounding box.center)},thick,scale=0.6, every node/.style={scale=1.0},
  arrow/.style={
    line width=1.2pt, 
    postaction={decorate},
    decoration={markings, mark=at position 0.6 with {\arrow{latex}}}
  }
  ]
  \def\gap{2}     
  \def\h{1.5}     
    \fill[black] (0.5*\h,0) circle (3pt); 
    \fill[black] (-0.5*\h,0) circle (3pt); 
    \fill[black] (0,0.5*\h) circle (3pt); 
    \draw[thick] (0,0) -- (0,\h);  
    \node[right] at (0+0.1,0.5*\h) {$b_i$};  
  \draw[thick] (0,0) -- (\h,0); 
    \node[below] at (\h/2,-0.1) {$a_i$};
    \draw[thick] (-\h,0) -- (0,0);  
    \node[below] at (-\h/2,-0.1) {$a_{i-1}$};
\end{tikzpicture} }
=\sum\ket{\begin{tikzpicture}[baseline={([yshift=-.5ex]current bounding box.center)},thick,scale=0.6, every node/.style={scale=1.0}]
  \def\h{1.5}     
    \fill[black] (0.5*\h,0) circle (3pt); 
    \fill[black] (-0.5*\h,0) circle (3pt); 
    \fill[black] (0,0.5*\h) circle (3pt); 
    \draw[thick] (0,0) -- (0,\h);  
    \node[right] at (0+0.1,0.5*\h) {$h^{(3)} b_i$};  
  \draw[thick] (0,0) -- (\h,0); 
    \node[below] at (\h/2+0.4,-0.1) {$h^{(2)} a_i $};
    \draw[thick] (-\h,0) -- (0,0);  
    \node[below] at (-\h/2-0.4,-0.1) {$a_{i-1}S(h^{(1)})$};
\end{tikzpicture} },
\end{align}
where $h^{(i)}$ are obtained from the iterated comultiplication of the Haar integral $h$ as explained above. Similar to the last section, the symmetry operators are given by the MPO supported on the horizontal edges, and the Thouless pump operators are given by ribbon operators supported on the vertical edges,
\beq
    U_F^{(k)} = \prod_i R_k^{\text{vertical }(i)},
\eeq
where $R_k^{\text{vertical }(i)}\equiv \sum_{b_i}\ket{b_i S(k)}_{\text{vertical }(i)}\bra{b_i}$ for each grouplike element $k\in H$. The set of grouplike elements
\beq
    \mathcal{G}(H) = \{a\in H\backslash \{0\}|\Delta(a) = a\otimes a\}
\eeq
hence classifies distinct classes of Thouless pumps of the Rep($H$) SPT states. Similar to the Rep($G$) symmetry, a minimal lattice model can be obtained by disentangling the vertical degrees of freedom~\cite{jia2024generalized}. 

\end{widetext}

\bibliography{ref}

\end{document}
%